\documentclass[12pt]{article}
\usepackage{latexsym}
\usepackage{epsf}
\usepackage{amssymb}
\input{epsf}
\def\hybrid{\topmargin -20pt    \oddsidemargin 0pt
        \headheight 0pt \headsep 0pt
        \textwidth 6.35in
        \textheight 9.65in
        \marginparwidth .875in
        \parskip 5pt plus 1pt   \jot = 1.5ex}

\catcode`\@=11
\newcount\hour
\newcount\minute
\newtoks\amorpm
\hour=\time\divide\hour by60
\minute=\time{\multiply\hour by60 \global\advance\minute by-\hour}
\edef\standardtime{{\ifnum\hour<12 \global\amorpm={am}%
        \else\global\amorpm={pm}\advance\hour by-12 \fi
        \ifnum\hour=0 \hour=12 \fi
        \number\hour:\ifnum\minute<10 0\fi\number\minute\the\amorpm}}
\edef\militarytime{\number\hour:\ifnum\minute<10 0\fi\number\minute}

\def\marginnote#1{}
\def\draftlabel#1{{\@bsphack\if@filesw {\let\thepage\relax
   \xdef\@gtempa{\write\@auxout{\string
      \newlabel{#1}{{\@currentlabel}{\thepage}}}}}\@gtempa
   \if@nobreak \ifvmode\nobreak\fi\fi\fi\@esphack}
        \gdef\@eqnlabel{#1}}
\def\@eqnlabel{}
\def\@vacuum{}
\def\draftmarginnote#1{\marginpar{\raggedright\scriptsize\tt#1}}

\def\draft{\oddsidemargin -.2truein
        \def\@oddfoot{\sl preliminary draft \hfil
        \rm\thepage\hfil\sl\today\quad\militarytime}
        \let\@evenfoot\@oddfoot \overfullrule 3pt
        \let\label=\draftlabel
        \let\marginnote=\draftmarginnote
   \def\@eqnnum{(\theequation)\rlap{\kern\marginparsep\tt\@eqnlabel}%
\global\let\@eqnlabel\@vacuum}  }
\def\preprint{\twocolumn\sloppy\flushbottom\parindent 2em
        \leftmargini 2em\leftmarginv .5em\leftmarginvi .5em
        \oddsidemargin -.5in    \evensidemargin -.5in
        \columnsep .4in \footheight 0pt
        \textwidth 10.in        \topmargin  -.4in
        \headheight 12pt \topskip .4in
        \textheight 6.9in \footskip 0pt
        \def\@oddhead{\thepage\hfil\addtocounter{page}{1}\thepage}
        \let\@evenhead\@oddhead \def\@oddfoot{} \def\@evenfoot{} }
\def\numberbysection{\@addtoreset{equation}{section}
        \def\theequation{\thesection.\arabic{equation}}}
\def\underline#1{\relax\ifmmode\@@underline#1\else
        $\@@underline{\hbox{#1}}$\relax\fi}
\def\titlepage{\@restonecolfalse\if@twocolumn\@restonecoltrue
\onecolumn
     \else \newpage \fi \thispagestyle{empty}\c@page\z@
        \def\thefootnote{\fnsymbol{footnote}} }
\def\endtitlepage{\if@restonecol\twocolumn \else \newpage \fi
        \def\thefootnote{\arabic{footnote}}
        \setcounter{footnote}{0}}  
\catcode`@=12
\relax
\hybrid
\makeatletter
\@addtoreset{equation}{section}
\makeatother
\def\lefthook{{\vrule height5pt width0.4pt depth0pt}}
\def\righthook{{\vrule height5pt width0.4pt depth0pt}}
\def\leftrighthookfill{$\mathsurround=0pt \mathord\lefthook
     \hrulefill\mathord\righthook$}
\def\underhook#1{\vtop{\ialign{##\crcr$\hfil\displaystyle{#1}\hfil$\crcr
      \noalign{\kern-1pt\nointerlineskip\vskip2pt}
      \leftrighthookfill\crcr}}}

\def\SING#1{\underhook{#1}}
\begin{document}
\def\be{\begin{equation}}
\def\ee{\end{equation}}
\def\bea{\begin{eqnarray}}
\def\eea{\end{eqnarray}}
\def\tfrac#1#2{{\textstyle{#1\over #2}}}
\def\r#1{{(\ref{#1})}}
\def\half{\tfrac{1}{2}}
\def\quart{\tfrac{1}{4}}
\def\nn{\nonumber}
\def\na{\nabla}
\def\ms{M_{\rm string}}
\def\y{\'{\i}}
\def\gs{g_{\rm string}}
\def\pd{\partial}
\def\a{\alpha}
\def\b{\beta}
\def\bz{\bar{z}}
\def\g{\gamma}
\def\d{\delta}
\def\m{\mu}
\def\n{\nu}
\def\t{\tau}
\def\l{\lambda}
\def\mh{\hat\m}
\def\nh{\hat\n}
\def\rh{\hat\rho}
\def\th{\vartheta}
\def\s{\sigma}
\def\e{\epsilon}
\def\ap{\alpha'}
\def\al{\alpha}
\font\mybb=msbm10 at 12pt
\def\bb#1{\hbox{\mybb#1}}
\def\Z{\bb{Z}}
\def\R{\bb{R}}
\def\Z{\bb{Z}}
\def\R{\bb{R}}
\def\C{\bb{C}}
\def\square{\hbox{{$\sqcup$}\llap{$\sqcap$}}}
\def\dslash{{\partial\hspace{-7pt}/}}
\hyphenation{re-pa-ra-me-tri-za-tion}
\hyphenation{trans-for-ma-tions}
%
%
\begin{titlepage}
\begin{center}
\hfill IFT-UAM/CSIC-98-2\\
\vskip 2cm
{\Large\bf String Primer }
\vskip .8in
{\bf Enrique Alvarez}\footnote{e-mail: enrique.alvarez@uam.es}
\hspace{.3cm}{\bf and}\hspace{.3cm}
{\bf Patrick Meessen}\footnote{e-mail: patrick.meessen@uam.es}
\vskip .1in
{\em  Instituto de F\'{\i}sica Te\'orica, C-XVI\\
Universidad Aut\'onoma, 28049 Madrid, Spain}
\end{center}
\vskip 1in
\begin{abstract}
This is the written version of a set of introductory lectures to string
theory.\footnote{The lectures were given at 
the Universidad Aut\'onoma de Madrid in the semester 1997/98 and at
the {\it VI Escuela de Oto\~no de F\'{\i}sica Te\'orica},
held in Santiago de Compostela (10-23 september 1998).}
\end{abstract}
\vskip 5cm
\begin{flushleft}
IFT-UAM/CSIC-98-2\\
September 1998\\
\end{flushleft}
\end{titlepage}

\setcounter{page}{1}
\setcounter{footnote}{1}
\renewcommand{\theequation}{\thesection.\arabic{equation}}

\tableofcontents

\section{Motivation}
Strings, in a broad sense, is a topic studied by a sizeable 
fraction of the particle physics community since the mid-eighties. 
In this interval
it has gotten the reputation, among some, of belonging to the limbo of 
unfalsifiable theories, sharing this place with Inflation, Quantum Gravity
{\it et cetera}.
\par
To date it is difficult to argue that phenomenological predictions are 
around the corner
and it is fair to say that there does not yet appear any physically appealing
guiding principle (something similar to the equivalence principle 
and general covariance in General
Relativity) in the new developments.
\par
And yet, the new developments are fascinating. 
There is a renewed (and deeper) sense 
in which it can
be claimed that all five string theories are manifestations of some unique
{\em M-theory}, described at long wavelengths by 11-dimensional supergravity.
Conformal Field Theory, and String Perturbation Theory are now stored waiting 
for a corner of parameter space in which they could make useful physical 
predictions.
In a sense, the situation has some similarities with the late seventies, when 
the non-perturbative structure of the QCD vacuum started to being appreciated.
In strings, non-perturbative effects are known to be important (in particular,
all sorts of 
extended objects spanning p spacelike dimensions, {\em p-branes}), and plenty of
astonishing consistency checks can be made, without meeting any clear 
contradiction
(so far). To the already seemingly miraculous correlation between world-sheet 
and spacetime phenomena, one has to add, no less surprising, interrelations
between physics on the world-volume of a D(irichlet)-brane 
(described by supersymmetric 
Yang-Mills) and physics on the bulk of spacetime (including gravity).
\par
Many properties of supersymmetric {\em field} theories can easily be understood
by engineering appropriate brane configurations. Also, the classical string 
relationship {\em closed} = {\em open} $\times$ {\em open} seems to be valid,
at least for S-matrix elements, also for field theory, in the sense that
{\em gravity} = {\em gauge} $\times$ {\em gauge} \cite{bern}.
\par
The  implementation of the Montonen-Olive conjecture by 
Seiberg and Witten in theories with only N=2 supersymmetry
led to the first concrete Ansatz embodying confinement
in field theory to date.
\par
Unfortunately, in many aspects the situation is even worse than in QCD.
The structure of the vacuum and the symmetries of the theory
are still unknown.
The status of p-branes with respect to quantum mechanics is still unclear for
$ p > 1$. That is, it is not known whether membranes and higher branes
are fundamental objects to be quantized, or only passive topological defects
on which strings (corresponding to $p=1$) can end. Besides, the amount
of physical observables which can be computed
has not increased much with respect to the pre-duality period.
\par
Still, it can be said, paraphrasing Warren Siegel \cite{siegel}, that this is 
{\em the best time 
for someone to read a book on the topic and the worst time for someone to write 
one}. (He presumably meant it to encourage people to work in
open topics such as this one). 
The aim of these lectures (written under duress) is quite modest: 
To whet
the appetite of some students for these matters,
and to direct them to the study
of the original papers, or at least, to books and reviews written by the authors
who made the most important contributions, 
many of them cited in the
bibliography \cite{lag,gh,GSW,LT,peskin,joepboek,polya,sen,siegel,vv}.
\section{Maximal supergravity, p-branes and electric/magnetic duality 
for extended objects}
It has been emphasized many times before why supersymmetry is a fascinating 
possibility. Besides being the biggest possible symmetry of the S-matrix
(the Haag-{\L}opuszanski-Sohnius theorem),
it can solve many phenomenological {\em naturalness}
problems on the road to unification, and, at the very least, provide
very simple ({\it i.e.} finite) quantum field theories (the analogue of the
harmonic oscillator in quantum mechanics) from which more elaborate examples
could hopefully be understood.
\par
Supergravities in all possible spacetime dimensions have been classified by
Nahm \cite{nahm}. The highest dimension in which it is possible to build
an action with highest spin two\footnote{It is not known how 
to write down consistent actions
with a finite number of fields containing spin 5/2 and higher.} 
is (N=1 supergravity in) 
d=11, and this was done in a classic
paper by Cremmer, Julia and Scherk \cite{cjs}. 
Upon (toroidal) dimensional reduction
this theory leads to N=8 supergravity in d=4, giving in the process a set of 
theories in different dimensions with 32 (real) supercharges.
\par
Giving the fact that this is, in a sense, the most symmetric of all 
possible theories
we can write down, let us examine the hypothesis that it is also 
the most fundamental,
in a sense still to be clarified.
\subsection{$N=1$ supergravity in 11 dimensions}
The action can be written as
\bea
&&S = 
\int d^{11}x \left\{ -\frac{e}{4\kappa^2} R(\omega) - 
i \frac{e}{2}\bar{\psi}_M \Gamma^{MNP}D_N(\frac{\omega + 
\hat{\omega}}{2}) \psi_P \right. \nn\\
&&-\frac{e}{48} F_{MNPQ}F^{MNPQ} +
\frac{2\kappa}{(144)^2}\epsilon^{A_1 \ldots A_{11}} 
   F_{A_1 \ldots A_4} F_{A_5 \ldots
A_8} A_{A_9 \ldots A_{11}} \nn\\
&&+\left. \frac{\kappa e}{192}\left( \bar{\psi}_{A_1 }
     \Gamma^{A_1 \ldots A_6} \psi_{A_2}
     +12 \bar{\psi}^{A_3}\Gamma^{A_4 A_5} \psi^{A_6}\right)
       \left( F_{A_3 \ldots A_6} + 
         \hat{F}_{A_3 \ldots A_6}\right)
      \right\}
\eea
Here $e$ is the determinant of the {\em Elfbein} representing the graviton
(with zero mass dimension); $\psi_M$ represents the gravitino (of mass 
dimension 5), taken as a $C_{-}$
Majorana \footnote{We define, following \cite{pvn},
$C_{\pm}$ Majorana spinors as those obeying $\psi^{T} C_{\pm} = \alpha
\psi^{+}
\gamma_0$, with $\gamma_{\mu}^{T} = \pm C_{\pm} \gamma_{\mu} C_{\pm}^{- 1}$,
and $\a $ is a phase.}
Rarita-Schwinger vector-spinor
(A Majorana spinor in $D=11$ has 32 real components); and $A_{MNP}$ is a 
(mass dimension $\frac{9}{2}$) three-form
field, a kind of three-index Maxwell field. 
\par
The Lorentz connection is given in terms of the Ricci rotation coefficients
and the contorsion tensor as
\be
\omega_{M a b} = \omega^{Ricci}_{M a b} + K_{M a b} \; ,
\ee
and the contorsion tensor itself is given by
\be
K_{M a b} = \frac{i \kappa^2 }{4} \left( - \bar{\psi}_{\alpha}
\gamma_{M a b}^{\alpha \beta}
\psi_{\beta} + 2( \bar{\psi}_{M} \gamma_b \psi_a - 
\bar{\psi}_{M} \gamma_a \psi_b + \bar{\psi}_b \gamma_{\m} \psi_a )
\right) \; .
\ee
The {\em supercovariant} connection and field strength are given by
\be
\hat{\omega}_{M a b}\equiv \omega_{M a b} + 
\frac{i \kappa^2 }{4} \bar{\psi}_{\alpha}\gamma_{M a b}^{\alpha\beta} 
\psi_{\beta} \; ,
\ee
and
\be
\hat{F}_{MNPQ} = F_{MNPQ}  - 3 \kappa \bar{\psi}_{[ M} 
\gamma_{NP} \psi_{Q ]} \; .
\ee
On shell, the graviton corresponds
to the (2,0,0,0) representation of the little group SO(9), with 44 real states;
the three-form to the (1,1,1,0) of SO(9), with 84 real polarization states; and, 
finally, the gravitino lives in the (1,0,0,1), yielding 128 polarizations which
matches the bosonic degrees of freedom.
\par
The Chern-Simons-like coupling in the preceding action suggests a 12-dimensional
origin but, in spite of many attempts, there is no clear understanding 
of how this could come about.
\par
There are a couple of further remarkable properties of this theory (stressed,
in particular by Deser \cite{deser}). First of all, there is no
globally supersymmetric matter (with highest spin less than 2), which means 
that there are no sources. Furthermore, it is the only theory which 
forbids a cosmological constant because of a symmetry ({\it i.e.} it is not 
possible to extend the theory to an Anti-de Sitter background, 
although this is an active field of research).
Let us now concentrate on the three-form $A_{(3)} \equiv \frac{1}{3!}
A_{MNP}
dx^{M}\wedge dx^{N}\wedge dx^{P}$. From this point of view, the Maxwell 
field
is a one-form $A_{(1)} \equiv A_{M}d x^{M}$, which couples minimally 
to a point
particle through 
\be
e \int_{\gamma} A \; ,
\ee
where the integral is computed over the trajectory 
$\gamma : x^{\mu}= x^{\mu}(s)$
of the particle. A Particle is a zero dimensional object, so that its
world-line has one dimension more, that is, it is a one-dimensional
world-line. It is very appealing to keep 
the essence of this coupling
in the general case, so that a general (p+1)-form would still couple in 
exactly the same
way as before, except that now $\gamma$ must be a 
(p+1)-dimensional region of
spacetime. If we want to interpret this region
as the world-volume of some object,
it would have to be a p-dimensional extended object, a {\em p-brane}.
\par
In this way we see that just by taking seriously the geometrical
principles of minimal coupling we are led to postulate the existence of
{\em two-branes} ({\em membranes}) naturally associated to the three-form
of supergravity. 
\par
On the other hand, as has been stressed repeteadly by Townsend \cite{paul},
the maximally extended (in the sense that it already has 528 (=
$\frac{32 \times 33}{2}$) 
algebraically independent charges, the maximal amount possible) 
supersymmetry algebra in d=11 is
\be
\{Q_{\alpha},Q_{\beta}\} = (C\Gamma^M ) P_M +(C\Gamma_{MN})_{\a\b} Z_{(2)}^{MN}
+(C\Gamma_{M_1 \ldots M_5})_{\a\b}Z_{(5)}^{M_1 \ldots M_5} \; .
\ee
Clearly the first term on the r.h.s. would be associated to the graviton,
the second one to the membrane, and the last one to the fivebrane.
\par
This fact (given our present 
inability to quantize branes in a consistent way) in turn
suggests that 11-dimensional supergravity can only be, at best, the long
wavelength limit of a more fundamental theory, dubbed {\em M-Theory}.
We shall return to this point later on.
\subsection{The Dirac monopole}
Many of the properties of charged extended objects are already
visible in the simplest of them all: Dirac's magnetic monopole in ordinary
four dimensional Maxwell theory ({\it cf.} \cite{dirac,go}).
Although Dirac's magnetic monopole is pointlike, we shall see
that one needs to introduce an extended object in order to have 
a gauge-invariant description of it.
\par
We {\em assume} that there is a pointlike magnetic monopole, with magnetic 
field given by
\be
\vec{B}_m \equiv \frac{g}{4\pi r^2}\hat{r}
\ee
($\hat{r}\equiv \frac{\vec{r}}{r}$). In quantum mechanics, 
minimal coupling demands the existence of a vector potential
$\vec{A}$ such that $\vec{B}_m = \vec{\nabla} \times \vec{A}_m$.
Unfortunately, this is only possible when $\vec{\nabla}\cdot\vec{B}_m = 0$, 
which is not 
the case, but rather $\vec{\nabla}\cdot\vec{B}_m = g \delta^3 (x)$.
Dirac's way out was to introduce a {\em string}, (along the negative z-axis, 
although its position is a gauge-dependent concept) with magnetic field
$\vec{B}_s = g \theta(- z)\delta(x)\delta(y) \hat{z}$, such that the total 
magnetic field $\vec{B}_m + \vec{B}_s$ is divergence-free.
\par
It is quite easy to compute the vector potential of the monopole, $\vec{A}_m$.
The flux through the piece of the unit sphere with polar angle,
parametrized by $\theta^{\prime}$ say,
smaller than $\theta$, which
will be called $S_{(+)}$ is given by (using Stokes' theorem and the spherically 
symmetric {\em Ansatz}, $\vec{A}_m = A(r, \theta ) \,\partial_{\phi}$  ),
\be
\Phi(S_{(+)}) = \int_{S_{(+)}}\vec{B_m}.d\vec{S} = 
\int_{C\equiv \partial S_{(+)}}\vec{A}_m . d\vec{l} = 
2\pi A r^2 \sin^2 \theta \; .
\ee
On the other hand, knowing that the total flux through the sphere is $4\pi g$,
we could write $\Phi(S_{(+)})$ as the solid angle subtended by $S_{(+)}$,
\be
\Phi(S_{(+)}) = \frac{g}{4\pi}\Omega(S_{(+)}) = \frac{g}{4\pi}\int_{0}^{\theta}
d\theta'\sin \theta' d\phi' = \frac{g}{2} (1 - \cos \theta) \; .
\ee
This yields
\be
\vec{A}_m^{(+)} = \frac{g}{4\pi r^2 \sin^2 \theta} 
(1 - \cos \theta)\partial_{\phi} \; .
\ee
\begin{figure}[!ht]
\begin{center}
\leavevmode
\epsfxsize= 6cm
\epsffile{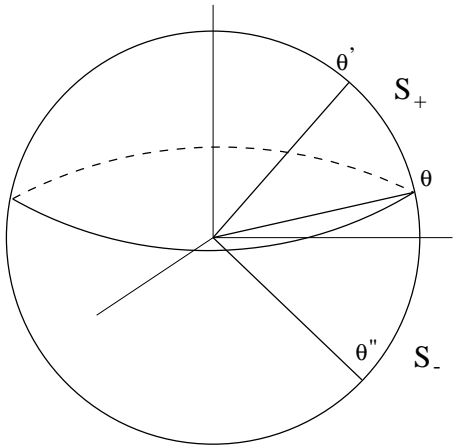}
\caption{Sphere surrounding a Dirac Monopole}
\label{Figure 1}
\end{center}
\end{figure}
There is a certain ambiguity because of
$\partial S_{(+)} = \partial
S_{(-)}$, where $S_{(-)}$ is the complementary piece of the unit sphere defined
by $\theta^{\prime\prime} > \theta$. Using Stokes theorem on the lower 
piece, the flux is given by
\be
\Phi_{(-)}(C) = (1 - \frac{\Omega_{(+)}}{4\pi})g = \frac{g}{2}(1 + \cos\theta)
= - \int_{\partial S_{(-)}}\vec{A_m}d\vec{l} \; ,
\ee
yielding 
\be
\vec{A}_m^{(-)} = - \frac{g}{4\pi r^2 \sin^2 \theta}
(1 + \cos\theta)\,\partial_{\phi} \; .
\ee
In both cases $\vec{\nabla}\times\vec{A}_m^{(+)} =
\vec{\nabla}\times \vec{A}_m^{(-)} =
\frac{g}{4\pi r^2} \hat{r}$.
The corresponding covariant vectors, expressed as one-forms, are
\be
A^{(\pm )}_m = \frac{g}{4\pi}\frac{1}{2r}\frac{x dy - y dx}{z \pm r} = 
\frac{g}{4\pi}(\pm 1 - \cos\theta ) d\phi \; .
\ee
In this language it is obvious that the two 
possible determinations of the gauge
potential of the monopole differ by a {\em gauge} transformation
\be
A^{(+)} - A^{(-)} = d\Lambda \equiv - \frac{g}{2\pi}d\phi \; .
\ee
Had we included the string in the computation, the $A^{(+)}$ would remain
unaffected, and $\vec{A}^{(-)}_{m-s} = \vec{A}^{(+)}_{m-s} 
= \vec{A}^{(+)}_m$,\footnote{It should be clear that this whole argument fails
at the origin} such
that the one-form potential assotiated to the string in the said configuration 
is the closed, but not exact one-form $A_s \equiv \frac{g}{2\pi}d\phi$.
\par
Demanding that the gauge transformation connecting the two potentials
is single valued acting on fields minimally charged, that is 
$e^{i e \Lambda(\phi = 0)} =e^{i e \Lambda(\phi = 2\pi)}$ imposes Dirac's
quantization condition
\be
\frac{e g}{2\pi} \in\mathbb{Z} \; .
\ee
\subsection{Extended poles}
%
%
The only purely geometrical action for a ($p-1$)-brane with a classical
trajectory, parametrized by 
\begin{equation}
X^{\mu}\;=\; X^{\mu}\left( \xi^{0},\ldots \xi^{p-1}\right) \; ,
\end{equation}
is the $p$-dimensional world-volume induced on the trajectory by the
external $d$ dimensional metric
\begin{equation}
{\cal S} = -T_{p}\, \int_{W_{p}}\; d\left( Vol\right) \; , 
\ee
where the Riemannian volume element is given in terms of the determinant of the
world-volume metric $h\equiv \det \left( h_{ij}\right) $, by
\be
d\left( Vol\right) = d\xi^{p-1}\wedge\ldots\wedge d\xi^{0}
                        \, \sqrt{| h|} \; .
\ee
The metric on the world-volume is the one induced from the spacetime metric
by the imbedding itself, namely
\be
h_{ij} = \partial_{i}X^{\mu}\, \partial_{j}X^{\nu}\;
           g_{\mu\nu}\left( X\right) \; .
\label{eq:imbedmetr}
\end{equation}
Classically, this is equivalent to the Polyakov-type action
\begin{equation}
{\cal S}\;=\; T_{p}\int_{W_{p}}d^{p}\xi\, \sqrt{| h|}\left[
-\textstyle{\frac{1}{2}}h^{ij}
 \partial_{i}X^{\mu}\partial_{j}X^{\nu}g_{\mu\nu}(X)
 \;+\;\textstyle{\frac{1}{2}}\left( p-2\right)
\right]\; .
\end{equation}
\par
In this action the two-dimensional metric is now a dynamical field as well as
the imbeddings. On shell, the equations of motion 
for the two-dimensional metric
force it to be equal to the induced metric (\ref{eq:imbedmetr}), 
but off-shell this is not the case.
\par
Given an external $p$-form field $A_{p}$, there is a natural 
(``Wess-Zumino") coupling to the $(p-1)$-brane
\begin{equation}
S_{int} \;=\; e_{p}\int_{W_{p}}\; \tilde{A}_{p} \; ,
\end{equation}
where the induced form on the world-volume is given by
\begin{equation}
\tilde{A}_{p} \;=\; \textstyle{\frac{1}{p!}}
              A_{\mu_{1}\ldots\mu_{p}}\left( X\right)
              \partial_{i_{1}}X^{\mu_{1}}\ldots\partial_{i_{p}}X^{\mu_{p}}
              \; d\xi^{i_{1}}\wedge\ldots\wedge d\xi^{i_{p}} \; .
\end{equation}
%
%
In any {\em Abelian} theory of p-forms with gauge invariance
\be
A_{p}\rightarrow A_{p} + d \Lambda_{p-1} \; ,
\ee
the standard definition of   the field strength $ F_{p+1} \equiv d A_{p}$,
implies the {\em Bianchi identity},
to wit
\be
d F_{p+1} = 0 \; .
\ee
%
%
The minimal ``Maxwell" action for Abelian p-forms is
\be
{\cal S}\;=\; \int_{V_{d}}d\left( Vol\right)\; 
              \left( *F\right)_{d-p-1}\wedge F_{p+1} \; .
\ee
The equations of motion of the p-form itself can be written as
\be
\frac{\delta S}{\delta A_{p}} = d (*F)_{d-p-1} = (*J^{(e)})_{d-p} \; ,
\ee
where the source $J^{(e)}$ is a (p)-form with support in $W_{p}$,
the world-volume spanned by the (p-1)-brane,
%
%
\be 
J_{p}^{(e)}\;=\; e_{p}\left\{
\int_{W_{p}}\textstyle{\frac{1}{p!}}
  \partial_{i_{1}}X^{\mu_{1}}\ldots\partial_{i_{p}}X^{\mu_{p}}
  d\xi^{i_{1}}\ldots d\xi^{i_{p}}
\right\}\, dX_{\mu_{1}}\wedge\ldots\wedge dX_{\mu_{p}} \; . 
\ee
Electric charges are naturally
defined as a boundary contribution in the subspace orthogonal to
the world-volume, $M^{\perp}_{d-p}$ (in which all the history of the brane
is just a point),
\be
e_{p} \equiv \int_{S_{d-p-1} = \partial M^{\perp}_{d-p}} (*F)_{d-p-1}
= \int_{M^{\perp}_{d-p}} (*J^{(e)})_{d-p} \; .
\ee
This means that $(*J^{(e)})_{d-p}$ is a Dirac current with support in
$M^{\perp}_{d-p}$, with charge $e_{p}$.
\par
In ordinary Maxwell theory, 
\be
A_1 = \frac{e_{1}}{4\pi r} dt \; ,
\ee
so that
\be
F_2 \equiv dA_1 = \frac{e_{1}}{4\pi r^3} \sum_i x^i dt \wedge dx^i \; .
\ee
The Hodge dual
is given by 
\be
(*F)_2 = \frac{e_{1}}{4\pi r^3} \sum_{i,j,k}x^i 
\epsilon_{ijk} dx^j\wedge dx^k \; ,
\ee
and indeed 
\be
d(*F)_2 = e_{1} \delta^{(3)} (x) d(vol) \; ,
\ee
so that
\be
J^{(e)}_1 = e_{1} \delta^{(3)}(x) dt \; .
\ee 

The string is geometrically the place on which there fails to exist a 
potential
for $*F$, (when there is an electric source) because if we write
\be
(*F)_{d-p-1}\equiv d\tilde{A}_{d-p-2} +(*\tilde{S})_{d-p-1},
\ee
consistency with the equations of motion  demands that
\be
d(*\tilde{S})_{d-p-1} = (*J^{(e)})_{d-p} \; .
\ee
Given a $(p-1)$-brane, then, coupling to an $A_p$, the {\em dual brane},
coupling to $\tilde{A}_{d-p-2}$ will be a ($\tilde{p}\equiv d-p-3$)-brane.
{}For example, in $d=4$ the dual of a $0$-brane is again a $0$-brane.
In $d=11$, however, the dual of a $2$-brane is a $5$-brane.
\par
We would like to generalize Dirac's construction to this case. This would mean
introducing a magnetic source such that
\be
dF_{p+1} = J^{(m)}_{p+2} \; ,
\ee
(And we do not have now an electric source, so that $d*F = 0$.)
which is incompatible with the Bianchi identity, unless we change 
the definition of $F_{p+1}$.
In this way we are led to (re)define
\be
F_{p+1}\equiv dA_{p} + S_{p+1} \; ,
\ee
where the Dirac (hyper)string is an object, with a 
$(p+1)$-dimensional world-volume, such that
\be
dS_{p+1} = J^{(m)}_{p+2} \; .
\ee
Under these conditions the magnetic charge is defined as
\be
g_{d-p-2} = \int_{\partial M_{p+2}} F_{p+1} = \int_{M_{p+2}} 
J^{(m)}_{p+2} \; .
\ee
This means that $J^{(m)}_{p+2}$ is a Dirac current with support in $M_{p+2}$
and charge $g_{d-p-2}$.
\par
Writing the (free) action without any coupling constant, the form field has
mass dimension $[A_{p}] = \frac{d}{2} - 1$, which implies that $[e_{p}] 
= p+1-\frac{d}{2}$
, and $[g_{d-p-2}] = \frac{d}{2}-p-1$.
\par 
Demanding now that the (hyper)string could not be detected
in a B\"ohm-Aharanov 
experiment using a (d-p-3)-brane imposes that the phase factor it picks up 
when it moves around the string is trivial \cite{nt},
\be
\exp{i e_{p} \int F_{p+1}}=\exp{i e_{p}g_{d-p-2}} \; ,
\ee
{\it i.e.}
\be
e_{p}g_{d-p-2} \;\in\;  2 \pi \mathbb{Z} \; .
\ee
\subsection{Bogomol'nyi-Prasad-Sommerfeld states}
In extended SUSY it is possible for some central charges to enter
the commutation relations between the supercharges, as predicted
by the Haag-{\L}opuszanski-Sohnius theorem.
The supersymmetry relations in that case put a restriction on the
lowest value for the energy in terms of the eigenvalues of this central charge.
When this bound is saturated, the states are called {\em BPS states}, and they
are stable by supersymmetry. Also, those states form supersymmetry
multiplets of dimension lower than non-BPS states, the so-called
{\em short multiplets}. This means that most of their physical properties,
like masses, charges {\it etc.},
are protected from quantum corrections and can be
computed at lowest order in perturbation theory.
Physically, this bound
in the most important cases takes the form
of $ M \geq k Q$, where $M$ is the mass of the state, $k$ is a
parameter of order unity, and $Q$ is a geometric mean of the charges of
the said state.
\par
Let us illustrate all this with a very simple quantum mechanical
example due to Polchinski \cite{polchinski}.
We are given two charges, such that
the commutation relations read
\begin{eqnarray}
\left\{ {\cal Q}_{1},{\cal Q}_{1}^{\dagger}\right\} &=& H\,+\, Z \; , \\
\left\{ {\cal Q}_{2},{\cal Q}_{2}^{\dagger}\right\} &=& H\,-\, Z \; ,
\end{eqnarray}
where $Z$ is a central charge, {\it i.e.} it commutes with
all the elements of the SUSY algebra.
\par
There is a $4$-state representation in a given $(h,z)$ sector of the
four-dimensional Fock space
\be
\begin{array}{lclclcl}
{\cal Q}_{1}\mid 0\, 0\rangle &=& 0
  &\hspace{.3cm},\hspace{.3cm}&
{\cal Q}_{2}\mid 0\, 0\rangle &=& 0 \; ,\\
 & & & & & & \\
{\cal Q}^{\dagger}_{1}\mid 0\, 0\rangle &=& \lambda_{1}\mid 1\, 0\rangle
 &,&
{\cal Q}_{2}^{\dagger}\mid 0\, 0\rangle &=& \lambda_{2}\mid 0\, 1\rangle
\; .
\end{array}
\ee
and, of course,
\be
{\cal Q}_{2}^{\dagger}\mid 1\, 0\rangle = \lambda_{3}\mid 1\, 1\rangle
\ee
By hypothesis we see that
\begin{eqnarray}
h+z &=& \langle 0\, 0\mid {\cal Q}_{1}{\cal Q}_{1}^{\dagger}\,+\,
           {\cal Q}_{1}^{\dagger}{\cal Q}_{1} \mid 0\, 0\rangle
   \,=\, \langle 1\, 0\mid 1\, 0\rangle\, |\lambda_{1}|^{2} \; , \\
h-z &=& \langle 0\, 0\mid {\cal Q}_{2}{\cal Q}_{2}^{\dagger}\,+\,
           {\cal Q}_{2}^{\dagger}{\cal Q}_{2} \mid 0\, 0\rangle
   \,=\, \langle 0\, 1\mid 0\, 1\rangle\, |\lambda_{2}|^{2} \; , \\
h-z &=& \langle 1\, 0\mid {\cal Q}_{2}{\cal Q}_{2}^{\dagger}\,+\,
           {\cal Q}_{2}^{\dagger}{\cal Q}_{2} \mid 1\, 0\rangle
   \,=\, \langle 1\, 1\mid 1\, 1\rangle\, 
         |\frac{\lambda_{3}}{\lambda_1}|^{2} \; ,
\end{eqnarray}
and it should be obvious that $h\geq |z|$.
Note that when $h=|z|$, $\lambda_2 =\lambda_3 = 0$ and
we have a two-state representation, generated by $ \mid 0\, 0\rangle\,$ and
$\mid 1\, 0\rangle\,$, because $ Q_2^{\dagger}\mid 0\, 0\rangle\ = 0 =
Q_1^{\dagger}Q_2^{\dagger}\mid 0\, 0\rangle\,= - Q_2^{\dagger}\lambda_1
\mid 1\, 0\rangle\,$.
\par
The {\em number} of BPS states is a sort of topological
invariant, which does not change under smooth variations of the parameters
of the theory (like coupling constants). This fact is at the root of the
recent successes in counting the states corresponding to configurations
which are in a sense equivalent to extremal black holes (See section (6.3)).
\subsection{Brane surgery}
P. K. Townsend \cite{paul} has shown how to get information on 
intersections of branes 
({\it i.e.} which type of brane can end on a given brane)
by a careful examination of the Chern-Simons term in the action.
\par
In d = 11 the 2-brane carries an electric charge
\be
Q_2 = \int_{S^7} (*F)_7 \; ,
\ee
where $S^7$ is a sphere surrounding the brane in the 8-dimensional 
transverse space
(in which the brane is just a point).
\par
The analogous expression for the 5-brane is
\be
Q_5 = \int_{S^4} F_4 \; .
\ee
The Bianchi identity for $F$ is $dF = 0$, meaning that charged 5-branes 
must be closed (otherwise one could slide off the $S^4$ encircling 
the brane, and contract it to a point, the Bianchi identity guaranteeing that
the integral is an homotopy invariant).
\par
This argument does not apply to the 2-brane, however, owing to the presence of the
Chern-Simons term in the 11-dimensional supergravity action, which modifies the 
dual Bianchi identity to
\be
d*F = - F\wedge F \; .
\ee
This fact implies that the homotopy invariant charge is
\be
\tilde{Q}_2\equiv \int _{S^7} *F + F\wedge A \; .
\ee
If the 2-brane had a boundary, the last term could safely
be ignored as long as
the distance L from the boundary to the $S^7$ is much bigger than the
 radius R of the sphere itself. If now we slide the $S^7$, keeping $L/R$ large
as $L\rightarrow 0$, as $L=0$ the sphere collapses to 
the endpoint (which must be
assignated to a nonvanishing value of the Chern-Simons, if a contradiction is
 to be avoided). At this stage the sphere can be deformed to the product 
$S^4\times S^3$, in such a way that the contribution to the charge is
\be
\tilde{Q}_2 = \int_{S^4} F \int_{S^3} A \; .
\ee
The first integral is the charge $Q_5$ associated to a 5-brane.
Choosing also $F_{\parallel}=0$ (so that $A = d V_2$ in the second integral),
we would have
\be
\tilde{Q}_2 = Q_5 \int_{S^3} dV_2 \; ,
\ee
namely, the (magnetic) charge of the string boundary of the 2-brane in the 
5-brane.
\par
We have learned from the preceding analysis 
that in d=11 a 2-brane can end in a
5-brane, with a boundary being a 1-brane. 
A great wealth of information can be 
gathered by employing similar reasonings to 10-dimensional physics.
\subsection{Dyons, theta angle and the Witten effect}
There are allowed configurations with both electric and magnetic charge 
simultaneously, called {\em dyons}, whose charges  will be denoted by 
$(e,g)$. Given two of them, the only possible generalization of 
Dirac's quantization condition compatible
with electromagnetic duality, called the Dirac-Schwinger-Zwanziger
quantization condition, is \cite{dirac}
\be
e_1 g_2 - g_1 e_2 = 2\pi \mathbb{Z} \; .
\ee
\par
E. Witten \cite{teta} pointed out that in the presence of a theta term
in the Yang-Mills action, the electric charges in the monopole sector
are shifted.
There is a very simple argument by Coleman, which goes as follows:
The theta term in the Lagrangian can be written as
\be
-\frac{\theta e^2}{32 \pi^2} F^a_{\m\n}*F^{a\m\n} \; ,
\ee
which for an Abelian configuration reduces to
\be
\frac{\theta e^2}{8 \pi^2} \vec{E}\cdot\vec{B} \; .
\ee
In the presence of an magnetic monopole one can write
\bea
\vec{E} &=& \vec{\nabla}A_0 \; ,\nn\\
\vec{B} &=& \vec{\nabla} \times \vec{A} + 
             \frac{g}{4\pi} \frac{\vec{r}}{r^2} \; ,
\eea
which when used in the action, yields
\bea
\delta S &=& \frac{\theta e^2}{8\pi^2} \int d^3 x\vec{\nabla}A_0 
\left(
\vec{\nabla} \times \vec{A} + \frac{g}{4\pi} \frac{\vec{r}}{r^2}\nn
\right) \\
&=& - \frac{\theta e^2 g}{8\pi^2} \int d^3 x A_0 \delta^3 (x) \; .
\eea
This last term is nothing but the coupling of the scalar 
potential $A_0$ to an electric charge $-\frac{\theta e^2 g}{8\pi^2}$ at $x=0$.
This means that a minimal charge monopole\footnote{Note that 
due to the possibility of coupling the theory to fields 
having half-integer charges, $e/2$,
the Dirac quantization condition reads $eg =4\pi\mathbb{Z}$ \cite{harvey}.}
with $eg = 4\pi$
has an additional electric charge $-\frac{e\theta}{2\pi}$.
\par
All this means that the explicit general solution to the quantization condition
in the presence of a theta term is
\bea
Q_m &=& \frac{4\pi n_m}{e} \; ,\nn\\
Q_e &=& n_e e - \frac{e n_m \theta}{2\pi} \; .
\eea
\par
Montonen and Olive \cite{monol} proposed that in a non-Abelian gauge theory 
(specifically, in an $SO(3)$ Yang-Mills-Higgs theory) there should
exist (at least in the BPS limit) an {\em exact} duality between electric
and magnetic degrees of freedom.
\par
It was soon realized by Osborn \cite{art:osborn}
that, in order for this idea to have any chance
to be correct, supersymmetry was necessary, and the simplest candidate model was
$N=4$ supersymmetric Yang-Mills in 4 dimensions, with 
Lagrangian given by
\bea
L &=& -\frac{1}{4}Tr\left( F_{\m\n}F^{\m\n}\right) 
+i\lambda_i\sigma^{\m}D_{\m} \bar{\lambda}^i 
+\frac{1}{2}D_{\m}\Phi_{ij}D^{\m}\Phi^{ij} \nn\\
&& + i\lambda_i 
[\lambda_j ,\Phi^{ij}] + i \bar{\lambda}^{i}[\bar{\lambda}^j ,\Phi_{ij}] +
 \frac{1}{4}[\Phi_{ij},\Phi_{kl}][\Phi^{ij},\Phi^{kl}] \; .
\eea
where the gauginos are represented by 
four Weyl spinors $\lambda_i$, transforming in the $\mathbf{4}$ of $SO(6)$,
and the six scalar fiels $\Phi_{ij}$ obey 
$(\Phi_{ij})^{\dagger}\equiv \Phi^{ij} = 
\frac{1}{2}\epsilon^{ijkl}\Phi_{kl}$.
\par
By defining the parameter
\be
\tau=\frac{\theta}{2\pi} + \frac{4\pi i}{e^2} \; ,
\ee
where in our case $\theta = 0$ and the coupling constant 
has been absorbed in the definition of the gauge field,
electric-magnetic duality (dubbed {\em S-duality} in this context) would be
an $SL(2,\mathbb{Z})$ symmetry
\be
\tau\rightarrow \frac{a\tau + b}{c\tau+d} \; ,
\ee
where $a,b,c,d\in\mathbb{Z}$, $ad-bc=1$.
Please note that this is a {\em strong-weak} type of duality, because
the particular element $\tau\rightarrow - \frac{1}{\tau}$ transforms (when
$\theta =0$, for simplicity) $\a \equiv \frac{e^2}{4\pi}$ into $\frac{1}{\a}$.
\par
S-duality is believed to be an exact symmetry of the full quantum field theory.
In spite of the fact that this theory is conformal invariant ($\beta =0 $)
and is believed to be finite
in perturbation theory, the hard evidence in favour of this conjecture
is still mainly kinematical ({\it cf.} Vafa and Witten in \cite{vw}).
(Dynamical interactions between monopoles are notoriously difficult to study
beyond the simplest approximation using geodesics in
moduli space \cite{manton}).
\par
This S-Duality in Quantum Field Theory is closely related
to a corresponding symmetry
in String Theory, also believed to be exact for
toroidally compactified heterotic strings
as well as for ten dimensional Type $IIB$ Strings.
\subsection{Kappa-symmetry, conformal invariance and strings}
It is not yet known what is to be 
the fundamental symmetry of fundamental physics.
For all we know, however, both kappa symmetry and conformal invariance are
basic for the consistency of any model, specifically for spacetime supersymmetry
and for the absence of anomalies.
\subsubsection{Kappa-symmetry}
When considering {\em branes}, the fact that those p-branes are embedded
in an {\em external} spacetime $M_d$
\be
\xi^i \rightarrow X^{\m}(\xi^i) \; ,
\ee
where $i = 0\ldots p$ and $\m = 0\ldots (d-1)$, 
is the root of  an interesting interplay
between {\em world-volume} properties ({\it i.e.} properties of the theory
defined on the brane, where the $X^{\m}$ are considered as  fields,
with consistent quantum properties when $p=1$)
and spacetime properties, that is, properties of fields living on the 
target-space, whose coordinates are the  $X^{\m}$
themselves. 
\par
One of the subtler aspects of this correspondence is the case of the fermions.
If the target-space ({\em spacetime}) theory is supersymmetric, the
most natural thing seems to use fields which are spacetime fermions 
to begin with. This is called the {\em Green-Schwarz} kind of actions.
\par
It so happens that it is also possible to start with 
world-sheet fermions, which are spacetime
vectors, and then reconstruct spacetime fermions through bosonization 
techniques
(the Frenkel-Ka\v{c}-Segal construction). This is called the
{\em Neveu-Schwarz-Ramond} action.
\par
The first type of actions are only imperfectly understood and, in particular,
it is not known how to quantize them in a way which does not spoil manifest
covariance. This is the reason why the NSR formalism is still the only 
systematic way to perform string perturbation theory.
\par
There is however one fascinating aspect of Green-Schwarz actions
worth mentioning: They apparently need, by consistency
the presence of a particular fermionic symmetry,
called $\kappa$-{\em symmetry} which allows to halve the number of 
propagating fermionic degrees of freedom.
\par
%
%
It was apparently first realized by Ach\'ucarro {\it et. al.} \cite{aetw} 
that in order 
to get a consistent theory one needs world-sheet supersymmetry 
realized linearly,
that is, with matching fermionic and bosonic degrees of freedom (d.o.f.).
The condition for equal number of bosonic and fermionic d.o.f.
after halving the (real) fermionic components of the minimal 
spinor,\footnote{That is $\tilde{d}_{F}=2^{[d/2]-1}$ for a
Majorana or Weyl spinor,
except in $d=2+8\mathbb{Z}$ because one can impose the Majorana and
Weyl condition at the same time, so that $\tilde{d}_{F}=2^{[d/2]-2}$.
For a general discussion of spinors
in arbitrary spacetime dimension and signature, the reader is kindly
referred to \cite{pvn}.} is \cite{aetw},
$N_{S}^{SUSY}\times
\textstyle{\frac{1}{2}}\left(\kappa\right)\times
\tilde{d}_{F}=d-p-1$.
\par
For example:
\begin{itemize}
\item[$p=0$] (Particles): Applying the above formulas one finds that
$N_{S}=4$ for $d=2$ and $N_{S}=3$ when $d=4$.
\item[$p=1$] (Strings): For $d=10$ one finds that $N_{S}=2$, corresponding
to the type IIA and type IIB theories. 
\item[$p=2$] (Membranes): for $d=11$ one finds that $N_{S}=1$, which is kosher.
\end{itemize}
%
\par
It seems to be generally true that exactly,
\be
\hspace{1cm}
 N_{SUSY}(\mbox{World-volume}) \;=\;
 \textstyle{\frac{1}{2}} N_{S}(\mbox{Spacetime}) \; .
\ee
To illustrate this idea in the simplest context, consider the
Lorentz-invariant
action for the superparticle
given by
\be
S=\frac{1}{2}\int d\tau \frac{1}{e}(\dot{x}^{\m} - i 
\bar{\theta}^A \Gamma^{\m} \dot{\theta}^A)\eta_{\m\n}(\dot{x}^{\n} - i 
\bar{\theta}^A \Gamma^{\n} \dot{\theta}^A) \; .
\ee
This action is supersymmetric in any dimension without
assuming any special reality properties for the target-space
spinors $\theta^{A}(\tau )$: 
The explicit rules are
\bea
\d \theta^A &=& \epsilon^A \; ,\\
\d x^{\m} &=& i \bar{\epsilon}^A \Gamma^{\m}\theta^A \; .
\eea
Please note that the presence of the {\em Einbein} $e$ is necessary, because
only then the action is reparametrization invariant; {\it i.e.} when
\bea
\tau&\rightarrow& \tau' \; ,\\
e&\rightarrow& e'\equiv\frac{\partial \tau}{\partial \tau'}\, e \; .
\eea
But precisely the equation of motion for $e$, $\frac{\delta S}{\delta e} = 0$,
implies that on-shell the canonical momentum associated to $x^{\m}$,
$\pi_{\m}=\frac{1}{e}(\dot{x}_{\m} - i 
\bar{\theta}^A \Gamma_{\m} \dot{\theta}^A)$, is a null vector, $\pi_{\m}
\pi^{\m} = 0$, so that as a consequence, the Dirac equation only couples
half of the components of the spinors $\theta^A$. This remarkable fact
can be traced to the existence of the ($\kappa$- or Siegel-)symmetry
\bea
\delta \theta^A &=& - i p_{\m}\Gamma^{\m} \kappa^A \; ,\\  
\delta x^{\m} &=& i \bar{\theta}^A \Gamma^{\m} \delta\theta^A \; ,\\
\delta e &=& 4 \, e \,\dot{\bar{\theta}}{}^A\kappa^A \; ,
\eea
where $\kappa^A$ is a (target-space) spinorial parameter.
The algebra of
$\kappa$-transformations closes on shell only, where
\be
\left[ \d (\kappa_1 ),\d(\kappa_2 )\right] = \d(\kappa_{12}) \; ,
\ee
with $\kappa_{12}\equiv - 4 \kappa_2 \dot{\bar{\theta}}{}^B \kappa_1^B
+ 4 \kappa_1 \dot{\bar{\theta}}{}^B \kappa_2^B $.
\par
The example of the superparticle is a bit misleading, however, because
one always has kappa-symmetry, and this does not impose any restrictions
on the spacetime dimensions.
\par
Historically, this kind of symmetry was first discovered in the Green-Schwarz
\cite{gs} action for the string, by trial and error.
Henneaux and Mezincescu \cite{henmez} interpreted the extra 
non-minimal term (to be introduced
in a moment) as a Wess-Zumino contribution, 
and Hughes and Polchinski \cite{hp}
emphasized that the minimal action is of the 
Volkov-Akulov type, representing
supersymmetry nonlinearly in the Nambu-Goldstone model. 
Kappa symmetry, from this
point of view, allows half of the supersymmetries to be realized linearly.
This fact has also been related \cite{paul} to the BPS property
of fundamental strings. Let us mention finally that there is another framework,
{\em doubly supersymmetric}, in which kappa-symmetry appears as a consequence
of a local fermionic invariance of the world-volume \cite{bstpv,hs}.
\par
In another important work, Hughes, 
Liu and Polchinski \cite{hlp} first  generalized this set up 
for 3-branes in $d=6$ dimensions.
%
%
\par
In order to construct a $\kappa$-symmetric Green-Schwarz action \cite{gs2}
for the string, moving on Minkowski space, we start, following \cite{henmez},
from the supersymmetric 1-forms
\begin{equation}
\left\{
\begin{array}{lcl}
\omega^{\mu} &=& dX^{\mu}\;-\; i\bar{\theta}^{B}\Gamma^{\mu}d\theta^{B} \; ,\\
d\theta^{A} \; ,
\end{array}
\right.
\end{equation}
where $\theta^{A}(\xi^{i})$, the $\xi^{i}$ are the coordinates on 
the worldsheet, are two $d=10$ MW fermions
and, at the same time, world-sheet scalars.
\par
The ``kinetic energy" part of the GS action is given by
\begin{equation}
\left\{
\begin{array}{lcl}
{\cal L}_{1} &=&  -\textstyle{\frac{1}{2}}\sqrt{|h|}h^{ij}
                \omega^{\mu}_{i}\omega^{\nu}_{j} \eta_{\mu\nu} \; ,\\
 & & \\
\omega^{\mu}_{i} &=& \partial_{i}X^{\mu} \;-\,
                 i\bar{\theta}^{A}\Gamma^{\mu}\partial_{i}\theta^{A} \, .
\end{array}
\right.
\end{equation}
As emphasized before, it is easy to check that this part by 
itself has supersymmetry realized in a nonlinear way. 
This fact can be interpreted
\cite{paul} as an indication that, generically, 
an extended object will break all
supersymmetries.
It turns out that there is, in addition, a closed (actually exact), 
Lorentz and SUSY invariant three-form in superspace, namely
\be
\Omega_{3}\;=\; i\left(
\omega^{\mu}\wedge d\bar{\theta}^{1}\Gamma_{\mu}\wedge d\theta^{1}\;-\;
\omega^{\mu}\wedge d\bar{\theta}^{2}\Gamma_{\mu}\wedge d\theta^{2}
\right)\; ,
\ee
with $\Omega_{3}=d\Omega_{2}$ and
\be
\Omega_{2} \;=\; -idX^{\mu}\wedge\left[
\bar{\theta}^{1}\Gamma_{\mu}d\theta^{1}\,-\,
\bar{\theta}^{2}\Gamma_{\mu}d\theta^{2}
\right]\,+\,
\bar{\theta}^{1}\Gamma^{\mu}d\theta^{1}\wedge 
     \bar{\theta}^{2}\Gamma_{\mu}d\theta^{2} \; .
\ee
$\Omega_{2}$ is SUSY invariant up to a total derivative.
The GS action is just
\be
{\cal S}_{GS} \;=\; \int\left[ {\cal L}_{1} \;+\; \Omega_{2}\right] \; ,
\ee
and can be shown to be invariant under the transformations
\be
\left\{
\begin{array}{lcl}
\delta\theta^{A} &=&  \epsilon^A \; ,\\
& & \\
\delta X^{\mu} &=& i \bar{\epsilon}^A \Gamma^{\m} \theta^A \; .
\end{array}
\right.
\ee
Now some of the supersymmetries are realized in a linear way, which physically
means that the extended object is BPS, and thus preserves half of 
the supersymmetries.
%
\par
Let us now turn our attention to the supermembrane.
In \cite{bst} the following GS-type action was proposed for a supermembrane
coupled to d=11 supergravity
\be
S = \int d^3\xi\left\{ \frac{1}{2}\sqrt{-g} g^{ij}E_{i}^{A}E_{j}^{B} \eta_{AB}
+ \epsilon^{ijk} E_{i}^A E_{j}^B E_{k}^C B_{CBA} - \frac{1}{2} \sqrt{-g}
\right\} \, .
\ee
Here the $\xi^i$ (i=0,1,2) label the coordinates of the bosonic
world-volume, and
the (target-space) superspace coordinates are denoted by $Z^M (\xi)$.
The action just represents the embedding of the three dimensional world-volume
of the membrane, in eleven dimensional superspace.
Lowercase latin indices will denote
vectorial quantities; Lower case greek indices will denote spinorial quantities.
Capital indices will include both types. Frame indices are denoted by the first letters of
the alphabet, whereas curved indices will be denoted by the middle letters.
On the other hand,
$E_{i}^A \equiv {E^{A}}_B \partial_i Z^B$, and the super-three form
$B$ is the one needed for the superspace description of $d=11$ supergravity.
\par
Bergshoeff, Sezgin and Townsend  imposed invariance under
$\kappa$-symmetry, that is, under
\bea
\delta E^a &=& 0 \; ,\\
\delta E^{\a} &=& (1 + \Gamma)^{\a}{}_{\b} \kappa^{\b} \; ,\\
\delta g_{ij} &=& 2 X_{ij} - g_{ij} X^k_k \; ,
\eea
where $\kappa^{\a}(\xi)$ is a Majorana spinor and a world-volume scalar,
$\delta E^A \equiv \delta Z^B E_B^A$ and $\Gamma^{\a}_{\b}\equiv
\frac{1}{6}\sqrt{g} \epsilon^{ijk}E_i^a E_j^b E_k^c (\gamma_{abc})^{\a}{}_{\b}$.
$X_{ij}$ is a function of the $E_i^A$
which should be determined by demanding invariance of the action.
They found that for consistency they had to impose the constraints
\bea
H_{\a\b\gamma\d} &=& H_{\a\b\gamma d}= 
    H_{\a abc} = \eta_{c(a}T^c_{b)\a} = 0 \; ,\\
T^a{}_{\a\b} &=& (\Gamma^a)_{\a\b} \; ,\\
H_{\a\b ab} &=& -\frac{1}{6} (\Gamma_{ab})_{\a\b} \; ,
\eea
where $H_{\ldots}$ and ${T^{.}}_{..}$ are the components
of the super-fieldstrength of 
$B$ and the super-torsion {\it resp.}.
It is a remarkable fact that these constraints 
(as well as the Bianchi identities) are solved by the superspace
constraints of d=11 supergravity as given by Cremmer and Ferrara in \cite{cf}.
\par
This is the first of our encounters with some deep relationship between
world-volume and spacetime properties: By demanding $\kappa$-symmetry
(a world-volume property),
we have obtained some spacetime equations of motion
which must be satisfied for the world-volume symmetry to be possible
at all.
\subsubsection{Conformal invariance}
There seems to be something special about the case {\em p = 1}
({\em strings}). We have then,
as we shall see, some extra symmetry, {\em conformal invariance}, which
allows for the construction of a seemingly consistent perturbation theory.
\par
There are no strings in d=11: 
{}From our present point of view the most natural way
of introducing them is through {\em double dimensional reduction}
of the 11-dimensional membrane ({\em M-2-brane}). If we start with the 
(bosonic
part of the) previous
action of \cite{bst} for the latter, namely:
\be
S_3 = T_3 \int d^3 \xi \left[ \frac{1}{2}\sqrt{- \gamma}\g^{ij}\partial_i X^M 
\partial_j X^N G_{MN}^{(11)}(X)
- \frac{1}{2}\sqrt{ - \g} +\frac{1}{3!} \e^{ijk}\partial_i X^M 
\partial_j X^N \partial_k X^P A_{MNP}^{(11)} (X)\right] \; ,
\ee
and follow \cite{dhis}, in assuming that there are isometries both
in the spacetime generated by $\partial_Y$, and in the world-volume as well,
generated by $\partial_{\rho }$. (We  label spacetime  coordinates as 
$X^M = (X^{\m },Y)$
where $M = 0\ldots 10$ and $\m = 0\ldots 9$; World-volume coordinates 
as $\xi^i = ( \xi^a, \rho )$, where i = 0, 1, 2 ; a = 0, 1).
We now identify the two ignorable coordinates ({\em static gauge})
\be
\rho = Y \; ,
\ee
and perform stardard Kaluza-Klein reduction, namely
\begin{eqnarray}
G^{(11)}_{\m\n} &=&  e^{-\frac{2\phi}{3}}(G^{(10)}_{\m\n} \,+\,
                   e^{2\phi}A_{\m}A_{\n}) \; , \nonumber \\
G^{(11)}_{\m Y} &=&  -e^{\frac{4\phi}{3}}A_{\m} \; , \nonumber \\
G^{(11)}_{Y Y} &=&   e^{\frac{4\phi}{3}} \; ,
\end{eqnarray}
and for the three-form
\be
A^{(11)}_{\m\n\kappa} = A^{(10)}_{\m\n\kappa}
\hspace{1cm},\hspace{1cm}
A^{(11)}_{\m\n Y} = B^{(10)}_{\m\n} \; .
\ee
This then leads to the action for a ten-dimensional string, namely
\be
S_2 = T_2 \int d^2 \xi \left[ \frac{1}{2} \sqrt{-\g} \g^{ab}
\partial_a X^{\m} \partial_b X^{\n} G^{(10)}_{\m\n}(X)
+\frac{1}{2} \epsilon^{ab} \partial_a X^{\m} \partial_b X^{\n} 
B^{(10)}_{\m\n}(X) \right] \; .
\ee
This is a remarkable action,\footnote{Were we to
reduce the kappa-symmetric supermembrane action
we would have found the kappa-symmetric Green-Schwarz string action in d=10.}
which enjoys both two-dimensional 
reparametrization invariance and ten-dimensional invariance under the isometry
group of the target-manifold (including the appropriate torsion) and, most
importantly, under Weyl transformations
\be
\gamma_{ab}\rightarrow e^{\psi(\xi )}\gamma_{ab} \; .
\ee
\par
On the other hand, a well-known mathematical theorem ensures that, locally,
any two-dimensional (Euclidean) metric can be put in the form
\be
\gamma_{ab} = e^{\sigma}\delta_{ab} \; .
\ee
Owing to Weyl invariance, the trace of the energy-momentum tensor vanishes
\be
0 = T^{a}_{a}\equiv \gamma^{ab}\frac{2}{\sqrt{\gamma}}\frac{\delta}
{\delta \gamma^{ab}} S[\gamma] = \frac{\delta}{\delta\sigma} S[\gamma] \; .
\ee
This, in its turn, means that the two-dimensional 
action is conformally invariant
in flat space: That is, invariant under conformal Killing transformations
$\delta \xi^a = k^a$, with
\be
\partial_a k_b + \partial_b k_a = \delta_{ab}\,\partial_c k^c \; ,
\ee
with reduces to
\be
\partial_1 k_1 = \partial_2 k_2
\hspace{1cm},\hspace{1cm}
\partial_1 k_2 = - \partial_2 k_1 \; ,
\ee
which in turn implies
\be
\Box k_a = 0 \; ,
\ee
an infinite group; In terms of the natural coordinates 
$\xi^{\pm}\equiv \xi^0 \pm \xi^1$
the general two-dimensional conformal transformation is:
\be
\delta \xi_a = f_a(\xi^{+}) + g_a(\xi^{-})
\ee
with arbitrary functions $f_a$ and $g_a$. This infinite conformal 
symmetry is the root
of many aspects of the physics of strings.
\subsection{The string scale and the string coupling constant}
If the radius of the eleventh dimension is R, and we denote the M-2-brane 
tension by $T_3 \equiv l_{11}^{-3}$, the string tension (traditionally
denoted by $\alpha^{\prime}$) 
will be given by $T_2 = \frac{R}{l_{11}^3} \equiv 
\frac{1}{l_s^2}\equiv \frac{1}{\alpha^{'}}$. This gives the string length as
\be
l_s = \frac{l_{11}^{3/2}}{R^{1/2}} \; .
\ee
\par
The mass of the first Kaluza-Klein excitation with one unit of momentum
in the eleventh direction is $ M(KK)\equiv R^{-1}$. 
As we shall see later on, this state
is interpreted, from the 10-dimensional point of view, 
as a D0-brane, and its mass
could serve as a {\em definition} of the {\em string coupling constant},
$M(D0)\equiv \frac{1}{g_s l_s}$. Equating the two expressions gives
\be
g_s = \frac{R}{l_s} = \left(\frac{R}{l_{11}}\right)^{3/2} \; .
\ee
\par
This formula is very intriguing, because it clearly suggests that the string
will only live in 10 dimensions as long as the coupling is small. The historical
way in which Witten \cite{ew} arrived to this result
was exactly the opposite, by
realizing that the mass of a D0 brane (in 10 dimensions) goes to zero at 
strong coupling, and 
interpreting this fact as the opening of a new dimension. Although some partial
evidence exists on how the full O(1,10) can be implemented in the theory
(as opposed to the O(1,9) of ten-dimensional physics), there is no clear
understanding about the r\^ole of conformal invariance (which is equivalent
to BRST invariance, and selects the critical dimension) in eleven dimensional
physics. We shall raise again some of these points in the section devoted
to the strong coupling limit.
\par
The radius could also be eliminated, yielding the beautiful formula
\be
g_s = \left( \frac{l_{11}}{l_s} \right)^3 \; .
\ee
An inmediate consequence is that 
\be
R=\frac{l_{11}^3}{l_s^2}
\ee
On the other hand, the eleven-dimensional gravitational coupling
constant is defined by
\be
\kappa_{11}\equiv l_{11}^{9/2}
\ee
so that the ten-dimensional gravitational coupling constant is given by
\be
\kappa_{10}\equiv \kappa_{11}{R^{1/2}}= g_s l_s^4 \; .
\ee

\section{Conformal field theory}
Starting from the classic work of Belavin,
Polyakov and Zamolodchikov \cite{bpz}
the study of two-dimensional conformally invariant quantum field theories (CFT)
has developed into a field of study on its own (See for example the textbooks
\cite{bk:ketov,bk:francesco}),
with applications in Statistical
Physics \cite{ginsparg}. From the point of view of Strings, the imbeddings
$x^{\m}(\xi)$ are to be considered as two-dimensional quantum fields.
%
\subsection{Primary fields and operator product expansions}
%
In {\em any} $d$, the group of (Euclidean) conformal transformations,
$C(d)\sim O(1,d+1)$,
is defined by all transformations $x^{\mu}\rightarrow x^{\prime\mu}(x)$
such that 
\be
\delta_{\mu\nu}dx^{\prime\mu}dx^{\prime\nu} \;=\;
  \Omega^{-2}(x)\, \delta_{\mu\nu}dx^{\mu}dx^{\nu} \; .
\ee
Given a conformal transformation we may define a corresponding 
local transformation by
\be
{\cal R}_{\mu\alpha}\;\equiv\; \Omega (x)
  \frac{\partial x^{\prime}_{\mu}}{\partial x_{\alpha}} 
\hspace{.5cm},\hspace{.5cm} 
{\cal R}_{\mu\alpha}{\cal R}_{\nu\alpha}\;=\; \delta_{\mu\nu} \; .
\ee
A {\em quasi-primary} field \cite{ms}  ${\cal O}^{i}$, (where $i$ denotes the 
components
in some space on which some representation of $O(d)$ acts) , is
defined to transform as
\be
{\cal O}^{\prime i}(x^{\prime}) \;=\; \Omega^{h}(x)
 {D^{i}}_{j}\left( {\cal R}\right) {\cal O}^{j}(x) \; ,
\ee
where $h$ is called the scale dimension of the field.
A quasi-primary field is called a {\em primary} field if it transforms as a
scalar under the action of $O(d)$.
\par
We have previously seen that in $d=2$ conformal
transformations are of the type
\be
\delta \xi_a = f_a(\xi^{+}) + g_a(\xi^{-}) \; .
\ee
\par
We will frequently be interested in CFT on the {\em cylinder},
$S^1\times \mathbb{R}$, where
the two-dimensional Lorentzian coordinates $(\tau,\sigma )$
are such that $\sigma = \sigma +2\pi$. Performing a two-dimensional 
Wick rotation $\t \rightarrow -i \t$, $\xi^{\pm}\equiv \tau \pm \sigma
\rightarrow -i (\t \pm i \sigma)$, the coordinate $ z \equiv \t - i \sigma$
describes the (Wick rotated) cylinder. 
\par
One can now perform a conformal transformations (physics should be 
insensitive to this) to the Riemann sphere (the extended complex plane),
$z\rightarrow e^z$. Quite frequently, coordinates on the Riemann sphere 
will also be denoted by $z$.
Translations in $\tau\equiv \xi^{0}$, $\delta \tau = \epsilon$, map on the
complex plane into $|z|\rightarrow e^{\epsilon} |z|$. Regular time evolution
in $\tau$ on the cylinder then maps onto radial evolution from the origin of 
the complex plane (corresponding to the point $\tau = - \infty$ on the 
cylinder). In order to emphasize this, quantization on the complex plane
is sometimes refered to as {\em radial quantization}.
The energy momentum tensor $T_{ab}$ represents the response of the action to a 
variation of  the two-dimensional metric. Given any Killing vector field,
$k_a$, the currents $T_{ab}k^b$ are conserved. This includes in particular
all conformal transformations.
%
%
\par
For open strings (with $0\leq\sigma\leq\pi$) this conformal transformation
maps the strip $(\sigma ,\tau )$ into the upper half of the complex plane.
A further conformal transformation could be used to map it into the
unit disc; For example
\be
  z \;\rightarrow\; \frac{z-i}{z+i} \; ,
\ee
maps the origin ($\tau =-\infty$) to the point $-1$ and semi-circles
around the origin into arcs corresponding to circles centered in the 
real axis. The region $\tau\rightarrow\infty$ is mapped onto the
single point $+1$.
%
%
\par
In complex coordinates the $2d$ metric locally reads $ds^{2}=dzd\bar{z}$, and one
can see that the tracelesness and the conservation of the stress tensor
read
\begin{eqnarray}
{T^{\alpha}}_{\alpha} &=& {T^{z}}_{z}\,+\, {T^{\bar{z}}}_{\bar{z}} \;=\; 
 2T_{z\bar{z}} \;=\; 0 \; , \\
\partial_{\mu}{T^{\mu}}_{z} &=& \partial_{\bar{z}}T_{zz} \,+\,
  \partial_{z}T_{\bar{z}\, z} \;=\; 0 \; .
\end{eqnarray}
The last equation then means that
\be
\partial_{\bar{z}}T_{zz} \;\equiv\; \bar{\partial}T\;=\; 0 \; ,
\ee
where, from now on, we will display the holomorphic part only.
\par
The action for a massless scalar field, such as any of the $d$
imbedding functions $x^{\m}(z)$, is given by\footnote{The transformation
between the coordinates is given by $z=x+iy$,
$\partial =\textstyle{\frac{1}{2}}\left(\partial_{x}-i\partial_{y}\right)$ and
$\bar{\partial}=
\textstyle{\frac{1}{2}}\left(\partial_{x}+i\partial_{y}\right)$.}
\be
{\cal S} \;=\; \frac{1}{2\pi}\int d^{2}z\, \partial\phi\bar{\partial}\phi
\;=\; \frac{1}{4\pi i}\int d^{2}x\, \partial_{\mu}\phi\partial^{\mu}\phi \; ,
\ee
and the equation of motion reads
\be
\partial\bar{\partial}\phi \;=\; 0 \; .
\ee
This means that {\em on-shell}
\be
\phi(z,\bar{z}) = \frac{1}{2}\left(\phi(z) + \bar{\phi}(\bar{z}) \right) \; .
\ee
The propagator must solve the differential equation
\be
\pd\bar{\pd}\langle T\phi (z_{1},\bar{z}_1)\phi (z_{2},\bar{z}_2)\rangle
= - 2\pi \delta^{(2)}(z_1 - z_2 ,\bar{z}_1 - \bar{z}_2) \; ,
\ee
which after using the formula
\be
\bar{\pd} \frac{1}{z} = \pi \delta^{(2)}(z,\bar{z}) \; ,
\ee
results in
\be
\langle T\phi (z_{1},\bar{z}_1)\phi (z_{2},\bar{z}_2)\rangle = -
log( |z_1 - z_2|^2) \; .
\ee
It is customary to omit corresponding expressions for the anti-holomorphic
part, and write down explicitely the holomorphic part only
\be
\langle T\phi (z_{1})\phi (z_{2})\rangle = -
log (z_1 - z_2) \; .
\ee
Wick's theorem ensures that the $T$-product is expressible as
\be
T\phi (z_{1},\bar{z}_1)\phi (z_{2},\bar{z}_2)= \;
:\phi (z_{1},\bar{z}_1)\phi (z_{2},\bar{z}_2): + 
\phi\SING{(z_{1},\bar{z}_1)\phi}(z_{2},\bar{z}_2) \; .
\ee
Clearly by construction
\be
\langle :\phi (z_{1},\bar{z}_1)\phi (z_{2},\bar{z}_2): \rangle \;=\; 0 \; ,
\ee
and the normal-ordered product obeys the classical equation of motion, without
source terms
\be
\pd\bar{\pd} :\phi (z_{1},\bar{z}_1)\phi (z_{2},\bar{z}_2): \; =\;  0 \; .
\ee
This means that there is a na\"{\i}ve Operator Product Expansion (OPE),
given simply by the Taylor expansion whose holomorphic part is
\be
\phi(z)\phi(w) = - log (z-w) \,+\, :\phi\phi:(w)
\,+\,  (z-w):\phi\pd\phi:(w) \, +\, \ldots
\ee
Contractions then represent the singular part of the OPE. Correlators in free
theories (such as the most interesting examples in String Theory)
are given by
the general form of Wick's Theorem
\be
\langle : A_1(z_1)\ldots A_n(z_n):\ldots :D_1(w_1)\ldots D_m(w_m):\rangle
\ee
is given by the sum of all possible pairings, excluding those corresponding to
operators inside the same normal ordering.
\par
Using the above rules we obtain the OPE
\be
\pd \phi(z)\pd\phi(w)\;\sim\;  -\frac{1}{(z-w)^2} \; .
\ee
\par
The energy momentum tensor corresponding to a scalar field
coupled minimally to the two-dimensional metric is given by
\be
T(z) \; =\;  -\frac{1}{2}\, :(\pd \phi )^{2}:(z) \; .
\ee
Given a (conformal) Killing, $k^a$, there is an associated conserved
current, given by $j_a\equiv T_{ab}k^b$. Its conserved charge is given 
on the cylinder by
\be
Q(k)\;\equiv\; \int_{\tau =cons.}T_{0b}k^b d\sigma  \; .
\ee
A conformal transformation $z\rightarrow z+\xi (z)$ is associated 
on the plane to the charge
\be
Q(\xi)\;\equiv\; \oint \frac{dz}{2\pi i}T(z)\xi (z) \; .
\ee
The corresponding transformation of a field $\pd\phi$ will be given by
\be
\delta\left[ \partial\phi(z)\right] \equiv
\langle\left[ Q(\xi),\partial\phi(z)\right]\rangle  \; .
\ee
The fact that path integrals in the plane are automatically radially ordered
allows for a simple representation of correlators in terms of Cauchy integrals
\begin{eqnarray}
&=& \int {\cal D}\phi\; e^{-{\cal S}}
    \left[
      Q_{\xi}(|z|+\epsilon )\partial\phi (z)\,-\,
      \partial\phi (z) Q_{\xi}(|z|-\epsilon )
    \right] \nonumber \\
&=& \oint_{C_z}\frac{dw}{2\pi i}\xi (w)\langle T(w)\partial\phi (z)\rangle \; ,
\end{eqnarray}
%
%
\begin{figure}[t]
\begin{center}
\leavevmode
\epsfysize= 4cm
\epsffile{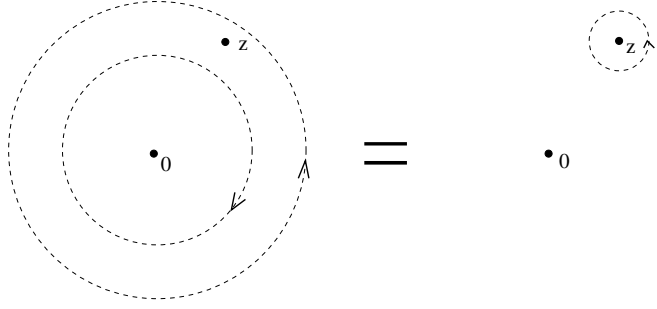}
\caption{\label{fig:countour1}The needed contour deformation.}
\end{center}
\end{figure}
where the two contours are deformed as in fig.(\ref{fig:countour1})
into a single one $C_z$ around the privileged point $z$, using the fact that
the conserved charge, $Q(\xi)(w)$ is independent of $w$, which is implemented
mathematically by the fact that we can deform the contour of definition 
of $Q$ as long as we do not meet any singularities, which is 
possible only at $z$.
Since we are dealing with a free theory, we can use Wick's theorem
to evaluate the last v.e.v., {\it i.e.}
\begin{eqnarray}
&=& -\frac{1}{2}\oint\frac{dw}{2\pi i}\, 
   \langle :\partial\phi\partial\phi:(w) \partial\phi (z)\rangle 
    \xi (w) \nonumber \\
&=& \oint\frac{dw}{2\pi i}\,\frac{1}{(w-z)^{2}}\partial\phi (w)\xi (w)
    \nonumber \\
&=& \oint\frac{dw}{2\pi i}\,\left[
      \frac{\partial\phi (z)}{(z-w)^{2}}\,+\, 
      \frac{\partial^{2}\phi (z)}{w-z} \; +\ldots
    \right]
    \left[
      \xi (z)\,+\, \partial\xi (z)\left( w-z\right) \, + \ldots
    \right] \nonumber \\
&=& \partial\phi (z)\partial\xi (z) \;+\; \partial^{2}\phi (z) \xi (z) \; .
\end{eqnarray}
In general, for a field, $\phi^{(\lambda )}$, of arbitrary 
scaling dimension $\lambda$, we would have
\be
T(z)\phi^{(\lambda )}(w) \;=\; \frac{\lambda \phi^{(\lambda )}(w)}{(z-w)^{2}} 
        \,+\, \frac{\partial\phi^{(\lambda )} (w)}{z-w} \; .
\ee
This is telling us that the scaling dimension of $\pd\phi$ is one,
\be
h(\pd\phi) = 1 \; .
\ee
A conformal, dimension 1, field can be expanded in Fourier modes as
\be
\pd\phi(z)=\sum \frac{\a_m}{z^{m+1}} \; .
\ee
Note that we have written $z^{m+1}$ instead of $z^m$; This is an effect of
the conformal mapping from the cylinder to the plane.
\par
Other important primary fields associated to a scalar field are the
vertex operators $V_{\alpha}(z)=:e^{i\alpha\phi (z)}:$. 
It is a simple exercise to show that
\begin{eqnarray}
T(z)V_{\alpha}(w) &=& \frac{\alpha^{2}V_{\alpha}(w)}{2 (z-w)^{2}} \,+\,
                      \frac{\partial V_{\alpha}(w)}{z-w} \; ,\\
\langle V_{\alpha}(z)V_{-\alpha}(w)\rangle &=& 
             \left( z-w\right)^{-\alpha^{2}} \; , \\
:e^{i\alpha\phi (z)}:\, :e^{i\beta\phi (w)}: &=& (z-w)^{\alpha\beta}
   :e^{i\alpha\phi (z)\,+\, i\beta\phi (w)}: \; ,
\end{eqnarray}
meaning that the scaling dimension of a vertex operator $V_{\alpha}$
is $h \left(V_{\alpha}\right) =\alpha^{2}/2$.
\par
Note that derivatives of primary Fields are not primary Fields (BPZ calls them
secondary fields) \cite{bpz}.
\par
In CFT there is a natural mapping from operators to states, given by the path 
integral with an operator insertion, 
in terms of boundary values of this operator
on the unit circle.
\be
 \Theta \longrightarrow 
 \int_{\phi (|z|=1)=\phi_{B}}
      {\cal D}\phi\, \Theta\left[\phi\right]e^{-S[\phi ]}
 \,=\, \Psi \left( \phi_{B}\right) \;\equiv |\Theta > .
\ee
The in-vacuum (the state at $\tau = - \infty$, that is $z=0$ on the plane),
corresponds to the unit operator.
\par
In order to study scattering states,
let $\mid 0\rangle$ be an asymptotic, $\tau\rightarrow -\infty$, state
without any insertion. Then if the action of the operator at the origin is to
be well defined,
\be
\phi(z)\mid 0\rangle \;=\; \sum \phi_{n}z^{-n-h}
\mid 0\rangle \; ,
\ee
it is necessary that
\be
\phi_{n}\mid 0\rangle \;=\; 0 \; ,
\ee
for $ n + h > 0$.
\par
Let us next consider states constructed out of vertex operators 
of the form $e^{ip_{\mu}X^{\mu}(0)}\mid 0\rangle$.
They represent the
asymptotic state of a ground-state string at momentum $p^{\mu}$.
It should be easy to prove that 
\be 
\alpha_{m}^{\mu} e^{ip_{\mu}X^{\mu}(0)}\mid 0\rangle \;=\; 
  \delta_{n,0}p^{\mu}\, e^{ip_{\mu}X^{\mu}(0)}\mid 0\rangle \; .
\ee
Other excited states are represented by composite operators of the type
\be
\partial X^{\mu}e^{ip_{\mu}X^{\mu}}\; ,\; 
\partial X^{\mu}\partial X^{\nu} e^{ip_{\mu}X^{\mu}} \;, \;  et\; cetera 
\ee
where the $p^{2}$ has to be chosen, such that the conformal dimension
of the operator equals 1 (See section (4.4)).
\par
Let us now consider the $\t \rightarrow +\infty$ behaviour:
Expand an arbitrary dimension $h$ field $\phi$ as 
$\phi =\sum_{n}\phi_{n}z^{-n-h}$ and have a look at the 
{\em out} state defined by
\be
\langle\phi\mid \;=\; \lim_{w\rightarrow \infty }\langle 0\mid\phi (w) \;=\;
  \lim_{z\rightarrow 0}\langle 0\mid\phi (\textstyle{\frac{1}{z}})\, z^{-2h} 
\equiv \langle 0\mid \phi^{+}
  \; .
\ee
The BPZ adjoint is then defined by
\be
 \phi^{\dagger} \;\equiv\; \sum_{n}\phi_n z^{n+h}z^{-2h} \;=\;
 \sum_{n}\phi_{n}^{\dagger}\, z^{-n-h} \; ,
\ee
Summarizing, then, the {\em in} and {\em out} vacua obey
\be
 \phi_{n}\mid 0\rangle\;=\; 0 \; (n>-h)\hspace{.5cm},\hspace{.5cm}
 \langle 0\mid\phi_{n}\;=\; 0 \; (n<h) \; .
\ee
\subsection{The Virasoro algebra}
Classically conformally invariant theories do not, in general, preserve this
property at the quantum level, because of the well-known {\em trace anomaly}
\cite{bk:birrell}.
\par
Given any Conformal Field Theory with conformal anomaly coefficient $c$,
the manifestation of this anomaly can be altered somewhat by local counterterms.
There is a definition of the energy-momentum tensor
such that it is conserved, but
there is a trace anomaly {\em sensu stricto}
\bea
&&T_{(trace)\m}^{\m} = -\frac{c}{6} R(g)\nn \; ,\\
&&\nabla_{\m} T_{(trace)}^{\m\n} = 0 \; .
\eea
Another definition in such a way that it is traceless, but
not conserved, so that there is now a gravitational
anomaly
\bea
&&T_{(grav)\m}^{\m} = 0\nn\; ,\\
&&\nabla_{\m} T_{(grav)}^{\m\n} = \frac{c}{12} \nabla^{\n} R(g) \; .
\eea
\par
The Virasoro algebra in OPE notation reads
\be
T(z)T(w) \;=\; \frac{c/2}{(z-w)^{4}} \,+\,
               \frac{2T(w)}{(z-w)^{2}}\,+\,
               \frac{\partial T(w)}{z-w} \; .
\ee
It is an easy exercise to show that the energy momentum of the scalar field
obeys the above equation with $c=1$.
\par
The most important property of the mapping from the cylinder to the plane is 
\be
T_{cylinder}(z^{\prime}) \;=\; z^{2}T_{plane}(z) \,-\, \frac{c}{24} \; ,
\ee
which one can deduce using the above rules.
\par
Expanding $T(z)$ in Fourier series
\be
 T(z) \;=\; \sum_{n\in\mathbb{Z}} \, L_{n}z^{-n-2} \; ,
\ee
so that 
\be
  L_{n} \;=\; \oint\frac{dz}{2\pi i}\, z^{n+1}T(z) \; ,
\ee
one can compute the commutators using OPEs, leading to
\be
\left[ L_{n},L_{m}\right] = (n-m)L_{n+m}\;+\; \frac{c}{12}n(n+1)(n-1)
\delta_{m,-n} \; .
\ee
Direct inspection shows that
$\left\{ L_{0},L_{\pm 1}, \bar{L}_{0}, \bar{L}_{\pm 1} \right\}$
generate the algebra $Sl(2,\mathbb{C})$, and that $L_{0}$ generates
dilatations $\xi (z)=z$.
\par
It is also possible to show, using the definition of conformal weight,
and expanding an arbitrary field $\phi^{(\lambda )}$
as $\phi^{(\lambda )} (z)=\sum_{n} \phi^{(\lambda )}_{n}z^{-n-\lambda}$ that
\be
\left[ L_{m}, \phi^{(\lambda )}_{n}\right] \;=\; 
\left[ (\lambda -1)m -n\right] \phi^{(\lambda )}_{n+m} \; .
\ee
\subsection{Non-minimal coupling and background charge}
Although the {\em minimal} coupling of a scalar field to the two-dimensional
metric consists simply in writing covariant 
derivatives instead of ordinary ones,
there are more complicated ({\em non-minimal} ) possibilities.
One of the most interesting involves a direct coupling to the two-dimensional
scalar curvature \cite{art:dotsfat}
\be
{\cal S}_{Q} \;=\; \frac{1}{8\pi}\int d^{2}z\,
\left[
\partial\phi\bar{\partial}\phi\;-\; 2Q\sqrt{g}R^{(2)}\phi
\right]\; ,
\ee
for some $Q$. 
\par
It is possible to show that the holomorphic stress tensor reads
\be
   T(z) \;=\; -\frac{1}{2}:\partial\phi\partial\phi:(z)\,+\,
              Q\partial^{2}\phi (z) \; .
\ee
\par
Introducing the, formerly conserved, current
$j=-\partial\phi$, and using the fact that the 
propagator for $\phi$ does not change, we can calculate
\begin{eqnarray}
T(z)j(w) &=& \frac{-2Q}{(z-w)^{3}}\,+\, 
             \frac{j(w)}{(z-w)^{2}} \,+\,
             \frac{\partial j(w)}{z-w} \; , \\
j(z)e^{q\phi (w)} &=& \frac{q}{z-w}\, e^{q\phi (w)} \; ,\\
T(z)e^{q\phi (w)} &=& -\frac{q(q+2Q)}{2(z-w)^{2}}e^{q\phi (w)}\,+\,
                      \frac{\partial e^{q\phi (w)}}{z-w} \; ,
\end{eqnarray}
as well as
\be
T(z)T(w)= \frac{1+12Q^{2}}{2(z-w)^4} + \frac{2 T(w)}{(z-w)^2} 
+\frac{\pd T (w)}{z-w}
\ee
showing  that $T$ generates a Virasoro algebra with $c=1+12Q^{2}$.
Note that the current behaves as an anomalous primary field, 
obviously due to the non-minimal coupling.
The equation of motion, written in terms of the current $j$,
already entails the occurrence of the anomaly, {\it e.g.}
\be
   \bar{\partial}j \;=\; Q\sqrt{g}R^{(2)} \; .
\ee
\par
Using the anomalous transformation law for $j$ one can show that 
under the transformation $z\rightarrow w=z^{-1}$, we have
\be
  \partial\phi (z) \;=\; -w^{2}\partial\phi (w)\,+\, 2Q\, w\; .
\ee
Using this, we get on the Riemann sphere, putting $Q=-i\alpha_{0}$
\be
Q(z=0) \,=\, \oint\frac{dz}{2\pi i}\partial\phi (z)\,=\,
\oint\frac{dw}{2\pi i}\partial\phi (w)\,-\,
\oint\frac{dw}{2\pi i}\, \frac{2Q}{w}\,=\,
Q(z=\infty )\,-\, 2Q\; .
\ee
\par
If we define $\mid\alpha\rangle =V_{\alpha}(0)\mid 0\rangle$,
then
\be
  \langle\alpha\mid \;\equiv\; \lim_{z\rightarrow\infty}
\langle 0\mid V_{\alpha}(z)\, z^{2h_{\alpha}} \; ,
\ee
with $2 h_{\alpha}=\alpha (\alpha -2\alpha_{0})$.
We can calculate
\be
\langle\alpha\mid\beta\rangle \;=\; 
\langle 0\mid V_{\alpha}(w^{-1})\, w^{-\alpha (\alpha -2\alpha_{0})}
V_{\beta}(z)\mid 0\rangle \;\sim\;
w^{-\alpha (\alpha -2\alpha_{0})}\, (w^{-1}-z)^{\alpha\cdot\beta} \; ,
\ee
which has a smooth and non-vanishing limit for $w\downarrow 0$
iff $\beta =2\alpha_{0}
-\alpha$.

\subsection{$(b,c)$ systems and bosonization}
%
It is very interesting to consider a system of (anti-)commuting analytical 
fields of conformal weights $j$ and $1-j$,
usually called $b_{j}$ and $c_{1-j}$.
These systems appear, in particular, when fixing the (super)conformal
gauge invariance.
It is possible to consider both cases simultaneously by introducing a parameter
$\epsilon$, valued $+1$ in the anticommuting case, and $-1$ in the commuting case.
The action for these fields, first order in derivatives, is
\be
S\equiv \frac{1}{2\pi} \int b \bar{\pd} c
\ee
It follows that
\begin{equation}
\left\{
\begin{array}{lcl}
b(z)c(w) &=& \frac{\epsilon}{z-w} \; , \\
& & \\
T(z) &=& -j\, :b\partial c:\,+\, (1-j):\partial b \cdot c: \; .
\end{array}
\right.
\end{equation}
(Please note that $c(z)b(w) = \frac{1}{z-w}$).
\par
The above stress tensor satisfies
\be
T(z)T(w) \;=\; -\epsilon \frac{ 6j^{2}-6j+1}{(z-w)^{4}} \,+\,
               \frac{2T(w)}{(z-w)^{2}}\,+\,
               \frac{ (\partial T)(w)}{z-w} \; .
\ee
This means that the conformal anomaly for the $(b,c)$ system reads
\be
 c\;=\; -2\epsilon\left[ 6j(j-1)+1\right] \; ,
\ee
\par
The physical $(b,c)$ systems needed in the quantization of
superstrings are an anticommuting system of ghosts
, denoted $(b,c)$, due to the gauge fixing of the diffeomorphisms, and
a commuting system, denoted $(\beta ,\gamma )$, due to the gauge fixing
of local supersymmetry. Their characteristics are
\be
\begin{array}{rclclcl}
(b,c) &\hspace{.3cm}:\hspace{.3cm}& c&=& -26 
      &\hspace{.5cm},\hspace{.5cm}& (j=2) \; ,\\
(\beta ,\gamma ) & : & c&=& 11
      & , & (j=\textstyle{\frac{3}{2}}) \; .
\end{array}
\ee 
\subsubsection{Bosonization}
In two dimensions there is no essential difference between bosons and fermions,
and, in particular, it is possible to bosonize fermionic expressions.
All our relationships are to be understood, as usual, valid inside 
correlators only.
\par
A free boson is a $c=1$ CFT and as such one 
can  show that it is equivalent to two minimal spinors. The operator
correspondence is generated by
\be
   \psi_{1}\,+\, i\psi_{2} \;=\; \sqrt{2}\, e^{i\phi } \; .
\ee
This is the simplest instance of bosonization: 
The starting point of the whole construction.
\par
The $(b,c)$ system on the other hand, is equivalent to a non-minimally
coupled boson, denoted $\sigma$.
Identification of the conformal anomaly $c_{bc} = -26$
leads to, using the formula for a non minimally coupled scalar $c=1+12Q^{2}$, 
a value for the background charge of $Q=-i\textstyle{\frac{3}{2}}$.
Using this fact, one can show that
\be
  T(z) e^{i\alpha\sigma} \;=\; 
     \left( 
       \frac{\alpha^{2}}{2}\,-\,
       \frac{3\alpha}{2}
     \right)
     \frac{1}{(z-w)^{2}}\, e^{i\alpha\sigma} \, + \ldots 
\ee
This suggests that the correct mapping of fields is given by
\be
\begin{array}{lclrl}
b(z)&=& e^{-i\sigma} &\hspace{2cm}:& (j=2) \; ,\\
c(z)&=& e^{i\sigma} & :& (j=-1) \; .
\end{array}
\ee
\par
Although we are going to be quite schematic about it, it is also 
possible to bosonize the (already bosonic) $(\b,\gamma)$ system. Actually,
we write the $(\beta ,\gamma )$ system as a $c=13$ non-minimally
coupled boson $\phi$ with background charge $Q=1$,
and another $j=0$ $(b,c)$-system with $\epsilon = +1$
(that is, anticommuting), the $(\xi ,\eta)$ system,
carrying a conformal anomaly of $c_{(0,1)}=-2$. 
The total stress tensor reads
\be
T(z) \;=\; -\textstyle{\frac{1}{2}}:\partial\phi\partial\phi :\,+\,
            \partial^{2}\phi\,+\, :\left(\partial\xi\right)\eta : \; ,
\ee
and the resulting central charge is $c=13-2 = 11$ as it ought to be.
The explicit bosonization rules are then
\be
\beta\;=\; \partial\xi\, e^{-\phi} \hspace{.5cm},\hspace{.5cm}
\gamma\;=\; \eta\, e^{\phi} \; .
\ee
\subsection{Current algebras and the Frenkel-Ka\v{c}-Segal construction}
There is a kind of non-Abelian generalization of the Virasoro algebra, 
called
Ka\v{c}-Moody algebras, and is associated to a Lie algebra $[T^{a},T^{b}]=
if^{abc}T^{c}$.
{}From our point of view, they are characterized by the OPE
\begin{equation}
J^{a}(z)J^{b}(w) \;=\; \frac{k\delta^{ab}}{(z-w)^{2}}\;+\;
                      if^{abc}\frac{J^{c}(w)}{z-w} \;+ \ldots
\end{equation}
where $k$ is the so-called level (``central element") of the
Ka\v{c}-Moody algebra. The Sugawara construction \cite{art:sugawara,bk:fuchs} 
of the stress
tensor stands for, in case of simple Lie algebras,
\be
T(z)\;=\; \frac{1}{2k+ c_{2}}\sum_{a}\, :J^{a}J^{a}:(z) \; ,
\ee
where $c_{2}$ is the value of the quadratic Casimir in the adjoint
representation, which for simply laced groups ($A_n,D_n, E_6,E_7,E_8$) 
is given by:
\begin{equation}
  c_2 = 2\left(\frac{dim(G)}{rank} - 1\right)
\end{equation}
\par
Computing $T(z)T(w)$ yields:
\begin{equation}
   c \;=\; \frac{ 2k\, dim(G)}{2k+c_{2}} \; .
\end{equation}
(This implies, in particular, that $c(SU(2)_{k=1}) = 1$).
The value of the conformal anomaly lies between the rank of the group
(the minimal possible value) and its dimension.
\par
The simplest physical representation of a KM algebra is through
a system of $2N$ two-dimensional fermions, satisfying
\be
\psi^{\mu}(z)\psi^{\nu}(w) \;=\; -\delta^{\mu\nu}\; \frac{1}{z-w} \; ,
\ee
such that the currents, the $T^{a}_{\mu\nu}$ are the generators
of $SO(2N)$ in the vector representation,
\be
   j^{a}(z) \;=\; \textstyle{\frac{1}{2}}
                  :\psi^{\mu}T^{a}_{\mu\nu}\psi^{\nu}:(z) \; ,
\ee
generate an $SO(2N)_{k=1}$ current algebra, as can easily be checked using
the above OPEs.
We can relabel the indices, using $SU(N)\subset SO(2N)$,
\be
 \psi^{\pm a} \;=\; \frac{1}{\sqrt{2}}\left(
                      \psi^{a}\pm \psi^{a+1}
                    \right) \; , \hspace{2cm}a=1\ldots N \, ,
\ee
in such a way that 
\be
 \psi^{+a}(z)\psi^{-b}(w) \;=\; - \frac{\delta^{ab}}{z-w} \; ,
\ee
\par
This system can be bosonized, {\it i.e.} written in terms of $N$ bosonic
fields $\phi_{a}$ by a technique very similar to the one 
used in the previous paragraph, {\it i.e.}
\begin{equation}
 \psi^{a} \;=\; i{\cal C}(a)\, e^{i\alpha_{a}\cdot\phi} \; ,
\end{equation}
where $\alpha_{a}^{i}=\delta_{a}^{i}$ is a weight of the vector representation
of $O(2N)$
and we are forced to introduced the quantities ${\cal C}$, called cocycles,
which satisfy
\be
\left\{
\begin{array}{lclcl}
{\cal C}(a){\cal C}(b) &=& -{\cal C}(b){\cal C}(a) \hspace{2cm}&,& a\neq b \\
{\cal C}(a){\cal C}(a) &=& 1 &,&
\end{array}
\right.
\ee
This immediatly yields an exceedingly useful
representation of the currents in terms
of the vertex operators associated to the scalar fields
\be
\begin{array}{lclclcl}
j^{+ab} &=& {\cal C}(a){\cal C}(b)e^{i\alpha_{ab}\cdot\phi}
   &\hspace{.5cm},\hspace{.5cm}&
\alpha_{ab}^{i} &=& \delta_{a}^{i}+\delta_{b}^{i} \, , \\
j^{+a\bar{b}} &=& {\cal C}(a){\cal C}(b)e^{i\alpha_{a\bar{b}}\cdot\phi}
&,& 
\alpha_{a\bar{b}}^{i} &=& \delta_{a}^{i}-\delta_{b}^{i} \, , \\
j^{a\bar{a}} &=& i\partial \phi^{a} &,
\end{array}
\ee
On the plane the corresponding charges are defined through
\be
M_{a\bar{a}} = \textstyle{\frac{1}{2\pi i}}\oint j_{a\bar{a}}(z) \; ,
\ee
This procedure is known as the Frenkel-Ka\v{c}-Segal (FKS) 
construction, although
in the particular case of $SU(2)$ it was anticipated by Halpern \cite{halpern}.
\par
It is plain that all the preceding can be generalized to an arbitrary
representation. Actually,
for an arbitrary weight $\alpha^{i}$, we have
\be
j_{a\bar{a}}(w) e^{i\alpha_{s}\cdot\phi (z)}  \;=\; 
\frac{\alpha_{s}^a}{w-z}\cdot e^{i\alpha_{s}\cdot\phi (z)} \; ,
\ee
as well as
\be
\psi^{a}e^{i\alpha_{s}\cdot\phi} \;\sim\; (z-w)^{\alpha^{a}_{s}} \,
  :i{\cal C}(a)e^{i(\alpha_{a}+\alpha_{s} )\cdot\phi}: \; .
\ee
In the particular case when $\alpha_{s}$ is the weight 
vector corresponding to a
spinor representation, {\it i.e.} $(\pm\textstyle{\frac{1}{2}}, \ldots,
\pm\textstyle{\frac{1}{2}})$, the 
preceding OPE has a characteristic square root singularity,
and is then called a spin operator, because it transforms
as an $O(2N)$ spinor. 
\par
This process is quite remarkable: Starting with two-dimensional spinors,
which are also spacetime vectors, we have constructed, by bosonization, and
vertex operators, a set of spacetime fermions.
To be specific
\begin{equation}
S_{A}(z) \;=\; {\cal C}(A)e^{i\alpha_{A}\cdot\phi (z)} \; ,
\end{equation}
and the cocycles can be chosen such that
\begin{equation}
j^{ab}(z)S_{A}(w) \;=\; \frac{1}{z-w}\, 
                  \left(\frac{1}{4}\gamma^{ab}\right)_{AB}\, S_{B}(w) \; .
\end{equation}
\section{Strings and perturbation theory}
\begin{figure}[t]
\begin{center}
\leavevmode
\epsfysize=8cm
\epsffile{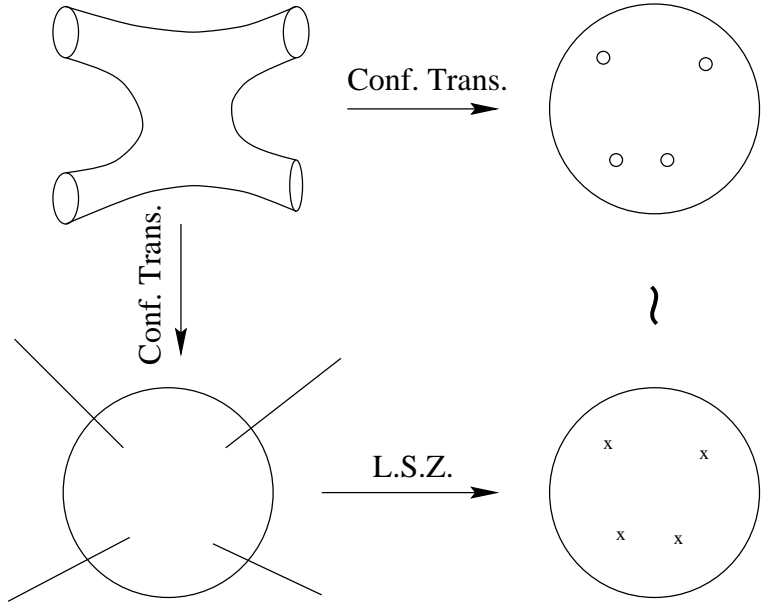}
\caption{\label{fig:vertex} Picture showing the equivalence between the 
punctured sphere (punctures being the small circles) and the sphere
with the insertion of the vertex operators, as depicted through the 
crosses.}
\end{center}
\end{figure}

In string perturbation theory, Feynman diagrams are, as was to be suspected,
thick versions of the usual Feynam diagrams.\footnote{One might say that 
the propagator is replaced by a cylinder.} One can then use conformal
invariance (in the simplest closed string case) 
to map the diagram to a Riemann surface of genus $g$, with 
punctures on which we have to insert the string wavefunctions. 
This process is depicted in fig. (\ref{fig:vertex}) for a
tree level scattering of four closed strings.
If we then apply the Lehman-Symanzik-Zimmermann reduction-technique
to this diagramm
we get a compact surface but with the insertion of some local operators,
called again vertex operators, bearing the quantum numbers 
of the external string
states \cite{d'h&p}.
Strings have been studied from the vantage point of CFT in a
classic paper by
Friedan, Martinec and Shenker (\cite{fms}), where previous work is summarized.
This is a highly technical subject and we can only give here a flavour of it.
There is a very good review by E. and H. Verlinde (\cite{vv}).
\subsection{The Liouville field: Critical and non-critical strings}
\label{sec:liouville}
Polyakov \cite{polya} apparently was the first person to take
seriously covariant 
path integral techniques
to study string amplitudes. The zero point amplitude in the simplest
closed string case is organized as
\be
Z \equiv \sum_{g=0}^{\infty} \int_{\Sigma_{g}}
\mathcal{D}\left[ g_{ab}\right] 
e^{-S_{matt}(g) - \lambda \chi(\Sigma)} \; ,
\ee
where $\Sigma_{g}$ is a two dimensional closed surface without boundary, 
with Euler characteristic $\chi(\Sigma_{g}) = 2 g - 2$ , $g$ being the genus 
($g=0$ for the sphere,
$S^2$, $g=1$ for the torus $T^2$, {\it etc.}, and 
\be
S_{matt}(g)\equiv \int_{\Sigma_{g}} d(vol)g_{ab}\partial^{a}\vec{X}
\cdot\partial^{b}\vec{X} \; .
\ee
This action is classically invariant under both two-dimensional diffeomorphisms
and Weyl rescalings. This means three parameters, which allows for a complete
gauge fixing.
In a somewhat symbolic notation, we can always reach the 
{\em conformal gauge} that is, we can write
\be
g_{ab} = e^{2\phi} e^{\xi} \hat{g}_{ab}(\tau) \; ,
\ee
where $\phi$ generates the Weyl transformation, $\xi$ the diffeomorphism, and
$ \hat{g}_{ab}(\tau)$ is a fiducial metric on $\Sigma_{g}$. To be specific, 
locally the gauge
\be
g_{zz} = g_{\bz\bz} = 0 \; ,
\ee
can always be reached through diffeomorphisms, ($\delta g_{ab} \equiv
\nabla_{a}\xi_{b} + \nabla_{b}\xi_{a}$), leaving $g_{z\bz}$ to be traded
for a Weyl rescaling.
The path integral is then reduced to
\be
Z\sim \int \mathcal{D}g \mathcal{D}X \; e^{- S_{matt}} \delta(g_{zz})
\delta(g_{\bz\bz})
\det\frac{\delta g_{zz}}{\delta \xi}\cdot
\det\frac{\delta g_{\bz\bz}}{\delta \xi} \; .
\ee
The Faddeev-Popov determinants can as usual be represented by
a ghost integral namely
\be
e^{- W_{ghost}}\equiv \int \mathcal{D}c^{z}\mathcal{D}b_{zz}\mathcal{D}c^{\bz}
\mathcal{D}b_{\bz\bz}
e^{- \frac{1}{\pi}\int d^2\sigma c^{\bz} \nabla_{z} b_{\bz\bz} + 
c^{z} \nabla_{\bz} b_{zz}} \; .
\ee
\par
It is quite useful to keep in mind that the only non-vanishing
Christoffel symbols for the metric
\be
ds^2 = e^{2\phi} dzd\bz \; ,
\ee
are $\Gamma_{zz}^z = 2 \pd \phi$ and $\Gamma_{\bz\bz}^{\bz} = 2 \bar{\pd}\phi$.
This means that some covariant derivatives are just equivalent to the
holomorphic derivative operator; that is
\be
 \nabla_{\bz} t_{z_1\ldots z_n} = \bar{\pd} t_{z_1\ldots z_n} \; ,
\ee
or
\be
 \nabla_{\bz} t^{z_1\ldots z_n} = \bar{\pd} t^{z_1\ldots z_n} \; .
\ee
Other more complicated cases can be easily worked out. Finally, let us remark
that the two-dimensional curvature is just 
\be
R^{(2)} = - 2 e^{-2\phi}\pd\bar{\pd}\phi \; .
\ee
There are some small subtleties with this path integral. First of all, there 
could be 
ghost zero modes, {\it i.e.} solutions of the equation
\be
\bar{\partial} c^{z} = 0 \; .
\ee
They are called {\em conformal Killing vectors}, and are related to 
diffeomorphisms
which are equivalent to Weyl transformations. Their number is $C_0 = 3$ for the
sphere, $C_1 = 1$ for the torus, and $C_g = 0$ for $g >1$.
%
%
To understand their meaning, let us note that under a reparametrization
the metric changes as
\be
\delta g_{\bz\bz} \equiv 2\nabla_{\bz}\xi_{\bz} = 
2 \nabla_{\bz} g_{\bz z}\xi^z =
2 g_{\bz z}\nabla_{\bz} \xi^z\; .
\ee
This then 
shows that $c$'s zero modes yield reparametrizations 
that are equivalent to a Weyl rescaling.
\par
There could also be antighost zero modes (called
{\em holomorphic quadratic differentials} by mathematicians),
{\it i.e.} solutions of
\be
\bar{\partial} b_{zz}=0 \; .
\ee
To understand what this means, let us look at the action
\be
 {\cal S}\;=\; \int_{\Sigma}\left| 
                    \delta h_{zz}\,-\, \nabla_{z}\xi_{z}
               \right|^{2} \;\neq\; 0 \; .
\ee
Minimizing this action leads to $\nabla_{\bz}b_{zz}=0$. This should hopefully 
make plausible the fact  that antighost zero modes are related to deformations
of the metric with non-vanishing action $S$ as above.
The physical meaning of 
them then lies in the fact that there are metrics on some Riemann surfaces
not related by any gauge transformation
(either Weyl or diffeomorphism)
described by the so called {\em Teichm\"uller parameters}.
There is none for the 
sphere, $B_0 = 0$, one for the torus, $B_1 = 1$, and $B_g = 3 g - 3$ for $g>1$.
The {\em Beltrami differentials} $\mu$, are defined from an 
infinitesimal variation
in such a way that
\be
\delta g_{zz} = \sum_i \delta \tau_i \mu_{i zz} + \nabla_z \xi_z
\ee
%
%
The necessity to soak up zero modes means that it is neccessary to include
a factor of 
\be
\prod _j |<\mu_j|b>|^2
\ee
and to divide by the volume of the conformal Killing vectors,
$Vol(CKV)$ in lower
genus.
\par 
At any rate we shall write the effective action as
\be
W(g)\equiv W_{matt}(\hat{g},\phi) + W_{ghost}(\hat{g},\phi) \; .
\ee
Under a Weyl transformation $\delta \phi$ the conformal anomaly implies that
\be
\delta W = \frac{26 - c}{12\pi} \int \sqrt{g} R(g)\delta \phi
\;+\;  \int\frac{\mu^{2}_{0}}{2\pi}\sqrt{g}\delta\phi \; ,
\ee
which can also be written as
\be
\frac{26 - c}{ 12 \pi} \int \sqrt{\hat{g}}(R(\hat{g}) + \Delta_{\hat{g}}\phi)
\delta\phi
\;+\; \frac{\mu^{2}_{0}}{2\pi}\int e^{2\phi}\delta\phi \; .
\ee
This is easily  integrated, yielding
\be
W(\hat{g},\phi) = \frac{26 - c}{12\pi} \int \sqrt{\hat{g}}
\left(\frac{1}{2} 
\phi \Delta_{\hat{g}}\phi + R(\hat{g}) \phi\right)\, +\, 
\int \frac{\mu_{0}^{2}}{4\pi} e^{2\phi} \; ,
\ee
where $\mu_0^2$, the world-sheet cosmological constant, comes from any explicit
violation of conformal invariance in the trace of the energy-momentum tensor,
${T^{\alpha}}_{\alpha}= \textstyle{\frac{c}{6}}R+\mu_{0}^{2}$.
\par
This action was called by Polyakov the {\em Liouville action}. It clearly 
shows the difference between {\em critical strings} 
(which in the purely bosonic
case we are considering would mean $c=26$), in which the Liouville action
appears with zero coefficient, and {\em non-critical strings}, for all 
other values of c. In spite of a tremendous effort, 
in particular by the group at the
Landau Institute, it is fair to say that our understanding 
of non-critical strings is still rather limited.
%
%
\par
It is not difficult to rewrite the full Liouville action as a non-local $R^2$
term, which is sometimes useful
\begin{eqnarray}
{\cal S}_{L} &=& \int \left[
                   \mu^{2}e^{2\phi}\,-\, 4\phi\partial^{2}\phi 
                 \right] d^{2}\xi \nonumber \\
&=& \int\mu^{2}e^{2\phi} \,-\,
    \int d^{2}\xi d^{2}\xi^{\prime} 
      e^{2\phi}e^{2\phi^{\prime}}e^{-2\phi}e^{-2\phi^{\prime}}
      4\partial^{2}\phi\partial^{2}\phi^{\prime}
      \frac{e^{ik(\xi -\xi^{\prime})}}{k^{2}}\,
        \frac{d^{2}k}{(2\pi )^{2}} \; .
\end{eqnarray}
Using then the Fourier representation of the propagator
\be
\Box^{-1}(\xi ,\xi^{\prime}) \;=\; -\int\, 
\frac{e^{ik(\xi -\xi^{\prime})}}{k^{2}}\,
        \frac{d^{2}k}{(2\pi )^{2}} \; , 
\ee
we can rewrite the Liouville action as
\be
{\cal S}_{L} \;=\; \int d^{2}\xi d^{2}\xi^{\prime}
\sqrt{g(\xi )}\sqrt{g(\xi^{\prime})}
R(\xi ) \Box^{-1}(\xi ,\xi^{\prime}) R(\xi^{\prime} ) 
\,+\, \int \mu^{2}\sqrt{g}\, d^{2}\xi \; .
\ee
Once more, what is local and what is non-local depends on the variables used
to describe the system.
\par
One of the major difficulties in understanding Liouville comes from the fact
that the {\em line element} implicit in the path integral 
measure $\mathcal{D}\phi$
is $||\delta\phi||^2\equiv\int e^{2\phi}\delta\phi^2$, which is not 
translationally invariant
(As it would have been, had we used $\sqrt{\hat{g}}$
instead of  $\sqrt{g}$).
David, Distler and Kawai \cite{ddk} made the assumption that all the 
difference can be summarized by
a renormalization of all the parameters in the action, as well as a 
rescaling of the Liouville field itself
\be
{\cal D}_{g}\phi =\;
{\cal D}_{\hat{g}}\phi\cdot
e^{-\textstyle{\frac{1}{2\pi}}\int \left(
\partial\phi\bar{\partial}\phi
-\textstyle{\frac{1}{4}}\tilde{Q}\sqrt{\hat{g}}R(\hat{g})\phi
+\mu_1^2\sqrt{\hat{g}}e^{\alpha\phi}
\right)} \; .
\ee
We can always fine-tune $\mu_{0}$ so that $\mu_{1}=0$.
\par
The original theory only depends on $g$, so we have a fake symmetry
\be
\left.
\begin{array}{lcl}
\hat{g} &\rightarrow& e^{\sigma}\hat{g} \\
\phi &\rightarrow& \phi -\frac{\sigma}{\alpha}
\end{array}
\right\} \hspace{1cm}
e^{\alpha\phi}\hat{g} \rightarrow 
  e^{\alpha\phi -\sigma}\cdot e^{\sigma}\hat{g} \; ,
\ee
Implementing it in the full path integral leads to:
\begin{eqnarray}
{\cal D}_{e^{\sigma}\hat{g}}\left(\phi -\frac{\sigma}{\alpha}\right)
{\cal D}_{e^{\sigma}\hat{g}}\left( b\right)
{\cal D}_{e^{\sigma}\hat{g}}\left( c\right)
{\cal D}_{e^{\sigma}\hat{g}}\left( X\right)
e^{-{\cal S}\left( \phi -\frac{\sigma}{\alpha}, e^{\sigma}\hat{g}\right)}
& &\nonumber \\
 && \hspace{-5cm}=\;
{\cal D}_{\hat{g}}\left(\phi\right)
{\cal D}_{\hat{g}}\left( b\right)
{\cal D}_{\hat{g}}\left( c\right)
{\cal D}_{\hat{g}}\left( X\right)
e^{-{\cal S}\left(\phi ,\hat{g}\right)} \nonumber \\
&& \hspace{-5cm}=\;
{\cal D}_{e^{\sigma}\hat{g}}\left(\phi^{\prime}\right)
{\cal D}_{e^{\sigma}\hat{g}}\left( b\right)
{\cal D}_{e^{\sigma}\hat{g}}\left( c\right)
{\cal D}_{e^{\sigma}\hat{g}}\left( X\right)
e^{-{\cal S}\left(\phi^{\prime} ,\hat{g}\right)} \; .
\end{eqnarray}
The total conformal anomaly must vanish by consistency
\be
0\,=\, c_{tot}\;=\; c(\phi )+d-26\;=\; 1+3\tilde{Q}^{2}+d-26 
\ee
This leads to
\be
\tilde{Q}\;=\; \sqrt{\frac{25-d}{3}} \; .
\ee
On the other hand, the vertex operator 
$e^{\alpha\phi }$ must be a $(1,1)$ conformal field in order for it to be 
integrated invariantly. This fixes the conformal weight
\be
\Delta\left( e^{\alpha\phi} \right) \;=\; 
 -\frac{1}{2}\alpha\left( \alpha -\tilde{Q}\right) \;=\; 1 \; ,
\ee
which in turn determines $\tilde{Q}$ in terms of $\a$;
\be
\tilde{Q}\; =\; \a \,+\, \frac{2}{\a}
\ee
Unfortunately, this shows that Liouville carries a central charge of {\em at 
least} $25$, so that the only matter which can be naively coupled to it has
$c<1$, which is not enough for a string interpretation.
\subsection{Canonical quantization and first levels of the spectrum}
The two-dimensional locally supersymmetric action generalizing the one
used above for the bosonic string reads (once the auxiliary fields have been
eliminated)
\bea
S &=-\frac{1}{8\pi}\int d^2\xi \sqrt{h} &\left[
h^{ab}\pd_a X^{\m}\pd_b X^{\n}\eta_{\m\n} 
+ 2 i \bar{\psi}^{\m}\gamma^a\pd_a\psi^{\n}\eta_{\m\n}\right. \nn\\
&&- i \left. \bar{\chi}_a \gamma^b \gamma^a \psi^{\m} 
   (\pd_b X_{\m} - \textstyle{\frac{i}{4}}
   \bar{\chi}_b\psi_{\m}) \right] \; .
\eea
This action includes a scalar supermultiplet $(X^{\m},\psi^{\m},F^{\m})$,
where $F^{\m}$ are auxiliary fields, and the two-dimensional  gravity 
supermultiplet $(e^a,\chi_a,A)$, where again $A$ is an auxiliary field.
\par
The gravitino $\chi_a$ is a world-sheet vector-spinor.
Using all the gauge symmetries of the action (reparametrizations, local
supersymmetry and Weyl transformations) it is formally possible to reach
the {\em superconformal gauge}
where $h_{ab}=\delta_{ab}$ and $\chi_a = 0$. Off the critical dimension,
however, there are obstructions similar, although technically more involved,
to those present already in the bosonic string.
\par
In this gauge, and using again complex notation, the action reads
\be
{\cal S}\;=\; \frac{1}{2\pi}\int d^{2}z\left\{
\partial_{z}X^{\mu}\partial_{\bz}X_{\mu}\,+\, 
i\left(
\psi_{z}^{\mu}\partial_{\bz}\psi^{\mu z}\,+\,
\psi_{\bz}^{\mu}\partial_{z}\psi_{\mu\bz}
\right)
\right\} \; .
\ee
The energy-momentum tensor reads
\be
T(z)\equiv T_{zz}\,=\, \frac{1}{2}\partial_{z}X^{\mu}\partial_{z}X_{\mu}
            +\frac{i}{2}\psi_{z}^{\mu}\partial_{z}\psi_{z\mu} \; ,
\ee
and is holomorphic due to its conservation, $\partial_{\bz}T_{zz} = 0$.
\par
The supercurrent (associated to supersymmetry) reads
\be
T^{F}_z\,=\, \frac{1}{2}\psi(z)^{\mu}\partial_{z}X_{\mu} \; .
\ee
This is again a holomorphic quantity, $ \partial_{\bz}T^{F}_z = 0$.
\par
We saw earlier that $\pd X^{\m}$ were conformal fields of weight $h=1$, and as 
such admit a Fourier expansion
\be
\pd X^{\m}(z) = \sum \a^{\m}_m z^{-m-1} \; .
\ee
The anti-holomorphic part enjoys a similar expansion
\be
\pd X^{\m}(\bz) = \sum \bar{\a}^{\m}_m \bz^{-m-1}\; .
\ee
Similarly, the fermionic coordinates, 
being conformal fields with $h=\frac{1}{2}$,
can be expanded as
\be
\psi^{\m}(z)=\sum b^{\m}_n z^{-n-1/2} \; .
\ee
\par 
For open strings we fix arbitrarily at one end
\be
\psi_{+}(0,\tau)=\psi_{-}(0,\tau) \; ,
\ee
and the equations of motion then allow for two possibilities at the
other end
\be
\psi_{+}(\pi,\tau)=\pm \psi_{-}(\pi,\tau) \; .
\ee
The two sectors are called Ramond (for the $+$ sign) and
Neveu-Schwarz (for the $-$ sign).
\par 
In the closed string case, fermionic fields need only
be periodic up to a sign.
\be
 \psi_{\m}(e^{2\pi i} z ) \;=\; \pm \psi_{\m}(z ) \; .
\ee
Antiperiodic fields are said to obey the R(amond) 
boundary conditions; periodic ones
are said to obey N(eveu)-S(chwarz) ones.
Please note that, owing to the half-integer conformal
weight of these fields,
periodic fields in the plane
correspond to antiperiodic fields in the cylinder.
\par
This leads, in the closed string sector, to four possible combinations
(for left as well as right movers), namely:
(R,R), (NS,NS), (NS,R), (R,NS). 
\par
The Fourier components of the energy-momentum tensor (that is
the generators of the Virasoro algebra) can be computed to be
\be
L_{m}\,=\, \frac{1}{2\pi i}\oint dz\,
 T(z) z^{m+1} \,=\, 
\textstyle{\frac{1}{2}}\left[ 
\sum_{n}:\alpha_{-n}^{\mu}\alpha_{m+n}^{\mu}: \,+\,
\sum_{r\in\mathbb{Z},\mathbb{Z}+\textstyle{\frac{1}{2}}}
(r+\textstyle{\frac{m}{2}}):b_{-r}^{\mu}b_{m+r}^{\mu}:
\right] \; ,
\ee
and the modes of the supercurrent can be similarly shown to be equal to
\be
G_{r} \;=\; \frac{1}{2\pi i}\oint dz \, T_{F} z^{r+1/2} \,=\,
\sum_{n}\, \alpha_{-n}^{\mu}b_{r+n}^{\mu} \; .
\ee
The realitiy conditions then imply as usual
\be
  L_{n}^{\dagger} \,=\, L_{-n} \hspace{.5cm},\hspace{.5cm}
  G_{r}^{\dagger} \,=\, G_{-r} \; .
\ee
In terms of the generators of the Virasoro algebra, the Hamiltonian (that is,
the generator of dilatations on the plane, or, translations in $\tau$ on
the cylinder)
reads
\be
 H\;=\; L_{0}\,+\, \bar{L}_{0} \; .
\ee
\par
The unitary operator $U_{\delta}\equiv e^{i \delta (L_0 - \bar{L}_0 )}$
implements
spatial translations in $\sigma$ in the cylinder
\be
U^{\dagger}_{\delta} X^{\m}(\tau,\sigma)U_{\delta} 
\;=\;  X^{\m}(\tau,\sigma + \delta) \; .
\ee
This transformation should be immaterial for closed strings, which means that
in that case we have the further constraint
\be
L_{0} \;=\; \bar{L}_{0}  \; .
\ee
\par
Covariant, {\em old fashioned}, canonical quantization 
can then be shown to lead to the canonical 
commutators
\bea
\left[ x^{\mu}, p^{\nu}\right] &=& i\eta^{\mu\nu} \; , \nn\\ 
\left[ \alpha_{m}^{\mu},\alpha_{n}^{\nu}\right] &=& m\delta_{m+n}\eta^{\mu\nu}
\; ,\nn \\
\left[ b_{r}^{\mu}, b_{s}^{\nu}\right] &=& \eta^{\mu\nu}\delta_{r+s} \; ,
\eea
(all other commutators vanishing)
and similar relations for the commutator of the $\bar{\alpha}$'s.
\par
This means that we can divide all modes in two sets, positive and negative,
and identify one of them (for example, the positive subset) as annihilation
operators for harmonic oscillators.
On the cylinder, the modding for the NS fermions is half-integer
and integer for the R fermions.
We can now set up a convenient Fock vacuum 
(in a sector with a given center of mass
momentum), $p^{\m}$, by
\be
\left\{
\begin{array}{lclr}
 \alpha_{m}^{\mu}\mid\, 0,p^{\mu}\rangle &=& 0 &\hspace{2cm}(m>0) \; ,\\
 b_{r}^{\mu}\mid\, 0,p^{\mu}\rangle &=& 0  & (r>0) \; ,\\
 P^{\mu}\mid\, 0,p^{\mu}\rangle &=& p^{\mu}\mid\, 0,p^{\mu}\rangle & .
\end{array}
\right.
\ee
There are a few things to be noted here:
The first one is that 
$\alpha_{-m}^{0}\mid 0\rangle$ ($m>0$) are negative-norm, 
ghostly, states, {\it i.e.}
$\langle 0\mid\alpha_{m}^{0}\alpha_{-m}^{0}\mid 0\rangle=
-m\langle 0\mid 0\rangle <0$.
The second thing to note is that, in the case of the R sector,
the zero mode operators span a Clifford algebra,
$\left\{ b_{0}^{\mu},b_{0}^{\nu}\right\} =\eta^{\mu\nu}$,
so that they can be represented in terms of Dirac $\gamma$-matrices.
\par
Recalling again that the Virasoro algebra reads
\be
\left[ L_{m},L_{n}\right] \;=\; (m-n)L_{m+n}\,+\, 
\frac{c}{12}m(m^{2}-1)\delta_{m+n} \; ,
\ee
where the central charge is equal to the dimension of the external spacetime,
$c=d$, it is plain 
that we cannot impose the vanishing of the 
$L_{n}$'s as a strong constraint. Instead we can impose them
as a weak constraint. 
Therefore we impose
\be
\begin{array}{lcl}
NS &:&
\left\{
\begin{array}{lclr}
L_{m}\mid Phys\rangle &=& 0 &\hspace{2cm}m>0 \\
\left( L_{0}-a\right)\mid Phys\rangle &=& 0 & \\
 G_r \mid Phys\rangle &=& 0 & \hspace{2cm} r\geq 1/2 \\
\left( L_{0}-\bar{L}_{0}\right)\mid Phys\rangle &=& 0 &
\end{array}
\right. \\
& & \\
R &:& \left\{
\begin{array}{lclr}
L_{m}\mid Phys\rangle &=& 0 &\hspace{2cm}m\geq 0 \\
G_{r}\mid Phys\rangle &=& 0 & r\geq 0 
\end{array}
\right.
\end{array}
\ee
%
%
Spurious states are by definition states of the form 
\be
  L_{-n}\mid\chi\rangle\;+\; \bar{L}_{-n}\mid\bar{\chi}\rangle
\ee
for $n>0$ since they are orthogonal to all physical states.
Now, all physical states which are also spurious are called null. This
then means that the observable Hilbert space is equivalent to
the physical states modulo null states (Because the latter decouple
from any amplitude).
Let us now work out, for illustrative purposes, the first levels of the 
bosonic open string spectrum. We shall repeat this exercise from different
points of view because each one illuminates a 
particular aspect of the problem.
\par
For open strings we impose Neumann conditions at
the boundary of the string world-sheet (meaning physically that 
no momentum is leaking
out of the string), {\it i.e.}
\be
X^{\prime}_{\mu} \,=\, 0 
 \hspace{1cm},\hspace{1cm} 
(\sigma =0,\pi ) \; .
\ee
The appropriate solution then reads
\be
 X^{\mu}(\sigma ,\tau )\;=\; 
x^{\mu} \,+\, \frac{1}{\pi T}p^{\mu}\tau \,+\,
\frac{i}{\sqrt{\pi T}}\sum_{n\neq 0} \alpha_{n}^{\mu}e^{-in\tau}
\cos\left( n\sigma\right) \; .
\ee
The momenta $p^{\mu}$ will determine the mass spectrum through $m^2\equiv - p^2$.
A calculation of the Hamiltonian then shows that in this case
\be
 H\,=\, L_{0}\,=\, \textstyle{\frac{1}{2}}
\sum_{n=-\infty}^{\infty}\, \alpha_{-n}^{\mu}\alpha_{n}^{\mu}\, .
\ee
%
%
When representing the open string in the upper-half plane, the
tangent projection of the energy-momentum tensor, $T_{ab}t^{b}$,
is still conserved. The condition that no energy-momentum should
flow out of the string is then that, on the boundary $(y=0)$
\be
    T_{ab}t^{a}n^{b} \;=\; 0 \; ,
\ee
or, using $t=\frac{\partial}{\partial x}$ and $n=\frac{\partial}{\partial y}$,
\be
  T_{zz} \;=\; \overline{T}_{\bar{z}\,\bar{z}} \hspace{.5cm}
(Im\, z\, =\, 0) \; .
\ee
%
%
The Fock ground state for the open string $\mid 0\, k\rangle$
satisfies
\be
  0\;=\; \left( L_{0}\,-\, a\right)\mid 0\, k\rangle \;=\; 
  \left( 2k^{2}\,-\, a\right) \mid 0\, k\rangle \; ,
\ee
which implies that $M^{2}=-k^{2}=-a/2=-a/\alpha^{\prime}$.
\par
The next level is given by the states $\mid e,k\rangle =
e_{\mu}\alpha_{-1}^{\mu}\mid 0\, k\rangle$. Imposing that it is
a physical state
\begin{eqnarray}
0 &=& \left( L_{0}\,-\, a\right)\mid e\, k\rangle \;=\;
      (2k^{2}+\alpha_{-1}\cdot\alpha_{1}-a)\mid e\, k\rangle \;=\;
      (2k^{2}+1-a)\mid e\, k\rangle \; ,\\
0 &=& L_{1}\mid e\, k\rangle \;=\;
      2k\cdot\alpha_{1}\mid e\, k\rangle \;=\;
      2k\cdot e\mid e\, k\rangle \; ,
\end{eqnarray}
so that the state has to satisfy $M^{2}=\textstyle{\frac{1-a}{\alpha^{\prime}}}$
and $k\cdot e=0$. The only available spurious state is obtained when $e\sim k$
\be
  L_{-1}\mid\, 0,k\rangle \;=\; 2k\cdot \alpha_{-1}\mid\, 0,k\rangle \; ,
\ee
which is also null if $k^{2}=0$, which happens only if $a=1$.
There are several possibilities:
\begin{itemize}
\item[$a<1$,] $M^{2}>0$: There are no null states and the constraint
$k\cdot e=0$ removes the negative norm timelike polarization. This 
corresponds to a massive vector.
\item[$a=1$,]  $M^{2}=0$: This means that $k^{\mu}=(\omega ,\vec{0},\omega )$.
The physical states are $e^{\mu}\sim k^{\mu}$, (the null states), and 
$D-2$ states of the form $ \vec{e}_{T}$.
\item[$a>1$,] $M^{2}<0$: This seems to be unacceptable.
\end{itemize}
We shall see momentarily that only for $a=1$ and $D=26$, the
{\em old coveriant quantization} coincides with BRST- and lightcone
quantization.
\subsection{Physical (non-covariant) light-cone gauge and GSO projection}
It is actually possible to solve all 
constraints (so that the remaining variables
are all physical) by going to the light-cone gauge in which the $x^{+}$
target-{\em time} is related to the world-sheet time variable $\t$ by
\be
X^{+} \;=\;
\alpha^{\prime}p^{+}\tau \; .
\ee
The $T_{ab}=0$ constraint is explicitly solved by
\bea
\partial_{\pm}X^{-} \;&=&\; 
\frac{1}{\alpha^{\prime}p^{+}}(\left(\partial_{\pm}X^{i}\right)^{2} +
i \psi_{\pm}^i \pd_{\pm}\psi_{\pm}^i)\;\\
\psi_{\pm}^{-}&=&\frac{1}{p^{+}}\psi_{\pm}^i\pd_{\pm} X^i .
\eea
This then means that both $X^{+}$ and $X^{-}$ are actually eliminated
in the Light-Cone gauge ($X^{+}$ by definition, and $X^{-}$ as a 
consequence of the above).
\par
The $ mass^2 $ operator reads (for closed strings)
\be
M^{2}\;=\; 2P^{+}P^{-}-P_{T}^{2}\;=\; 
\frac{2}{\alpha^{\prime}}
\left(
\sum_{n>0}
(\alpha^{i}_{-n}\alpha^{i}_{n}+
\bar{\alpha}^{i}_{-n}\bar{\alpha}^{i}_{n}) +
\sum_{r} r(b_{-r}^{i}b_{r}^{i}+ \bar{b}_{-r}^i\bar{b}_{r}^i) -2a
\right) \; ,
\ee
and the Hamiltonian reads
\be
H\;=\; P^{+}P^{-}\;=\; P_{T}^{2}\,+\, \frac{M^{2}}{2} \;=\;
L_{0}\,+\, \bar{L}_{0}-2a \; .
\ee
\subsubsection{Open string spectrum and GSO projection}
It is now neccessary to discriminate between the different sectors.
\par
{\bf NS sector}: The ground state, 
{\it i.e.} the oscillator vacuum, satisfies
$\alpha^{\prime}M^{2}\mid\, 0,p^{i}\rangle =-a\mid\, 0,p^{i}\rangle$.
The first excited state $b^{i}_{-1/2}\mid\, 0,p^{i}\rangle$ is 
a $(d-2)$ vector and Lorentz invariance then tells us that 
$M^{2}=0=1/2 -a_{NS}$, fixing the value for $a_{NS}$ to be $a_{NS}=1/2$.
As a consequence we see that the mass of the vacuum state is given by:
\be
   \alpha^{\prime}M^{2}_{vac} \;=\; -\frac{1}{2} \; .
\ee
Remembering that $a_{NS}$ was a normal ordering constant we can 
calculate
\be
a_{NS}\,=\, -\frac{d-2}{2}\left\{
\sum_{n=0}^{\infty} n\,-\, \sum_{r=1/2}^{\infty} r
\right\} \,=\, 
\frac{d-2}{16} \; ,
\ee
resulting in the well-known $d=10$.\footnote{In order to evaluate the 
normal ordering constant we used $\zeta$-
regularization, {\it i.e.} the vacuum energy is given by $E_{\pm}\equiv 
\pm \frac{d_T}{2}S(\a)$, where the upper sign stands for bosons, 
and the lower one for fermions, and
\be
S(\alpha )\,\equiv\,
\sum_{n=0}^{\infty}(n+\alpha ) \,=\, \zeta (-1,\alpha )\,=\,
-\textstyle{\frac{1}{2}}(\alpha^{2}-\alpha +1/6) \; ,
\ee
Hardy in his famous book \cite{art:hardy} on divergent
series starts from properties one would like for any series to hold:
Define
$\sum_{n=0}^{\infty} a_{n}=S(a)$, 
then what we want is \newline
{\tt 1)} $\sum ka_{n}=kS(a)$,
\newline
{\tt 2)} $\sum (a_{n}+b_{n})=S(a)+S(b)$ and \newline 
{\tt 3)} that if we split the sum 
we should have
$\sum_{n=1}^{\infty} a_{n}= S(a)-a_{0}$. 
In this case one can see that 1) and 3) are satisfied, but that the
second is not, since upon splitting one finds that 
\be
  \sum_{n=0}^{\infty} (n+a) \;=\; \sum_{n=0}^{\infty}n \,+\,
\alpha\sum_{n=0}^{\infty}\, 1 \;=\; 
-\frac{1}{12}\,+\, \alpha\frac{1}{2} \; ,
\ee
so that one misses out on the qudratic part.
The sum is however uniquely defined by
\be 
  S(0)\;=\; -\textstyle{\frac{1}{12}}\,=\, \zeta (-1) \hspace{.3cm},
\hspace{.3cm} S(\alpha -1)\,=\, S(\alpha )\,+\,\alpha\,-\, 1 \; .
\ee
}
%
%
%
%
\par
{\bf R sector}: Let $\mid a\rangle$ be a state such that
$b_{0}^{\mu}\mid a\rangle=\textstyle{\frac{1}{\sqrt{2}}}{\left(
\gamma\right)^{\m a}}_{b}\mid b\rangle$, meaning that it defines
an $SO(1,9)$ spinor with a priori $2^{5}= 32$ complex components,
which after imposing the Majorana-Weyl condition are reduced to $16$ real 
components (8 on shell). This number is exactly the number that can be
created with the oscillators $b_0^i$. The root of this fact is the famous
triality symmetry of $SO(8)$ between the vector and the two spinor
representations, the three of having dimension $8$.
\par
There are then two possible chiralities:
$\mid a\rangle$ or $\mid \bar{a}\rangle$,
and $\alpha^{\prime}M^{2}=0$, because oscillators do not contribute, and 
$a_R = 0$.
\par
We are free to attribute arbitrarily a given fermion number to the vacuum.
(this can be given a ghostly interpretation in covariant gauges)
\be
   (-)^{F}\mid 0\rangle_{NS}\;=\; -\mid 0\rangle_{NS}
\ee
This gives $(-)^F = -1$ for states created out of the NS vacuum by an
even number of fermion operators.
Gliozzi, Sherk and Olive (GSO) \cite{art:GSO} proposed
to truncate the theory, by eliminating all states with $(-)^F = - 1$.
It is highly nontrivial to show that this leads to a consistent theory, but
actually it does, moreover, it is spacetime supersymmetric.
We demand then that all states obey $(-)^{F}_{NS}=1$,  
thus eliminating the tachyon. This is called the GSO projection.
On the Ramond sector, we define a generalized chirality operator, such that
it counts ordinary fermion numbers and on the R vacuum,
\be
   (-)^{F}\mid a\rangle \,=\, \mid a\rangle \; ,\; 
   (-)^{F}\mid\bar{a}\rangle \,=\, -\mid\bar{a}\rangle \; ,
\ee
There is now some freedom: To be specific,
on the R sector we can demand either $(-)^{F}_{R}=1$
or $(-)^{F}_{R}=-1$. 
%
%
\par
There is a rationale for all this: The tachyon vertex
operator in two-dimensional superspace is
\be
V(p) = \int dz d\theta : e^{i p X(z,\theta)}:
\ee
which is {\em odd} with respect to $\psi \rightarrow -\psi$.
Instead, the vector vertex operator is given by:
\be
V_{\m} = \int dz d\theta : i \mathcal{D} X_{\m} e^{i p X(z,\theta)}:
\ee
which is even. To say it in other words: if we accept as physical the vector
boson state, GSO amounts to projecting away all states related to it through
an {\em odd} number of fermionic $\psi$-oscillators.
\subsubsection{Closed string spectrum}
The difference with the above case
is that one has to consider as independent 
sectors the left and right movers.
\par
{\bf (NS,NS) sector}: The composite ground state is the tensor product
of the NS vacuum for the right movers and the NS vacuum for the 
left-movers, and as such it drops out after the GSO projection.
The first states surviving the GSO projection, that is 
$(-1)^F=(1,1)$, are
\be
    \bar{b}^{i}_{-1/2}\mid 0\rangle_{L}\otimes
    b_{-1/2}^{j}\mid 0\rangle_{R} \; .
\ee
Decomposing this in irreducible representations of the little group $SO(8)$ 
yields
 $\mathbf{1}\oplus \mathbf{28}\oplus \mathbf{35}$
showing that it is equivalent
to a scalar $\phi$, the singlet, 
an antisymmetric 2-form field $B_{\mu\nu}$, the $\mathbf{28}$,
and a symmetric 2-tensor field $g_{\mu\nu}$, the $\mathbf{35}$.
\par
{\bf (R,R) sector, type IIA}: The massless
states are of the form $(-1)^F=(-1,1)$
\be
   \mid\bar{a}\rangle_{L}\otimes\mid b\rangle_R \; ,
\ee
and decompose as $\mathbf{8}_{v}\oplus \mathbf{56}_{v}$, corresponding to 
a vector field, a one-form $A_{1}$, and a 3-form field, $A_3$.
\par
{\bf (R,R) sector, type IIB}: The massless states, with $(-1)^F=(1,1)$ are
\be
     \mid a\rangle_{L}\otimes \mid b\rangle_{R} \; ,
\ee
and they decompose as $\mathbf{1}\oplus \mathbf{28}\oplus \mathbf{35}_{s}$
corresponding to
a pseudo scalar, $\chi$, a 2-form field, $A_{2}$, and a selfdual
4-form field, $A_4$.
\par
{\bf (R,NS) sector, Type IIA}: The first GSO surviving states, 
with $(-1)^F=(-1,1)$, are
\be
  \mid\bar{a}\rangle_{L}\otimes b^{i}_{-1/2}\mid 0\rangle_{R} \; ,
\ee
and they decompose as $\mathbf{8}_{s}\oplus \mathbf{56}_{s}$.
\par
{\bf (R,NS) sector, Type IIB}: The first GSO surviving states, with
 $(-1)^F=(1,1)$ are
\be
 \mid a\rangle_{L}\otimes b^{i}_{-1/2}\mid 0\rangle_{R} \; ,
\ee
and they decompose as $\mathbf{8}_{c}\oplus \mathbf{56}_{c}$.
\par
{\bf (NS,R) sector, Type IIA}: The first GSO surviving states, with
$(-1)^F=(1,-1)$ are
\be
 \bar{b}^{i}_{-1/2}\mid 0\rangle_{L}\otimes\mid\bar{a}\rangle_{R} \; ,
\ee
and decompose as $\mathbf{8}_{s}\oplus \mathbf{56}_{s}$.
\par
{\bf (NS,R) sector, Type IIB}: The first GSO surviving states, 
with $(-1)^F=(1,1)$,
are
\be
\bar{b}^{i}_{-1/2}\mid 0\rangle_{L}\otimes\mid\, a\rangle_{R} \; ,
\ee
and decompose as $\mathbf{8}_{c}\oplus \mathbf{56}_{c}$.
The $\mathbf{56}_{c}$ corresponds to 
two gravitinos.
\subsection{BRST quantization and vertex operators}
Let us first consider the bosonic string.
We know that the total conformal anomaly
is given by
\be
 c_{total} \;=\; c(X)\,+\, c(ghosts) \;=\; d-26 \; .
\ee
We define a classically consererved fermion number, the ghost number,
operator (and the corresponding definition of the ghost number of a field) 
through
\be
\left\{
\begin{array}{lcl}
j_{gh}(z) &=& -:bc: \; , \\
j_{gh}(z)\phi (w) &=& \frac{N_{g}}{z-w}\phi(w) \; .
\end{array}
\right.
\ee
The BRST charge is then defined as usual in gauge
theories (See for example \cite{bk:henteit})
\begin{equation}
{\cal Q}\,=\, \oint \frac{dz}{2\pi i}\, j_{BRST}(z) \,=\,
 \oint \frac{dz}{2\pi i}\, c(z)
       \left[ T(z)\,+\,
          \textstyle{\frac{1}{2}}T_{gh}(z) 
       \right] \; .
\end{equation}
\par
It is possible to show that ${\cal Q}^{2}=0$ iff $d=26$.
\par
{}From the point of view of the gauge-fixed covariant theory, physical 
states correspond to the BRST cohomology 
(that is: BRST closed states modulo BRST exactness).
\par
Now, it is not difficult to show that those correspond to 
bosonic primary fields of conformal dimension (1,1):
\be
\oint \frac{dz}{2\pi i} j_{BRST}(z)V(w) \;=\;
\oint \frac{dz}{2\pi i} c(z)\left[ 
  \frac{h V(w)}{(w-z)^{2}} \,+\, \frac{\partial V (w)}{z-w}
\right] \,=\, h\pd c V\,+\, c\pd V \; ,
\ee
which is kosher iff $h=1$, because in that case it is equal to
$\pd(cV)$, which vanishes upon integration
over the insertion point of the vertex operator on the Riemann surface 
representing the world-sheet of the string.
\par
%
%
The  usual $Sl(2,\mathbb{C})$ ghost vacuum is 
defined as for any conformal field
by
\be
\left\{
\begin{array}{lclclcl}
b_{n}\mid 0\rangle_{gh} &=& 0 &,& n\geq -1 &,& (b=\sum b_{n}z^{-n-2}) \; ,\\
c_{n}\mid 0\rangle_{gh} &=& 0 &,& n\geq 2  &,& (c=\sum c_{n}z^{-n+1}) \; .
\end{array}
\right.
\ee
We know that canonical quantization yields
\be
  b_{0}^{2}\,=\, c_{0}^{2}\,= 0 \hspace{.5cm},\hspace{.5cm}
  \left\{ b_{0},c_{0}\right\}\;=\; 1 \; .
\ee
On the other hand, for any conformal field 
\be
[L_{n},\phi_{m}]=[n(h-1)-m]\phi_{n+m}
\ee
so that in particular we can lower the $L_0$ value of the 
$SL(2,\mathbb{C})$ vacuum, $\mid 0\rangle_{gh}$, using
\be
[L_0,c_1] \;=\;  - c_1\; .
\ee
The eigenvalue of $L_0$ is preserved by $c_0$, {\it i.e.}
\begin{eqnarray}
[L_0,c_0] &=& 0  \; , \nonumber \\
L_{0}c_{1}\mid 0\rangle_{gh} &=&  -c_{1}\mid 0\rangle_{gh} + c_1 L_0 \mid 0
\rangle_{gh}  \; , 
\end{eqnarray}
All this implies that the true lowest weight states are
\be
\begin{array}{lclcl}
c_{1}\mid 0\rangle_{gh} &\equiv & c(0)\mid 0\rangle_{gh} 
  &=& \mid\downarrow\rangle \; , \\
c_{0}c_{1}\mid 0\rangle_{gh} &=& -c\partial c(0)\, \mid 0\rangle_{gh} 
  &=& \mid\uparrow\rangle \; .
\end{array}
\ee
It is not difficult to check that these new states are both of zero norm,
$\langle\uparrow\mid\uparrow\rangle =
\langle\downarrow\mid\downarrow\rangle = \,_{gh}\langle 0\mid c_{-1}c_1 \mid
0\rangle_{gh}=\,_{gh}\langle 0\mid c_{-1}c_1(b_0c_0 + c_0b_0) \mid
0\rangle_{gh}=0 $; but we can write
\be
\langle\downarrow\mid\uparrow\rangle \;=\;
\langle 0\mid c_{-1}c_{0}c_{1}\mid 0\rangle \;\equiv\; 1 \; .
\ee
A small calculation then shows that
\begin{eqnarray}
b_{0}\mid\uparrow\rangle &=& b_{0}c_{0}c_{1}\mid 0\rangle \;=\;
   c_{1}\mid 0\rangle \;=\; \mid\downarrow\rangle \; , \\
c_{0}\mid\downarrow\rangle &=& c_{0}c_{1}\mid 0\rangle \;=\;
   \mid\uparrow\rangle \; ,
\end{eqnarray}
{}from which one can infer that 
\be
N_{g}\left(\mid\uparrow\rangle\right)\,=\, \frac{1}{2} 
\hspace{.3cm},\hspace{.3cm}
N_{g}\left(\mid\downarrow\rangle\right)\,=\, -\frac{1}{2}
\hspace{.3cm},\hspace{.3cm}
N_{g}\left(\mid 0\rangle\right)\,=\, -\frac{3}{2} \; ,
\ee
that is: The vacuum carries three units of ghost number.
It is quite easy to prove that, denoting by $z_{ij}\equiv z_i - z_j$,
\be
\langle 0\mid c(z_{1})c(z_{2})c(z_{3})\mid 0\rangle \;=\; 
z_{23}z_{12}z_{13} \; .
\ee
The appropriate projector is\footnote{Please note that 
${\cal P}^{\prime}\equiv \mid 0\rangle\langle 0\mid$ is null, since 
$\langle 0\mid 0\rangle =\langle 0\mid c_{0}b_{0}+b_{0}c_{0}\mid 0\rangle 
=0$.}
\be
{\cal P} \;=\; \mid 0\rangle\langle 0\mid c_{-1}c_{0}c_{1} \; ,
\ee
which obviously obeys
\be
\mathcal{P}^2 = \mathcal{P} \; .
\ee
%
\par
It is instructive to rederive some facts of the mass spectrum, using BRST 
techniques, at least for the bosonic string, to avoid technicalities:
On physical states we need have $b_{0}\mid\psi\rangle = 0$, which
can be used to derive\footnote{We shall denote by $L$ the {\em level}
of a given state, {\it i.e.} the number of creation operators
needed for the creation of the state out of the vacuum.}
\be
\left\{ {\cal Q}, b_{0}\right\} \mid\psi\rangle \;=\;
\left( L_{0}^{X}+L_{0}^{gh}\right)\mid\psi\rangle \;=\; 
\left( 2k^{2}+L-1\right)\mid\psi\rangle \;=\; 0\; ,
\ee
implying that $M^{2}=\frac{L-1}{2}$.
\par
At oscillator level zero we can write:
\be
 0 \;=\; {\cal Q}\mid\downarrow\, 0\, k\rangle \;=\;
   (2k^{2}-1)c_{0}\mid\downarrow\, 0\, k\rangle \; ,
\ee
implying $k^{2}=\textstyle{\frac{1}{2}}$ so that this state
is the tachyon in the open string sector.
\par
At oscillator level $L=1$, there are 26+2 possible
states having $M^{2}=0$ and they can be
parametrized by
\be
\mid\psi\rangle \;=\; 
      \left(
         e\cdot a_{-1}\,+\, \beta b_{-1}\,+\, \gamma c_{-1}
      \right) \mid\downarrow\, 0\, k\rangle \; .
\ee
Imposing that they be physical states,
{\it i.e.} ${\cal Q}\mid\psi\rangle =0$,
leads to the constraints
\be
k^{2}\;=\; 0 \hspace{.5cm}, \hspace{.5cm}
k\cdot e\;=\; \beta \;=\; 0 \; ,
\ee 
thus reducing the number of independent components to 26.
In order to find the true number of independent states, we need 
to throw out the exact states
\be
{\cal Q}\mid\chi\rangle\;=\; 
2\left( 
  k\cdot e^{\prime}c_{-1}\,+\, \beta^{\prime}k\cdot \alpha_{-1}
\right)\mid\downarrow\, 0\, k\rangle
\ee
This means that $c_{-1}\mid\downarrow\, 0\, k\rangle$ is exact
and that $e_{\mu}\sim e_{\mu}+2\beta^{\prime}k_{\mu}$, yielding
24 positive norm states for a massless vector.
\par
BRST reduces to old covariant for ghosts in the ground state, {\it i.e.}
every cohomology class includes a state of this form.
\subsubsection{Superstrings}
The $N=1$ superconformal algebra is usually represented by the
quantity $\hat{c}(=\textstyle{\frac{2}{3}}c$), and
is generated by definition by the stress tensor $T$
and the supercurrent $T_{F}$,
with the OPEs
\begin{eqnarray}
T(z)T(w) &=& \frac{\textstyle{\frac{3}{4}}\hat{c}}{(z-w)^{4}} \,+\,
             \frac{2T(w)}{(z-w)^{2}} \,+\,
             \frac{\partial T(w)}{z-w} \; +\,\ldots \\
& & \nonumber \\
T(z)T_{F}(w) &=& \frac{\textstyle{\frac{3}{2}}T_{F}(w)}{(z-w)^{2}}\;+\;
             \frac{\partial T_{F}(w)}{z-w} \; +\, \ldots \\
& & \nonumber \\
T_{F}(z)T_{F}(w) &=& \frac{\textstyle{\frac{1}{4}}\hat{c}}{(z-w)^{3}}\;+\;
                     \frac{\textstyle{\frac{1}{2}}T(w)}{z-w} \; +\, \ldots
\end{eqnarray}
In the superstring there are two basic superconformal fields
\begin{eqnarray}
T(z) &=& -\textstyle{\frac{1}{2}}:\partial X^{\mu}\partial X_{\mu}:\,-\,
         \textstyle{\frac{1}{2}}:\partial\psi^{\mu}\cdot \psi_{\mu}:\; ,\\
T_{F} &=& -\textstyle{\frac{1}{2}}:\psi_{\mu}\partial x^{\mu}:\; ,
\end{eqnarray}
and the ghost action corresponds to the superfields $B=\beta +\theta b$
and $C=c +\theta\gamma$. This then means that the contribution of
the combined ghost system to the conformal anomaly is
$c_{gh}=-26+11=-15$:
\begin{eqnarray}
T_{gh} &=& -2:b\partial c:\,-\, :\partial b\cdot c:\,-\,
           \textstyle{\frac{3}{2}}:\beta\partial\gamma :\,-\,
           \textstyle{\frac{1}{2}}:\partial\beta\cdot\gamma : \; ,\\
T_{F,gh}&=& \textstyle{\frac{1}{2}}:b\gamma :\,-\,
            :\partial\beta\cdot c:\,-\,
            \textstyle{\frac{3}{2}} :\beta\partial c:\; .
\end{eqnarray}
For any superalgebra we can expand
\be
T_{F}(z) \;=\; \textstyle{\frac{1}{2}}
               \sum_{r\in\mathbb{Z}+NSR} z^{-\textstyle{\frac{3}{2}}-r}
               G_{r} \; ,
\ee
where NSR is $0$($\textstyle{\frac{1}{2}}$) for R (NS, resp.).
Any Ramond field is periodic on the cylinder, but on the plane
\be
 V^{R}\left( e^{2\pi i}z\right) \;=\; - V^{R}\left( z\right) \; .
\ee
\par
{\bf NS Sector}:
There exists now a finite subalgebra $OSp(1\mid 2)$, generated by
$\left[ L_{0}, L_{\pm 1}, G_{\pm 1}\right]$, with 
$G_{-\textstyle{\frac{1}{2}}}^2
= L_{-1}$ and its vacuum is defined by
\be
\begin{array}{lclccrlclcl}
L_{n}\mid 0\rangle &=& 0 &:& n\geq -1 &\hspace{.4cm},&
\langle 0\mid L_{n} &=& 0 &:& n\leq 1 \; , \\
G_{r}\mid 0\rangle &=& 0 &:& r\geq -\textstyle{\frac{1}{2}} &\hspace{.4cm},&
\langle 0\mid G_{r} &=& 0 &:& r\leq \textstyle{\frac{1}{2}} \; ,
\end{array}
\ee
\par
{\bf R Sector}:
The superconformal anomaly implies that
on the plane $G_{0}^{2}=L_{0}-\textstyle{\frac{\hat{c}}{16}}$
(whereas on the cylinder $G_{0}^{2}=L_{0}$
and besides, $\left[ G_{0}, L_{0}\right] = 0$, which means that
there are now two different ground states, $\mid h^{+}\rangle$ which
is degenerate with $\mid h^{-}\rangle=
G_{0} \mid h^{+}\rangle$:
\be
G_0\mid h^{-}\rangle \,=\, 0
\hspace{.5cm},\hspace{.5cm}
G_0^2 \mid h^{\pm}\rangle \,=\, 0 \; .
\ee
They both obey
\be 
L_{0}\mid h^{\pm}\rangle= \textstyle{\frac{\hat{c}}{16}}\mid h^{\pm}\rangle ,
\ee
such that
\be
\langle h^{-}\mid h^{-}\rangle \;=\; 
 \langle h^{+}\mid G_{0}^{2}\mid h^{+}\rangle \;=\; 0 \; .
\ee
We then infer by completeness the existence of spin fields
\be
 \mid h^{\pm} \rangle \;=\; S^{\pm}(0)\mid 0\rangle \; ,
\ee
which furthermore satisfy
\be
\hat{G}_{0} S^{+}(z) \;=\; S^{-}(z) \; .
\ee
The OPEs then, by consistency, neccessarily read
\begin{eqnarray}
T_{F}(w)S^{+}(z) &=& \frac{1}{2}\frac{1}{(w-z)^{3/2}} S^{-}(z) \; , \\
T_{F}(w)S^{-}(z) &=& \frac{1}{2}\left( h-\frac{\hat{c}}{16}
                     \right) \frac{1}{(w-z)^{3/2}} S^{+}(z) \; ,
\end{eqnarray}
These fields interpolate between the NS and the R sectors, because they
transform the NS groundstate into the R groundstate.
In a somewhat symbolic notation
\be
\phi^{NS}_{f}\left( e^{2\pi i}z\right) S^{\pm}(0) \;=\;
-\phi^{NS}_{f}\left( z\right)S^{\pm}(0) \; .
\ee
\par
Let us finally mention that the fact that the vacuum 
carries three units of ghost charge is related to the fact that in order to 
bosonize the $(b,c)$ system, we had to introduce a
background charge $Q(b,c)= -i\textstyle{\frac{3}{2}}$.
Had we done the same exercise for the superconformal $(\b ,\gamma)$ system,
we would have seen that
at tree level, the total superconformal ghost charge adds to $-2$,
$Q(\b,\gamma)= +1$,
but it can be traded between different vertex operators within a 
BRST invariant correlation function, symbollically
\be
\langle 0\mid e^{3\sigma (0)-2\phi (0)}\mid 0\rangle \;=\; 1 \; ,
\ee
where $\sigma$ bosonizes the $(b,c)$ and $\phi$ bosonizes the 
$(\beta ,\gamma )$ system.
This is the basic reason why it is neccessary to have a different
representative of each vertex operator in every ghost number sector, and to 
combine them in any correlator so that they match as above. This procedure
was called {\em picture changing} by Friedan, Martinec and Shenker. 
See \cite{vv,bk:ketov}
for further details.
\subsection{Scattering amplitudes and the partition function}
We are now prepared to compute some
amplitudes (For more detail see \cite{GSW,d'h&p}).
The simplest thing would be
the open string tachyon-tachyon scattering amplitude. 
The tree level (lowest order)  contribution
will be given by the correlator
\be
A_4\equiv \langle\int dz_3 c(z_1) V_1 c(z_2)V_2 V_3 c(z_4) V_4 \rangle \; ,
\ee
where the vertex operator for the tachyon is given by
\be
V_i \; \equiv\; :e^{i k^{(i)}_{\m}X^{\m}(z_i)}: \; .
\ee
The ghost factors are neccessary in order to cancel the ghost charge
of the vacuum. We can arbitrarily choose their
positions: It can be seen that this is equivalent to correctly taking into
account the three conformal Killing vectors of the sphere
\be
 CKV \; :\; V\;=\; \left(\alpha +\beta z+\gamma z^{2}\right)\partial \; .
\ee
This leads to the basic string amplitude 
\be
A_{4}\;=\; \int dz_{3}\, z_{12}z_{14}z_{24}
 \prod_{i<j}e^{p_{i}\cdot p_{j}\log z_{ij}} \; ,
\ee
where $z_{ij}=z_{i}-z_{j}$.
Since this should not depend on the positions
at which we have placed the ghosts,
we can choose
\be
\left.
\begin{array}{c}
z_{1}\rightarrow\infty \\
z_{2}\rightarrow 1\\
z_{4}\rightarrow 0
\end{array}
\right\} \hspace{.4cm}
 z_{12}^{p_{1}\cdot p_{2}+1}
 z_{14}^{p_{1}\cdot p_{4}+1}
 z_{24}^{p_{2}\cdot p_{4}+1}
 z_{13}^{p_{1}\cdot p_{3}}
 z_{23}^{p_{2}\cdot p_{3}}
 z_{34}^{p_{3}\cdot p_{4}} \;=\;
 \left( 1-z_{3}\right)^{p_{2}\cdot p_{3}} z_{3}^{p_{3}\cdot p_{4}} \; ,
\ee
where we have used the fact that $\sum_{i} p_{i}=0$ and $p_{i}\cdot p_{i}=2$.
As a result one ends up with the Veneziano amplitude:
\be
A^{(g=0)}_{4} \;=\; \int_{0}^{1}dz_{3}\, 
\left( 1-z_{3}\right)^{p_{2}\cdot p_{3}} z_{3}^{p_{3}\cdot p_{4}} \; .
\ee
\par
If we recall the usual definition of the
Mandelstam parameters, $s\equiv - (p_1 + p_2)^2$
$t\equiv -(p_2 + p_3)^2$ and remember that we are dealing with tachyons,
so that $p_i^2=2$, we can notice that the Veneziano amplitude can be written
as Euler's Beta function, or, in terms of Gamma functions, as
\be
A^{(g=0)}_{4}=\frac{\Gamma(-1 - t/2)\Gamma(-1 - s/2)}{\Gamma(-2 - (t + s)/2)}
\; . \label{eq:tsdual}
\ee
Now, it is well known that Euler's $\Gamma(z)$ has poles
for all negative integers,
$z\in \mathbb{Z}^{-}$.
Here this translates into poles in the Veneziano amplitude
at integer values of the {\em Regge trajectory}
\be
\a(s) = 1 + s/2
\ee
or, in the t-channel,
\be
\a (t) = 1 + t/2
\ee
The fact that the amplitude (\ref{eq:tsdual}) does not change
by interchanging $t$ and $s$
signals the 
(much sought for) property of {\em duality}
(in the old sense of the word)
for physical amplitudes. There is a superb historical introduction on this,
and related, matters in the first chapter
of \cite{GSW}.
\par
The one-loop (genus one in the closed string case, that is a torus)
 amplitude is called the partition function.
%
%
\begin{figure}[t]
\begin{center}
\leavevmode
\epsfysize=5cm
\epsffile{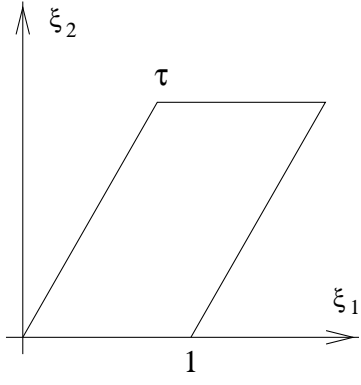}
\end{center}
\caption{\label{fig:torus}A general torus}
\end{figure}
\par
Any two-dimensional torus (That is $\mathbb{R}^2/\Lambda$, where
$\Lambda $ is a two-dimensional lattice) can be put,
by a conformal transformation,
in the canonical form represented in the figure,
where the lattice is generated by
the complex numbers $1=(1,0)$ and $\t =(\t_1,\t_2)$.
Let us use the coordinates
$z=\xi_{1}+i\xi_{2}$ and $ds^{2}=|dz|^{2}$.
The periodicity conditions on the bosonic fields are
\be
\left\{
\begin{array}{lcl}
\phi (\xi_{1}+1, \xi_{2}) &=& \phi (\xi_{1},\xi_{2}) \; ,\\
\phi (\xi_{1},\xi_{2}+\tau_{2}) &=& \phi (\xi_{1}+\tau_1,\xi_{2}) \; .
\end{array}
\right.
\ee
Instead of twisted boundary conditions we could as well use 
normal periodic ones,
but with a different metric, namely
\be
z\,=\, \xi_{1}+\tau\xi_{2} \hspace{.5cm},\hspace{.5cm}
ds^{2}\,=\, |dz|^{2} \hspace{.5cm},\hspace{.5cm}
g\,=\, \left(
\begin{array}{cc}
1 & \tau_{1} \\
\tau_{1} & |\tau |^{2}
\end{array}
\right) \; .
\ee
The total area is given by
\be
 A\;=\; \int\sqrt{g}\, d^{2}\xi \;=\; \tau_{2} \; .
\ee
The Beltrami differentials, corresponding to constant
deformations (zero modes) which can not be represented as derivatives
of periodic functions, are given by $\delta z=\delta\tau\xi_{2} =\delta\tau\, 
\frac{z-\bar{z}}{2i\tau_{2}}$. The non-gauge deformations of the metric are 
given by
\be
\delta g_{zz} \;=\; -\frac{\delta\tau}{2i\tau_{2}} \;=\; 
\frac{i}{2\tau_{2}}\delta\tau \;\equiv\; \mu_{zz}\delta\tau \; .
\ee
The quadratic differentials, on the other hand, are trivial constants:
$\phi = (dz)^{2}$, $(\phi ,\phi )=\tau_{2}$ and $(\mu ,\phi )=1$, so that the
operation of 
projecting the antighost zero modes (4.16) into an orthonormal basis of
quadratic differentials simply yields:
\be
\frac{ \left|\langle\phi\mid\mu\rangle\right|^{2}}{\langle\phi\mid\phi\rangle}
\;=\; \frac{1}{\tau_{2}} \; .
\ee
The CKV are given by the condition $\partial \bar{z}=0$, and 
the no-poles condition 
then means that
$\bar{z}=c$, where $c$ is a constant.
This means that
\be
  Vol\left( CKV\right) \;=\; \tau_{2} \; .
\ee
\par
The partition function then reads
(using $det'\Delta = \t_2 |\eta|^4)$:\footnote{The
easiest way to get this result is to perform a path integral
calculation in the light cone gauge using the normal mode
expansion
\be
X^i = \sum_{n_1,n_2\in\mathbb{Z}} X^i_{n_1,n_2} e^{2\pi i n_1 \xi_1} 
e^{2\pi i n_2 \xi_2}
\ee
and computing the ensuing determinant using $\zeta$-function 
techniques (which in this particular case lead to a simple Epstein function).}
\be
{\cal Z} \;=\; \int_{F} \frac{d^{2}\tau}{(\tau_{2})^{2}}\,
(\tau_{2})^{-12}\left|\eta (\tau )\right|^{-48} \;=\; 
\int_{F} \frac{d^{2}\tau}{(\tau_{2})^{2}}\, \chi (\tau ,\bar{\tau}) \; ,
\ee
where use has been made of the Dedekind function
\be
\eta (\tau )\;=\; q^{1/24}\prod_{n=1}^{\infty}\left( 1-q^{n}\right) 
\hspace{.5cm}:\hspace{.3cm} q\,\equiv\, e^{2\pi i\tau} \; .
\ee
There are, however, diffeomorphisms not connected to the identity
which we still have to factor out. They are generated by Dehn twists:
Once we identify two sides of the square to build a cylinder,
we twist one of the
boundaries by an integer multiple of $2\pi$ before identification.
It is not difficult to show that they are generated by
\be
S:\hspace{.3cm}\tau\; \rightarrow \tau^{\prime}\;\equiv\; - 1/\tau \; ,
\ee
and
\be
T:\hspace{.3cm}\tau\; \rightarrow \tau^{\prime}\;\equiv\; \tau + 1 \; .
\ee
That is, modular parameters related by any combination of the preceding
transformations, $(S,T)$ are diffeomorphic, and, as such, they should not
be counted twice. This set of transformations actually forms a group,
called the {\em modular group}, $\Gamma\equiv SL(2,\mathbb{Z})$.
All this physically means that we have to restrict the integration in the
 one-loop amplitude to a fundamental region, $F$, of the modular group, 
which is by
definition such that any $\tau\in H$ (the upper complex plane) can be mapped
into $F$ by a unique element of the modular group and, besides, no two different
elements in $F$ are gauge equivalent.
A convenient choice of $F$ is the intersection of
\be
-\frac{1}{2}\leq \tau_1 \leq +\frac{1}{2} \; ,
\ee
with
\be
|\tau| > 1 \; .
\ee
\par
Naturally, in order for this restriction to $F$ to be consistent, the 
{\em integrand} has to be modular invariant (Which is, let us recall, a 
remainder of two-dimensional general covariance).
This property is one of the most
important symmetries of string theory, lying at the root of its consistency,
in different incarnations: anomaly cancellation, spacetime supersymmetry,
finiteness of the amplitudes and so on.
This basic fact is quite easy to establish, using the fact that
the {\em Poincar\'e} measure $\frac{d^2 \tau}{(\tau_2 )^2}$ is modular
invariant by itself. Let us assume that the amplitude is given by some integral
over the upper half plane, $H$,
\be
A = \int_{H}\frac{d^2 \tau}{(\tau_2)^2} M(\tau) \; ,
\ee
If $M$ is modular invariant (and only then), we can write
\bea
A &=& \sum_{\gamma\in \Gamma} \int_{\gamma\, F}
 \frac{d^2 \tau}{(\tau_2 )^2}
M(\tau)\nn\\
&=& \sum_{\gamma\in \Gamma} \int_{ F}\frac{d^2 \tau}{(\tau_2 )^2}
M(\tau)= vol(\Gamma)\int_{ F}\frac{d^2 \tau}{(\tau_2 )^2}M(\tau) \; .
\eea
Please note that due to modular invariance the dangerous ultraviolet region
$\tau_2\rightarrow 0$ is excluded completely from the onset: Any possible
divergence can always be interpreted as an infrared one.
\par 
Under this $Sl(2,\mathbb{Z})$ transformation one finds that
\be
\begin{array}{lclclcl}
d^{2}\tau &\rightarrow& |c\tau +d|^{-4}d^{2}\tau &\hspace{.4cm},\hspace{.4cm}&
 \eta (\tau +1) &=& e^{i\pi /12}\eta (\tau ) \; , \\
\tau_{2}&\rightarrow& |c\tau +d|^{-2}\tau_{2} &,&
 \eta (-\frac{1}{\tau}) &=& \sqrt{ -i\tau}\, \eta (\tau )\; ,
\end{array}
\ee
showing that the partition function indeed enjoys modular invariance.
\par
There is a very important Hamiltonian interpretation of the preceding
corresponding to a propagation during a time $\tau_2$ and performing
an spatial twist of $\tau_1$. Remembering that $L_0+\bar{L}_0$ generates
time translations, and that $L_0-\bar{L}_0$ generates spatial translations,
we are led to
\be
\chi \;\sim\; Tr\, e^{2\pi i\tau_{1}\left( H_{R}-H_{L}\right)}
                   e^{-2\pi\tau_{2}\left( H_{R}+H_{L}\right)}
          \;=\; \int\frac{d^{24}p}{(2\pi )^{24}}\, 
            e^{-2\pi p^{2}\tau_{1}} e^{4\pi\tau_{2}} 
          Tr\, \bar{q}^{H_{L}}q^{H_{R}} \; .
\ee
In order to evaluate this, we calculate
\be
Tr\, \bar{q}^{N_{L}}q^{N_{R}} \,=\,
\frac{1}{(\tau_{2})^{12}}e^{4\pi\tau_{2}}\prod_{n=1}^{\infty}
\left( 1-\bar{q}^{n}\right)^{-24}\left( 1-q^{n}\right)^{-24} \,=\,
\frac{1}{(\tau_{2})^{12}} \mid\eta (\tau )\mid^{-48} \; .
\ee
The meaning of this becomes clear when we remember  that
\be
H_{L}\;=\; \frac{p_{T}^{2}}{2}+N_{L}-1 \hspace{.5cm},\hspace{.5cm}
H_{R}\;=\; \frac{p_{T}^{2}}{2}+N_{R}-1 \; ,
\ee
after which we can expand in a Laurent series
\be
\chi\;\sim\; \frac{1}{|q|^{2}} \,+\, \frac{24}{q}\,+\, \frac{24}{\bar{q}}
\,+\, 576 \, +\ldots
\ee
The quadratic pole represents the tachyon and the 576 
($= 299+276+1$) are the massless string states.
The whole string spectrum can be 
easily reconstructed in this way.
\subsubsection{Spin structures}
Seiberg and Witten \cite{sw} first realized that on a torus each fermion
can be characterized by the signs
it gets when coming back to the same point after
encircling one of the two homology cycles
\begin{eqnarray}
\psi (\xi_{1}+1,\xi_{2}) &=& \pm\psi (\xi_{1},\xi_{2}) \; , \nonumber \\
\psi (\xi_{1},\xi_{2}+1) &=& \pm\psi (\xi_{1},\xi_{2}) \; .
\end{eqnarray}
The corresponding contributions to the partition functions are given by:
\be
\begin{array}{lclclcl}
A^{++}(\tau )&=& \eta_{++}Tr\, e^{2\pi i\tau H_{R}}(-)^{F}
&\hspace{.5cm},\hspace{.5cm}&
A^{+-}(\tau )&=& \eta_{+-}Tr\, e^{2\pi i\tau H_{R}} \; ,\\
A^{--}(\tau )&=& \eta_{--}Tr\, e^{2\pi i\tau H_{NS}} &,&
A^{-+}(\tau )&=& \eta_{-+}Tr\, e^{2\pi i\tau H_{NS}}(-)^{F} \; ,
\end{array}
\ee
where the Hamiltonians are given by (remembering that the normal ordering 
constants can be computed by the general formula as 
$\frac{1}{3}=-\frac{d_T}{2}\zeta(-1,0)$ and $-\frac{1}{6}=-\frac{d_T}{2}
\zeta(-1,1/2)$)
\begin{eqnarray}
H_{R} &=& \sum_{m=1}^{\infty} mb^{i}_{-m}b^{i}_{m} \,+\, \frac{1}{3} 
      \; , \\
H_{NS} &=& \sum_{r=\textstyle{\frac{1}{2}}}^{\infty}\, rb^{i}_{-r}b^{i}_{r}
      \,-\, \frac{1}{6} \; .
\end{eqnarray}
Elliptic Theta functions are defined for arbitrary characteristics
 as \cite{munford}
\be
\Theta\left[\begin{array}{c}\theta\\ \phi\end{array}\right]
  \left( 0\mid \tau\right) 
\;=\; 
  \sum_{n=-\infty}^{\infty}\, 
  e^{
     i\pi\left( n+\theta\right)^{2}\tau
     +2\pi i\left( n+\theta\right)\phi
    } \; .
\ee
The Jacobi elliptic functions are particular cases of the above
\be
\begin{array}{lclclcl}
\Theta\left[\begin{array}{c} 1/2 \\ 1/2 \end{array}\right] 
  &=& \theta_{11}\equiv\theta_{1}
&\hspace{.6cm},\hspace{.6cm}&
\Theta\left[\begin{array}{c} 1/2 \\ 0\end{array}\right] &=& \theta_{10}
\equiv\theta_{2} \; ,\\
& & && & & \\
\Theta\left[\begin{array}{c} 0\\ 0\end{array}\right] &=& \theta_{00}\equiv
\theta_{3}
&,& 
\Theta\left[\begin{array}{c} 0\\ 1/2 \end{array}\right] &=& \theta_{01}\equiv
\theta_{4} \; ,
\end{array}
\ee
Using which we can rewrite the $A$'s as
\be
\begin{array}{lclclcl}
A^{++}(\tau )&=& \eta_{++}\frac{\theta_{1}^{4}(0\mid\tau )}{\eta^{4}(\tau )}
 \,=\, 0
&\hspace{.5cm},\hspace{.5cm}&
A^{+-}(\tau )&=& \eta_{+-}\frac{\theta_{2}^{4}(0\mid\tau )}{\eta^{4}(\tau )} \; ,\\
& & & & & & \\
A^{--}(\tau )&=& \eta_{--}\frac{\theta_{3}^{4}(0\mid\tau )}{\eta^{4}(\tau )}&,&
A^{-+}(\tau )&=& \eta_{-+}\frac{\theta_{4}^{4}(0\mid\tau )}{\eta^{4}(\tau )} \; .
\end{array}
\ee
$A^{++}$ is the odd spin structure and the others are the even spin structures
(where the label even or odd, stands for the number of zero modes of 
the Dirac operator ($+2\mathbb{Z}$)).
\par
Under the one-loop modular group  $SL(2,\mathbb{Z})$ theta functions transform
amongst themselves,\footnote{To be specific,
\bea
\theta_{ab}(z,\tau+1)&=&e^{\frac{\pi a i}{4}}\theta_{a,b+a+1}(z,\tau)\nonumber\\
\theta(z/\tau,- 1/\tau)&=&(-1)^{ab}\sqrt{- i\tau}e^{\pi i z^2/\tau}\theta_{ba}(z,\tau)
\eea
where the sum on the characteristics is made modulo $2$.}
so that fixing, for example, the phase
$\eta_{--}=1$, all other phases are
uniquely determined by modular invariance \cite{dixhar}
\begin{eqnarray}
A(\tau )&=& \frac{1}{2\eta^{4}}\left\{
            \theta_{3}^{4}\,-\, \theta_{4}^{4} \,-\, \theta_{2}^{4}
            \,+\, \eta_{++}\theta_{1}^{4}
            \right\} \nonumber \\
 &=& Tr\, e^{2\pi i\tau H_{NS}}
        \left(\frac{1}{2}\left( 1-(-1)^{F}\right)\right)
     \,-\,
  Tr\, e^{2\pi i\tau H_{R}}
        \left(\frac{1}{2}\left( 1-\eta_{++}(-1)^{F}\right)\right) \; ,
\end{eqnarray}
where one should note that the minus sign in the definition 
of the trace over the NS d.o.f. implies that the NS vacuum
has a negative value for the fermion number operator.
Jacobi's {\it Equatio identica satis abstrusa} ensures that this is
identically zero, and in this sense is equivalent to supersymmetry:
This allows for a reinterpretation of the GSO projection
as one-loop modular invariance.
\par
It is curious that if instead of asking for
separate L and R modular invariance,
we had summed over the same boundary conditions L and R, we would not
have gotten a spacetime supersymmetric action, but the bosonic
Dixon-Harvey model instead \cite{dixhar}.
\par
There is a nice argument by Polyakov \cite{polya}
on the relationship between
modular invariance and spacetime anomalies: The Ward identity associated
to conformal transformations reads
\be
\langle \delta S \phi(z_1)\ldots \phi(z_n)\rangle =
\sum_i \langle \phi(z_1)\ldots \delta\phi(z_i)\ldots \phi(z_n)\rangle \; ,
\ee
where 
\be
\delta S = \int T \bar{\pd}\epsilon \; ,
\ee
and
\be
\delta \phi = \epsilon\pd \phi + h \pd \epsilon\phi \; .
\ee
The basic origin of the OPE from this point of view stems from 
choosing $\epsilon$
in such a way that $\bar{\pd}\epsilon = \delta^{(2)}(z)$; that is, $\epsilon =
\frac{1}{z}$. On a torus, the best we can do is to choose $\epsilon = \zeta(z)$,
where the $\zeta$ function of Weierstra{\ss} is defined by $\zeta'(z) = - 
\mathcal{P}(z)$, and begins its Laurent expansion with a simple pole
at the origin. It is not doubly periodic however, but rather
\be
\zeta(z+2\omega+2\omega')=\zeta(z)+2\eta+2\eta' \; ,
\ee
where $\omega$ and $\omega'$ are the two half-periods, and
\be
\eta\omega' - \eta' \omega = \frac{\pi i}{2} \; .
\ee
This fact induces a change in the modular parameter
\be
\tau\rightarrow \tau +\frac{\pi i }{2\omega(\omega+\eta)} \; ,
\ee
which causes a suplementary term in the Ward identity proportional to
\be
\frac{\pd}{\pd\tau}\langle \phi_1\ldots \phi_n\rangle \; ,
\ee
giving rise to boundary terms when integrated, in case the correlator 
is not modular invariant. All two-dimensional anomalies are related: 
A conformal anomaly can be disguised as a gravitational anomaly by a 
local counterterm \cite{rds}.
\subsection{Spectral flow and spacetime supersymmetry}
The $N=2$ superconformal algebra is an extension of the preceding $N=1$
superconformal algebra, and is realized in the particular case when 
there is an $U(1)$ charge, generated by $J(z)$, with
$h[J]=1$, such that 
\begin{eqnarray}
T_{F}^{+}(z)T_{F}^{-}(w) &=& \frac{c/12}{(z-w)^{3}} \,+\, 
   \frac{\textstyle{\frac{1}{4}}J(w)}{(z-w)^{2}} \,+\,
   \frac{\textstyle{\frac{1}{4}}T(w)+
   \textstyle{\frac{1}{8}}\partial J(w)
   }{z-w} \; , \\
J(z)J(w) &=& \frac{c/3}{(z-w)^{2}} \; , \\
J(z)T_{F}^{\pm}(w) &=& \pm\frac{T^{\pm}_{F}(w)}{z-w} \; .
\end{eqnarray}
The Spectral flow of Schwimmer and Seiberg \cite{ss} is given by the following
transformations:
\begin{eqnarray}
T^{\eta}(z) &=& T(z) \,+\, \frac{\eta J(z)}{z} \,+\, 
\frac{c\eta^{2}}{6z^{2}} \; , \\
T^{\eta\pm}_{F}(z) &=& z^{\pm\eta}\, T^{\pm}_{F}(z) \; ,\\
J^{\eta}(z) &=& J(z) \,+\, \frac{c\eta}{3z} \; ,
\end{eqnarray}
where $\eta\in [0,1)$.
It is a fact that $T^{\eta}$ satisfies the Virasoro algebra
with the same $c$.
\par
This physically means that all fermionic boundary conditions yield
isomophic algebras. This is only possible because the supercurrent
is `real'
\be
T^{\pm}_{F}\left( e^{2\pi i}z\right) \;=\; 
-e^{\mp 2\pi i\eta}T^{\pm}_{F}\left( z\right) \; .
\ee
Spectral flow obviously corresponds to
\be
\left\{
\begin{array}{lclcl}
h &\rightarrow& h^{\eta}&=& h\,+\, \eta q\,+\, 
\textstyle{\frac{c}{6}}\eta^{2}  \\
q &\rightarrow& q^{\eta}&=& q\,+\, \textstyle{\frac{c}{3}}\eta
\end{array}
\right.
\ee
\par
This fact strongly suggests that in order to have spacetime supersymmetry
(That is, in order to be able to implement succesfully a GSO projection and
build spin operators) one needs $N=2$ supersymmetry on the world-sheet.
\subsection{Superstring Taxonomy}
We have now at hand all the necessary tools to build (super)strings,
{\em i.e.} a unitary CFT with $(c,\bar{c})=(0,0)$.\footnote{$c$ is the 
combined conformal anomaly of the matter and ghost sectors.}
The simplest possibility is the {\em bosonic string}, with $d=26$ in order
to cancel the ghost contribution to the conformal anomaly.
\par
If we insist on spacetime supersymmetry
with $(1,1)$ world-sheet superconformal theories we find that 
\be
c\,=\, \bar{c}\,=\, \textstyle{\frac{3}{2}}d\,-\, 26\,+\, 11 \; ,
\ee
where the `$11$' part comes from the commuting ghosts fixing local 
supersymmetry. Imposing the constraint of vanishing conformal 
anomaly then leads to $d_{crit}=10$.
\par
There are several related theories in d=10. The simplest ones enjoy
$N=2$ (that is, 32 supercharges) supersymmetry, and come in two versions:
One possibility is the so-called type
$IIA$ superstring, where the two gravitinos
have oposite chirality, so that there is no
chirality preferred, and the theory is non-chiral. This is the theory 
whose low energy limit is the  dimensional reduction of N=1 Supergravity  
in d=11 dimensions.
\par
Another possibility is that both gravitinos have the same chirality.
This is the
$IIB$ theory, a chiral one, whose low energy limit is
$IIB$ supergravity in d=10.
\par
There is also the possibility of having open strings.
Open boundary conditions break
the supersymmetry to $N=1$ only. These theories are in general anomalous,
unless a non-dynamical degree of freedom (a Chan-Paton index) is added to the
ends, such that it belongs to $SO(32)$. This is the Type I Superstring, with
gauge group $SO(32)$.
\par
Narain \cite{narain} showed that in
general one can have the dimensions of the left
and right momentum lattices different, $p$ and $q$ say,
with conformal invariance
putting the restrictions that the lattice $(k,\bar{k})$ is even, unimodular 
and self-dual, with a metric of signature $(p,q)$. (This means essentially that
$k\cdot k' - \bar{k}\cdot \bar{k}'\in 2 \mathbb{Z}$).
The number of parameters associated
to a Narain lattice is the dimension of the coset space 
$SO(p,q)/(SO(p)\otimes SO(q))$; that is, $pq$.
Narain, Sarmadi and Witten \cite{narain} further
showed that these lattices can be
interpreted as the effect of constant backgrounds for the spacetime metric
as well as for the two-index field.
\par
We can also cancel left and right anomalies in an independent way.
This will eventually lead to two further string theories also in d=10:
$E_8\times E_8$ Heterotic and $SO(32)$ Heterotic.
There are then altogether five
seemingly consistent ten dimensional string theories.
For the simplest $(1,0)$ `heterotic' theory we can do the same matching
\be
(c,\bar{c})\,=\, (-15,-26)_{ghosts} \,+\, (15,10)_{coord.}\,+\,
(0,16)_{extra} \; ,
\ee
where the extra part can be shown to be an $E_{8}\otimes E_{8}$ or
$SO(32)$ level 1 current algebra. 
\par
The Heterotic string has a 26 dimensional bosonic left moving 
sector, $X^{\mu}_{L}(\tau +\sigma )$, and 
an $N=1$ supersymmetric rightmoving sector, $\psi^{\mu}_{R}, X^{\mu}_{R}
(\tau -\sigma )$. Take $X^{I}_{L}(\tau +\sigma )$, $I=1..16$,
to be compactified, in the sense that 
\be
P_{L}\in \Gamma_{16}\; :\;
P^{I}_{L} \;=\; P_{i}\, {e_{i}}^{I} \hspace{.3cm},\hspace{.3cm}
P_{i}\in\mathbb{Z} \; .
\ee
The Hamiltonian turns out to be \cite{narain}
\begin{eqnarray}
H_{L} &=& \textstyle{\frac{1}{2}}P_{T}^{2} \,+\, N_{L}\,+\,
  \textstyle{\frac{1}{2}} P_{I}^{2} \,-\, 1 \; ,\\
H_{R} &=& \textstyle{\frac{1}{2}}P_{T}^{2} \,+\, N_{R} \,+\,
  H^{NSR} \; .
\end{eqnarray}
The bosonized Right sector then lives in a Lorentzian lattice $(D_{5,1})$,
putting together world-sheet fermionic imbeddings and ghosts
$(\psi^{\mu},\beta ,\gamma )$, with weights 
$w_{R}=\lambda_{R},q$ whereas the bosonic lattice is represented by 
$(\Gamma_{16})_{L}$.
\par
Vertex operators involve
\be
V \;=\; e^{iW_{L}\cdot X (\bar{z})}
 e^{i\lambda_{R}\cdot\phi (z)}
 e^{q\phi (z)} \; .
\ee
Locality, or rather the absence of branch cuts, 
\be
V_{1}(z)V_{2}(w) \;=\; (\bar{z}-\bar{w})^{W_{L_{1}}\cdot W_{L_{2}}}
(z-w)^{\lambda_{R_{1}}\cdot\lambda_{R_{2}}-q_{1}q_{2}}\, V_{1+2} \; ,
\ee
imposes that 
\be
-W_{L_{1}}\cdot W_{L_{2}} \,+\, 
\lambda_{R_{1}}\cdot\lambda_{R_{2}}-q_{1}q_{2} \in \mathbb{Z} \; ,
\ee
\be
\Gamma_{16;5,1} \;\equiv\; \left(\Gamma_{16}\right)_{L} \otimes
\left( D_{5,1}\right)_{R} \; .
\ee
This can further be elaborated to imply
that $\Gamma_{16}$ has to be an odd, selfdual Lorentzian
lattice.
\par
It is worth emphasizing that, although we do not have time to explain it in 
detail here, the fact that those string theories are anomaly free 
(which is related to conformal invariance) can also be studied from the 
effective field
theory point of view: To each string theory corresponds a consistent,
anomaly free supergravity \cite{GSW,agw}.
\par
The low energy, long wavelength limit of this theory is N=1 Supergravity
coupled to Yang-Mills in d=10.
The bosonic part
of the effective field theory of the heterotic string is
given by\footnote{When dealing with supergravities we will use the signature
$(+,-,\ldots ,-)$. Please also note that all fields are dimensionless.}
\be
S^{het}_{(d=10)} =\frac{1}{32\pi}\int d^{10}x
\sqrt{-g}e^{-\phi}\left[ R(g) - g^{\mu\nu}\partial_\mu\phi\partial_\nu \phi
+\textstyle{\frac{1}{12}} H_{\mu\nu\rho} H^{\mu\nu\rho}
-\textstyle{\frac{1}{4}} F^{I}_{\mu\nu}F^{I\mu\nu} \right] \; ,
\ee
where $I = 1\ldots 16$ represent the Abelian fields in the Cartan subalgebra
of either $E_8\times E_8$ or $SO(32)$, which are the only ones which 
{\em generically} will remain massless upon compactification
to four dimensions.
\par
On the other hand, the ordinary spacetime dimensional reduction of the
11-dimensional supergravity action gives the 10-dimensional $IIA$ supergravity
action, which can be written as
\begin{eqnarray}
S^{10}_{IIA} &=& \frac{1}{2\kappa_{10}^2}
\int d^{10}x \sqrt{-G} \left[
e^{-\phi} \left\{ 
        R(G) - (\partial_{\m} \phi)^2
+\textstyle{\frac{1}{2\cdot 3!}} (H_{\m\n\rho})^2 \right\}\right. \nonumber \\
&&-\left. 
\textstyle{\frac{1}{2\cdot 2!}} (F_{\m\n})^2
-\textstyle{\frac{1}{2\cdot 4!}} (J^{(4)}_{\m\n\rho\sigma})^2 
\right]
+\frac{1}{4\kappa_{10}^2}\int K^{(4)} \wedge K^{(4)}\wedge B^{(2)}
\end{eqnarray}
where $F^{(2)} \equiv dA^{(1)}$, $ H^{(3)} \equiv d B^{(2)}$,
$K^{(4)} 
\equiv dA^{(3)}$  and  $J^{(4)}\equiv K^{(4)}+A^{(1)}\wedge H^{(3)}$.
\par
One of the peculiar properties of supergravity is that there is another
ten dimensional theory, called IIB, which is chiral, in the sense that
the two gravitinos share the same chirality.
The covariant equations of motion involve a self-dual five-form,
and there is no known covariant action from which they can be derived.
In the understanding that the self-duality constraint has to be imposed
after the variation, a possible action is
\bea
S_{IIB} &=& \frac{1}{2\kappa_{10}^{2}}\left[ 
\int d(vol)\left\{ e^{- \phi}\left( R 
- (\partial_{\m}\phi)^2 +\frac{1}{2\cdot 3!} H_{\m\n\rho}^2\right) 
+ \textstyle{\frac{1}{2}} (\partial_{\m} l)^2 \right.\right. \nn\\
&&\left.\left. +\frac{1}{12}(H^{'}_{\m\n\rho} - l H_{\m\n\rho})^2 
+\frac{1}{60}F^{+ 2}_{5}\right\}
-\frac{1}{48}\int A^{+}_4\wedge H_3\wedge H'_3\right] \; .
\eea
{}From this action it is obvious, just by counting 
powers of the dilaton, that the
RR fields are $l\equiv A_0$, $A_2\equiv B'_2$ and $A^{+}_4$.
\par
It was known since a long time that there is, up to
field redefinitions, only one action 
for N=1 Supergravity
coupled to Super Yang Mills in d=10,
the one describing the low energy limit of Type $I$ open strings:
\be
S_{I} = \frac{1}{2\kappa_{10}^2} \int d(vol) \left\{ e^{-\phi}
\left[ R -(\partial \phi )^2\right]
{}- \textstyle{\frac{1}{4}}
e^{\phi /2} F^{a 2}_{\m\n} +\frac{1}{12}H^{' 2}_{\m\n\rho}\right\}  
\ee
Here the three different powers of the dilaton reflect the different 
geometrical origin: The first terms come from the spherical 
topology, the gauge term comes
from Chan-Paton factors attached to the disc (with $\chi = 1$), 
and the term with
no dilaton comes from a $B'_2(RR)$ when this theory is viewed (as we shall see) 
as an orientifold
of the $IIB$.
\subsection{Strings in background fields}

Up to now the strings have been propagating in ten-dimensional Minkowski space.
We physically expect, however, that some kind of string condensate
should explain spacetime curvature and, furthermore, that spacetime should
spontaneously compactify to $4$ dimensions. Unfortunately, these highly
interesting topics are very difficult to study
with first quantization techniques.
\par
A less ambitious problem is to determine, given a passive spacetime 
background, whether strings can consistently propagate in it.
Afterwards we could dream of including back-reactions in some self-consistent
approximation.
\par
Let us then assume that strings are propagating in a non-trivial background
of the massless (NS) fields.\footnote{To which we have also
included the tachyon
$T$, as well as an appropiate two-dimensional cutoff,
$\Lambda$ (\cite{tseytlin}).}
\begin{eqnarray}
{\cal S}\;=& \frac{1}{4\pi\alpha^{\prime}}\int d^{2}z&
\left\{
\sqrt{h}h^{ab}g_{\mu\nu}(X)\partial_{a}X^{\mu}\partial_{b}X^{\nu}
\;+\; 
\epsilon^{ab}b_{\mu\nu}(X)\partial_{a}X^{\mu}\partial_{b}X^{\nu}
+2 \Lambda^2 T(X) 
\right. \nonumber \\
&& \left.
+\textstyle{\frac{1}{2}}\alpha^{\prime}\sqrt{h}R^{(2)}\, \Phi (X)
\right\} \; .
\end{eqnarray} 
If we want to consider this as a $QFT_{2}$, the background fields have to be
considered as field-dependent coupling-constants
({\em cf.} \cite{friedan,cmpf,art:tseytlin,art:sigmasen}).
\par
By performing a covariant (using Riemann normal-coordinates) background
field expansion, the $\beta$-functions can be computed to be
\begin{eqnarray}
\beta_{\mu\nu}(g) &=& \a' R_{\mu\nu}
                 \,-\, \a' \nabla_{\mu}\nabla_{\nu}\Phi
                 \,+\, \textstyle{\frac{\a'}{4}} 
                            H_{\mu\kappa\rho} {H_{\nu}}^{\kappa\rho} 
                 \,+\, \textstyle{\frac{\a'}{4}}
                            \partial_{\mu}T \partial_{\nu}T
\; , \\
\beta_{\mu\nu}(b) &=& \a' \nabla^{\rho}H_{\mu\nu\rho}
                  \,-\, \a' \nabla^{\rho}\Phi\cdot H_{\mu\nu\rho} \; ,\\
\beta (\Phi )&=& \a' (\partial\Phi )^{2}
            \,-\, \a' \nabla^{2}\Phi
            \,-\, \textstyle{\frac{\a'}{6}}\, H^{2}
            \,-\, \textstyle{\frac{2(D-26)}{3}}
            \,+\, T^{2}
            \,-\, \textstyle{\frac{1}{6}} T^{3} \; , \\ 
\beta (T)&=& \a' \nabla^{2}T
        \,-\, \a' \partial_{\mu}\Phi\partial^{\mu}T 
        \,-\, 4T
        \,+\, T^{2} \; .
\end{eqnarray}
The conformal anomaly of the $\sigma$-model, $c\equiv \beta (\Phi )$,
must be constant by consistency. That this is actually true, can
be seen by noting that it is the integrated version
of the other two and the Bianchi identity (See {\it e.g.} \cite{cmpf}).
\par
It is a remarkable feat that the $\beta$-functions can be derived from
the action
\be
{\cal S}\;=\; \int d^{D}x\sqrt{g}\, e^{-\Phi}
\left[
\a'\left( R
    \,-\, (\partial\Phi)^2 
    \,+\, \textstyle{\frac{1}{2\cdot 3!}} H^{2} 
   \right)
    \,-\, \textstyle{\frac{2(D-26)}{3}} 
    \,+\, \textstyle{\frac{\a'}{4}}(\partial T)^2 
    \,+\, T^2 
    \,-\, \frac{1}{6}T^3
\right] \; .
\label{eq:sframeSEA}
\ee
This is perhaps the most important (together with the similar 
results we derived at the beginning for kappa symmetry) of all results 
linking spacetime physics with world-sheet properties. Classical solutions 
of the above spacetime action represent possible {\em string vacua}; ground
states of quantum strings.
\par
We can transform from the string frame to the Einstein frame,
by means of the field redefinition
\be
g_{\mu\nu} \;=\; e^{\frac{2}{D-2}\Phi}\, g^{(E)}_{\mu\nu} \; .
\label{eq:sigma2ein}
\ee
The action (\ref{eq:sframeSEA}) in the Einstein frame reads
\begin{eqnarray}
\hspace{-.5cm}{\cal S}_{D}\, &\hspace{-.5cm}=& 
\hspace{-.5cm}\int d^{D}x\sqrt{g^{(E)}}
     \left(
       \a' R(g^{(E)})\,+\,
       \textstyle{\frac{\a'}{D-2}}\left(\nabla^{(E)}\Phi\right)^{2}\,+\,
       \textstyle{\frac{\a'}{12}}
         e^{-\textstyle{\frac{4}{D-2}}\Phi} H^{2}_{(E)}
     \right. \nonumber \\
&& \hspace{2cm}\left.
   -\textstyle{\frac{2(D-26)}{3}} e^{\textstyle{\frac{2}{D-2}}\Phi}
   \,+\,\textstyle{\frac{\a'}{4}} (\nabla T)^2 \,+\,
     \left[
         T^2 - \textstyle{\frac{1}{6}}T^3
     \right] e^{\textstyle{\frac{2}{D-2}}\Phi}
   \right)    \; .
\end{eqnarray}
Classical solutions to this action give conformally invariant 
$\sigma$-models to ${\cal O}(\alpha^{\prime})$.
\par
If we write $\Phi =\langle\Phi\rangle +\hat{\Phi}$, any string amplitude
containing $\exp (-{\cal S})$ scales as
\be
e^{-\frac{\langle\Phi\rangle}{8\pi}\int \sqrt{h}R^{(2)}} \;=\; 
g_s^{2g-2} \; ,
\ee
where $g_s =e^{\langle\Phi\rangle /2}$ and Euler's theorem tells us that
\be
\frac{1}{4\pi}\int\sqrt{h}R^{(2)} \;=\; 2-2g \; .
\ee
This physically means that the vacuum expectation value (that is, the 
asymptotic value of the classical solution in most cases) of the dilaton field
gives directly the string coupling constant, which is then promoted to a 
dynamical field.
\par
Due to the presence of the background fields, the vertex operators
are changed correspondingly:
In general under a Weyl transformation $\gamma_{ab}\rightarrow e^{2\sigma}
\gamma_{ab}$ there will be operator mixing 
\be
\frac{\delta}{\delta\sigma}\langle V_{i}\rangle \;=\;
\sum_{k} \, \Delta_{ij} \langle V_{j}\rangle \; ,
\ee
where $\Delta$ is the anomalous dimension matrix.
The simplest example is the one corresponding to the tachyon determined
by Callan and Gan \cite{cg}.
They showed the anomalous dimension to be\footnote{This coincides with the
second variation of the spacetime effective action.} 
\be
\alpha^{\prime}\left( 
\nabla^{2}\,-\, \nabla_{\mu}\Phi\nabla^{\mu}
\right) T(X) \; .
\ee
Physical vertex operators need to have conformal dimension
(1,1), and are
correspondingly solutions of the equations 
\be
\left(\nabla^{2}\,-\, \nabla_{\mu}\Phi\nabla^{\mu}\right) T(X) \;=\;
\textstyle{\frac{4}{\alpha^{\prime}}}\, T(X) \; ,
\ee
recovering the old result that $T(X)=e^{ik\cdot X}$ in the simplest
case when the background is flat space.
\par
This can be put in a slightly different way by rescaling the metric
as in Eq. (\ref{eq:sigma2ein}), namely
\be
\left(
\nabla^{2}\;-\; \textstyle{\frac{4}{\alpha^{\prime}}}\, e^{\frac{2}{d-2}\Phi}
\right) T(X) \;=\; 0 \; .
\ee
Dilatons can then be thought of as locally rescaling the tachyon mass.
It is also possible to argue that the quadratic fluctuation operator of the 
effective action can be regarded as the anomalous dimension
operator for the massless state vertex operators.
\section{T-duality, D-branes and Dirac-Born-Infeld}
T-duality is the simplest of all dualities and the only one which can be shown
to be true, at least in some contexts. At the same time it is a very stringy
characteristic, and depends in an essential way on strings being extended
objects.  In a sense, the web of dualities
rests on this foundation, so that it is important to understand 
clearly the basic 
physics involved. Let us consider strings living on an external space with one 
compact dimension, which we shall call $y$, with topology $S^1$.
\subsection{Closed strings in $S^1$}
1.- Let us imagine that one of the dimensions of a given spacetime is a circle
$S^1$ of radius $R$. The corresponding field in the imbedding of the
string, which we shall call $y$
({\it i.e.} we are dividing the target-spacetime dimensions as $(x^{\m},y)$,
where $y$ parametrizes the circle), has then the
possibility of winding around it:
\be
y(\sigma + 2\pi,\tau) = y(\sigma, \tau) + 2\pi R m \; .
\ee
A closed string can close in general up to an isometry of
the external spacetime.
\par
The zero mode expansion of this coordinate (that is, forgetting about
oscillators) would then be
\be
y = y_{c} + 2 p_{c} \tau + m R \sigma \; .
\ee
Canonical quantization leads to $[y_{c},p_{c}] = i$, and single-valuedness
of the plane wave $e^{iy_{c}p_{c}}$ enforces as usual $p_{c}\in \mathbb{Z}/R$,
so that $p_{c} = \frac{n}{R}$.
\par
The zero mode expansion can then be organized into left and right movers in the
following way
\bea
y_L (\tau +\sigma)&=& y_c/2 + 
     \left(\frac{n}{R} + \frac{mR}{2}\right) (\tau + \sigma) \; ,\nn\\
y_R (\tau -\sigma)&=& y_c/2 + 
    \left(\frac{n}{R} - \frac{mR}{2}\right) (\tau - \sigma) \; .
\eea
The mass shell conditions reduce to
\bea
m_L^2 &=& \frac{1}{2}\left(\frac{n}{R} + 
          \frac{mR}{2}\right)^2 + N_L -1 \; ,\nn\\
m_R^2 &=& \frac{1}{2}\left(\frac{n}{R} - \frac{mR}{2}\right)^2 + N_R -1 \; .
\eea
Level matching, $m_L = m_R$, implies that there is a relationship between
momentum and winding numbers on the one hand, and the oscillator 
excess on the other
\be
N_R - N_L \;=\;  nm \; .
\ee
At this point it is already evident that the mass formula is invariant
under
\be
R\; \rightarrow\;  R^{*}\equiv 2/R \; ,
\ee
and exchanging momentum and winding numbers. This is the simplest instance of 
{\em T-Duality}.
\par
2.-The above transformation can be seen to lift 
to an automorphism to the CFT OPE
namely
\bea
y(z)&\rightarrow& y(z) \; ,\nn\\
\bar{y}(\bar{z})&\rightarrow & -\bar{y}(\bar{z}) \; .
\eea
\par
The total momentum of the scalar field is defined as
\be
\hat{p}.\equiv \frac{1}{4\pi}\oint(\partial\phi + \bar{\partial}\bar{\phi}).
\ee
so that vertex operators for momentum eigenstates are of the type
\be
:e^{i p (\phi + \bar{\phi})}: \; ,
\ee
because $\hat{p}\, :e^{ip (\phi +\bar{\phi})}: \, = 
p:e^{ip (\phi +\bar{\phi})}:$.
The total winding, on the other hand, is similarly written as
\be
\hat{w}.\equiv \frac{1}{4\pi}\oint(\partial\phi - \bar{\partial}\bar{\phi}),
\ee
so that vertex operators for winding eigenstates are of the type
\be
:e^{i k (\phi - \bar{\phi})}: \, ,
\ee
enjoying $\hat{w}\, :e^{i k (\phi - \bar{\phi})}: =
k\, :e^{i k (\phi - \bar{\phi})}:$.
\par
There is a simple argument showing that upon compactification, momentum 
eigenstates couple to the Kaluza-Klein gauge boson, whereas windings couple to
gauge bosons coming from the reduction of the Kalb-Ramond field.
Actually, considering the OPE
\begin{equation}
:\left( \partial X^{\m}\bar{\partial}y \pm
          \partial y\bar{\partial}X^{\m}
   \right)e^{i p\cdot X}:(z,\bar{z})\,: e^{i k\cdot X(w,\bar{w})}
e^{i l (y(w)-\bar{y}(\bar{w}))}: \,
\sim \, -k^{\mu}\,l\, \left( 1\mp 1\right) \; ,
\end{equation}
justifies the above claim.
\par
3.- Another point is that demanding the $(d-1)$-dimensional effective action
to be invariant under this transformation we are forced to assume that
\be
2\pi R^{*} e^{- 2 \phi^{*}} = 2\pi R e^{- 2 \phi } \; ,
\ee
leading to the necessity of transforming the dilaton, already in 
this simple setting.
\par
4.- The integrand of the partition function of the bosonic string compactified
on a circle (or a torus, for that matter) can be computed, for 
arbitrary  genus, and shown to enjoy T-duality. 
We can expand the holomorphic differentials, $\partial X$,
of the embedding in terms of the period matrix 
\begin{equation}
  \tau_{ij} \;=\; \int_{\beta_{j}}\, \omega_{i} \;\equiv\; 
  \left(\tau_{1}\right)_{ij}\,+\, i\left(\tau_{2}\right)_{ij} \; , 
\end{equation}
where $(\alpha_{i},\beta_{j})$, $i,j=1\ldots g$, is a canonical 
homology basis and the holomorphic differentials $\omega_{i}$
are normalized by
\begin{equation}
  \int_{\alpha_{i}}\, \omega_{j} \;=\; \delta_{ij} \; .
\end{equation}
We can then decompose the holomorphic differentials as
\begin{equation}
 \partial y \;=\; \chi \,+\, \sum_{n}\, C_{i}\omega_{i} \; ,
\end{equation}
where $\chi$ is the exact part and the $C$'s are determined 
by the number of times the string winds around the 
homology cycles, {\it i.e.}
\begin{equation}
\int_{a_{i}} dy \,=\, 2\pi n_{i}R \hspace{.5cm},\hspace{.5cm}
\int_{b_{i}} dy \,=\, 2\pi m_{i}R \; .
\end{equation}
This allows for the computation of the integrand with
the result \cite{atwit}
\begin{equation}
{\cal F}_{g}(R) \;=\; \theta \left[ 
  \begin{array}{cc} 0&0 \\ 0& 0 \end{array}
 \right]
\left( 0\mid \Omega \right) \; 
\Lambda_{g} \left( \tau ,\bar{\tau}\right) \; ,
\end{equation}
where $\Lambda_{g}$ is the integrand of the decompactified partition 
function, and Riemann's theta function is defined by
\begin{equation}
\theta \left[ \begin{array}{cc} 0&0 \\ 0& 0 \end{array}\right]
 \left( 0\mid \Omega \right) \;\equiv\;
\sum_{(n,m)\in\mathbb{Z}^{2g}}\, 
 e^{i\pi{}\, (m\, n)\Omega (n\, m)^{t}} \; ,
\end{equation}
and 
\begin{equation}
\Omega \;\equiv\; iR^{2}
\left(
\begin{array}{ccc}
\tau_{1}\tau_{2}^{-1}\tau_{1}+\tau_{2} &\hspace{.2cm}& -\tau_{1}\tau_{2}^{-1}\\
-\tau_{2}^{-1}\tau_{1} & & \tau_{2}^{-1} 
\end{array}
\right) \; .
\end{equation}
Now, under a symplectic transformation $\left(\begin{array}{cc}0&-1\\ 1&0
\end{array}\right)$, one finds \cite{ao}
\begin{eqnarray}
\theta\left( 0\mid -\Omega^{-1}\right) &=& 
 \det{}^{\frac{1}{2}}\left(\frac{\Omega}{i}\right)\, 
 \theta \left( 0 \mid \Omega\right)
 \; , \\
{\cal F}_{g}(R) &=& \left(\frac{1}{R^{2}}\right)^{g} \,
                 {\cal F}_{g}\left( R^{-1}\right)\; .
\end{eqnarray}
For the whole sum, this implies ($\tilde{\kappa}=\kappa /R$)
\begin{equation}
\sum_{g}\, \kappa^{2g-2}{\cal F}_{g}(R) \;=\;
R^{-2}\sum_{g}\, \tilde{\kappa}^{2g-2}{\cal F}_{g}(R^{-1}) \; .
\end{equation}
\par
5.- Something special happens at the self-dual radius $R = R^{*} = \sqrt{2}$: 
There are four extra massless vectors proportional to
\begin{equation}
 \partial x_{\mu}(z)e^{\pm i\sqrt{2}\bar{y}(\bar{z})}
\hspace{.5cm},\hspace{.5cm}
\bar{\partial}x_{\mu}(\bar{z})e^{\pm i\sqrt{2}y(z)} \; ,
\end{equation}
which, together with the two Kaluza-Klein vectors mentioned above,
generate $SU(2)_L\times SU(2)_R$.
Dine, Huet and Seiberg \cite{dhs} were the
first to realize that for generic values
of the compactification radius this group gets reduced to the Abelian part
$U(1)_L\times U(1)_R$. They interpreted this as a
stringy Higgs effect, and, as
a consequence, T-duality must be included in the full stringy gauge symmetry.
This was the first suggestion that T-duality ought to be exact, at least in 
perturbation theory, which was afterwards checked explicitly, at
least in some examples \cite{ao}.
\par
6.- In the supersymmetric case, owing to superconformal invariance,
the CFT mapping is
\bea
x\,\rightarrow\, x &\; ,\; &\bar{x}\rightarrow - \bar{x}\; ,\nn\\
\psi\, \rightarrow\, \psi & ,&\bar{\psi}\rightarrow - \bar{\psi} \; .
\eea
This means that chirality is reversed and that one goes from
IIA at radius R to IIB at radius $2/R$.
\par
7.- The contribution to the partition function of momentum states 
goes as $\frac{1}{R}$; whereas the contribution of winding modes is linear in $R$.
Let us consider a string moving in a p-dimensional torus, $T^p\equiv \mathbb{R}^p
/(2\pi\Lambda)$, where $\Lambda$ is a lattice.
The zero mode contribution to the Polyakov integral is
\be
\sum_{p,p'\in \Lambda} 
   e^{- 2\pi(\tau_2 p^2 + \frac{1}{\tau_2}(p' - \tau_1 p)^2)} \; .
\ee
We can now use the Poisson summation formula
\be
\sum_{\mathbb{Z}} e^{- m^2 R^2} = \frac{\sqrt{2}}{R} \sum_{\mathbb{Z}}
e^{- \frac{n^2\pi^2}{R^2}} \; ,
\ee
to rewrite it as
\be
\sum_{k,\bar{k},k-\bar{k}\in \Lambda^{*}} 
   e^{i\pi(\tau k^2 - \bar{\tau}\bar{k}^2)} \; .
\ee
It can be shown that $ (k,\bar{k})$ span a self-dual
lattice with signature $(p,p)$.
\par
8.- By compactifying a bosonic or superstring in $\Lambda(d,d)$ the full 
T-duality group is upgraded to $O(d,d;\mathbb{Z})$.
Representing by $E$ the
sum of the background metric plus the background Kalb-Ramond, 
$E\equiv G + B$, the group acts in a projective way \cite{gpr,aal}:
Given 
\bea
g\; \equiv \; \left (  \begin{array}{cc}a&b\\c&d\end{array}\right ) \; ,
\eea
then
\be
E\rightarrow \frac {a E + b}{ c E + d} \; .
\ee
This rather large discrete group is generated by the following transformations:
\newline
i){\tt Discrete translations on the Kalb-Ramond field},
\be
B_{\m\n}\rightarrow B_{\m\n} + \theta_{\m\n} \; .
\ee
That is, 
\bea
g \equiv \left(\begin{array}{cc}1&\theta\\0&1\end{array}\right) \; ,
\eea
where $\theta_{\m\n}\in \mathbb{Z}$.
\newline
ii) A {\tt change of basis on the lattice}, {\it i.e.}
\be
E' \; \equiv \; A E A^{t} \; ,
\ee
that is
\bea
g \equiv \left(\begin{array}{cc}A&0\\0&A^{-1 t}\end{array}\right) \; .
\eea
And, finally
\newline
iii) {\tt Factorized duality} (the analogous transformation 
to the duality we saw
earlier on $S^1$):
\bea
g \equiv \left(\begin{array}{cc}1 - e_i& e_i\\
e_i&1 - e_i\end{array}\right) \; ,
\eea
where 
\be
(e_i)_{jk}\equiv \delta_{ji}\delta_{ki} \; .
\ee
\par
In the case of the heterotic string, this gives $O(d+16,d;\mathbb{Z})$, 
because there is an extra $\Lambda_{16}$ for the left movers.
\subsection{T-duality for closed strings}
There is a known way of proving T-duality for
any string vacuum described by a sigma model such that the corresponding
target-spacetime enjoys at least one isometry. The classic work was
done by Buscher \cite{buscher},
but we shall follow the slightly different
gauging approach first introduced by Ro\v{c}ek and Verlinde \cite{rocver}.
Their formulation  starts with the $\sigma$-model (with
the Abelian isometry represented in adapted coordinates by
$\theta\rightarrow\theta +\epsilon$)
\be
S= \frac{1}{2\pi\a'}
\int d^2\xi \left[ {\sqrt{h} \left(
h^{ab}\partial_a x^{\m}\partial_b x^{\n}
+ R^{(2)}\phi(x)\right) +
i\epsilon^{ab} b_{\m\n}\partial_a x^{\m}\partial_b x^{\n}} \right]\; .
\ee
This means that in these adapted coordinates, $(\theta, x^i)$, the metric,
torsion and dilaton fields are $\theta $ independent.
The key point is to gauge the isometry by introducing some
gauge fields $A_a$ transforming as
$\delta A_{a}=-\partial_{a}\epsilon$.
Using a Lagrange
multiplier term, the gauge field strength is required to
vanish,
enforcing the constraint that the gauge field is
pure gauge. After gauge fixing the original model is then
recovered.
\par
Gauging the isometry  and adding the Lagrange
multiplier enforcing the condition that the gauge field strength vanishes
leads to
\bea
S_{d+1}&=&\frac{1}{4\pi\alpha^{'}}\int d^2\xi
[\sqrt{h}h^{ab}(g_{00}(\partial_a\theta+A_a)
(\partial_b\theta+A_b)+2g_{0 i}(\partial_a
\theta +A_a) \partial_b
x^{i} \nn\\
&&+g_{ij}\partial_a
x^{i}\partial_b x^{j})
+i\epsilon^{ab}(2b_{0i}(\partial_a \theta +A_a)
\partial_b
x^{i}+b_{ij}\partial_a
x^{i}\partial_b x^{j})
+2i \epsilon^{ab}{\tilde \theta} \partial_a A_b
\nn\\
&&+\alpha^{'}\sqrt{h}R^{(2)}\phi(x)] \; .
\eea
The dual theory is obtained integrating out the $A$ fields
\be
A_a=-\frac{1}{g_{00}}(g_{0i}\partial_{a}
x^{i}+i\frac{\epsilon_a\,^{b}}{\sqrt{h}}(b_{0i}
\partial_b x^{i}+\partial_{b}{\tilde \theta})),
\ee
and fixing $\theta=0$,
\bea
{\tilde S}&=&\frac{1}{4\pi\alpha^{'}}\int d^2\xi
[\sqrt{h}h^{ab}({\tilde g}_{00}\partial_a {\tilde \theta}
\partial_b {\tilde \theta}
+2{\tilde g}_{0i}\partial_a {\tilde \theta} \partial_b
x^{i}+{\tilde g}_{ij}\partial_a
x^{i}\partial_b
x^{j}) \nonumber\\
&&+i\epsilon^{ab}(2{\tilde b}_{0i}\partial_a
{\tilde \theta}\partial_b
x^{\alpha}+{\tilde b}_{ij}
\partial_a x^{i}\partial_b
x^{j})+\alpha^{'}\sqrt{h}R^{(2)}\phi(x)] \; ,
\eea
where
\bea
{\tilde g}_{00}&=&{1\over g_{00}} \; ,\nonumber\\
         {\tilde g}_{0i}&=&{b_{0i} \over g_{00}} \; ,
\qquad
{\tilde b}_{0i}={g_{0i} \over g_{00}} \; ,\nonumber\\
          {\tilde g}_{ij} &=& g_{ij} -
{g_{0i}g_{0j} - b_{0i} b_{0j}\over g_{00}} \; ,
\nonumber\\
        {\tilde
b}_{ij}&=&b_{ij}-{g_{0i}b_{0j}
         -g_{0j}b_{0i}\over g_{00}} \; .
\eea
\par
It so happens that Buscher's transformation
can not be  the whole story in presence of a nontrivial dilaton.
Indeed, the dual model
is not even conformally invariant in general, unless an
appropriate
transformation of the dilaton is included, namely
\be
\label{dil1}
{\tilde \phi}=\phi-\frac12\log{k^2},
\ee
where $k = \frac{\partial}{\partial\theta}$ is the Killing vector field;
in adapted coordinates, $k^2 = g_{00}$.
In the pathintegral approach the way to obtain the
correct dilaton
shift yielding a conformally invariant dual theory can
be seen as
follows.  In complex coordinates and on spherical
world-sheets we can parametrize $A=\partial\alpha$,
${\bar A}={\bar \partial}\beta$, for some 0-forms $\alpha, \beta$ on
the manifold $M$. The change of variables from $A, {\bar A}$
to $\alpha, \beta$ produces a factor in the measure
\be
{\cal D}A {\cal D}{\bar A}={\cal D}\alpha {\cal D}\beta
(\mbox{det} \partial) (\mbox{det} {\bar \partial})=
{\cal D}\alpha {\cal D}\beta (\mbox{det}\Delta) \; .
\ee
Substituting $A, {\bar A}$ as functions of $\alpha, \beta$
and integrating over $\alpha, \beta$, the
following
determinant emerges
\be
(\mbox{det}(\partial g_{00} {\bar \partial}))^{-1}
\equiv det \Delta_{g_{00}}^{-1} \; .
\ee
In particular, the integration on $\beta$ produces a
delta-function
\be
\delta ({\bar \partial}(g_{00}\partial\alpha+
(g_{0i}-b_{0i})
\partial x^{i}-\partial {\tilde \theta})) \; ,
\ee
which when integrated over $\alpha$ yields the factor in the
measure.
What we finally get in the measure is then
\be
\label{reg3}
\frac{\mbox{det}\Delta}{\mbox{det}\Delta_{g_{00}}} \; .
\ee
This formula provides a justification
for
Buscher's prescription for the
computation
of the determinant arising from the naive Gaussian
integration.
As we have just seen
some care is needed in order to correctly define the
measure of
integration over the gauge fields.
{}From the previous formula  the dilaton
shift  is obtained in the following way.
Writing $g_{00}$ as $g_{00}=1+\sigma\approx e^{\sigma}$ we have:
\be
\Delta_{g_{00}}=(1+\sigma)\Delta-h^{ab}
\partial_{a}\sigma\partial_{b} \; .
\ee
Plugging this into the infinitesimal variation of
Schwinger's formula
\be
\delta\log{\mbox{det}\Delta}=Tr \int_{\epsilon}^{\infty} dt
\delta\Delta
e^{-t\Delta} \; ,
\ee
we obtain
\be
\delta\log{\mbox{det}\Delta_{g_{00}}}=-\int d^2\xi
\sqrt{h}\Omega
\langle\xi|e^{-\epsilon
(\Delta+\sigma\Delta-h^{ab}\partial_{a}
\sigma\partial_{b})}
|\xi\rangle \; ,
\ee
where $\delta\Delta_{g_{00}}=-\Omega\Delta_{g_{00}}$
with $\delta
h_{ab}=\Omega \delta_{ab}$.
We can now use the standard heat kernel expansion \cite{art:gilkey}
\be
\langle\xi|e^{-\epsilon
D}|\xi\rangle=\frac{1}{4\pi\epsilon}+\frac{1}{4\pi}
(\frac16 R^{(2)}-V) \; ,
\ee
where
\be
D\equiv
\Delta-2ih^{ab}A_{a}\partial_{b}+\left( -\frac{i}{\sqrt{h}}
\partial_{a}(\sqrt{h}h^{ab}A_{b}\right)+
h^{ab}A_{a}A_{b})+V\, .
\ee
For
$D=\Delta+\sigma\Delta-h^{ab}\partial_{a}
\sigma\partial_{b}$
and after dropping the divergent term $1/4\pi\epsilon$
and the
quadratic terms in $\sigma$, we obtain
\be
\delta\log{\mbox{det}\Delta_{g_{00}}}=-\frac{1}{8\pi}
\int d^2\xi\sqrt{h}R^{(2)}\log{g_{00}} \; .
\ee
which inmediatly leads to
\be
\mbox{det}g_{00}=\exp{(-\frac{1}{8\pi}\int
d^2\xi\sqrt{h}R^{(2)}\log{g_{00}})} \, ,
\ee
implying ${\tilde \phi}=\phi-\frac12
\log{g_{00}}$.
\par
%
%
It is possible to interpret T-duality as a canonical transformation
with generating functional \cite{aal}
\be
{\cal F}\;=\; 
\textstyle{\frac{1}{2}}\int_{D,\partial D=S^{1}}
  d\tilde{\theta}\wedge d\theta \;=\; 
\textstyle{\frac{1}{2}}\int_{S^{1}}
 \left(
    \theta^{\prime}\tilde{\theta}\,-\,\theta\tilde{\theta}^{\prime}
 \right) \; .
\ee
In the operator formalism \cite{stringop}, it is implemented
by a Fourier transform of sorts, {\it i.e.}
\be
\Psi_{k}\left[ \tilde{\theta}(\sigma )\right] \;=\;
{\cal N}(k)\int {\cal D}\theta (\sigma )\,
  e^{i{\cal F}\left[ \theta ,\tilde{\theta}\right]}
  \, \Phi_{k}\left[ \theta (\sigma )\right] \; .
\ee
In the simplest free case, the Hamiltonians are
\begin{eqnarray}
H &=& \frac{1}{2R}\left(\frac{\delta}{\delta\theta}\right)^{2} 
      \;+\;
      \frac{R}{2}\left(\theta^{\prime}\right)^{2} \; , \\
\tilde{H} &=& \frac{R}{2}\left(\frac{\delta}{\delta\tilde{\theta}}\right)^{2}
      \;+\;
      \frac{1}{2R}\left(\tilde{\theta}^{\prime}\right)^{2} \; .
\end{eqnarray}
After a functional integration by parts, one obtains
\be
\tilde{H}\Psi_{k} \;=\; 
 {\cal N}(k)\int{\cal D}\theta\, e^{i{\cal F}}
 \left(
    \frac{1}{2R}\left(\frac{\delta}{\delta\theta}\right)^{2}
    \;+\;
    \frac{R}{2}\left(\theta^{\prime}\right)^{2}
 \right)\Phi_{k} \; .
\ee
\subsection{T-Duality for open strings and D-branes}
Let us begin with the simplest bosonic model, which,
while allowing for the most interesting physical phenomena,
is devoid of complications due to supersymmetry.
We shall consider open and closed  bosonic strings propagating
in an arbitrary $d$-dimensional 
metric and (Abelian) gauge field.
Wess-Zumino antisymmetric tensors are not consistent if
the theory is non-orientable, but we shall include
them nevertheless for the time being. In modern language (to be justified
momentarily), we have a
string interacting with a Dirichlet  $(d-1)$-brane, and we
consider non-trivial massless backgrounds in the longitudinal
directions.   
\par
In the neutral
case (that is, the charge is opposite in both ends of 
the string),
the action can be written as\footnote{We set $\alpha' =1$ throughout.}
\be
S = 
{1\over 4\pi
}\int_{\Sigma} (g_{\mu\nu}\eta^{ab} +i
\, b_{\mu\nu}\epsilon^{ab})\, \partial_a
x^{\mu}\partial_b x^{\nu} +
{i\over 2\pi}
\int_{\partial \Sigma} n_a A_{\mu}\partial_b x^{\mu} \epsilon^{ab} \; .
\label{eq:actopstr}
\ee
The classification of allowed world-sheet
topologies is much more complicated in the open case 
than in the more familiar closed one.
The Euler characteristic can be written, in a somewhat symbolic form, as
\be
\chi = 2 -2g -b -c,
\ee
where $g$ is the number of handles, $b$ the number
of boundaries, and $c$ the number of crosscaps.
To the lowest order in string perturbation theory $\chi = 1$,
only the disc $D_2$ and in the non-orientable case
the crosscap or, to be more precise, the two-dimensional real 
projective plane $P_2(\mathbb{R})$, contribute.
(To the following ``one loop" order,
corresponding to $\chi = 0$, we have  the annulus $A_2$,
the M\"obius band $M_2$, 
and the Klein bottle $K_2 $). In this section we shall only consider
the leading contributions  from the disc and the crosscap.  
\par
The action   will be invariant under a target isometry
with Killing vector $k^{\mu}$,  
\be
\delta_{\epsilon} x^{\mu} = \epsilon\,\, k^{\mu}(x) \; ,
\ee
provided a vector $\omega_{\mu}$ 
and a scalar $\varphi$ exist, such that
\bea
\mathcal{L}_k \, g_{\mu\nu} &=& 0 \; ,\nn\\ 
\mathcal{L}_k \, b_{\mu\nu} &=&
 \partial_{\mu}\omega_{\nu} - \partial_{\nu}\omega_{\mu} \; ,\nn\\  
\mathcal{L}_k \, A_{\mu} &=&  - 
\omega_{\mu} + \partial_{\mu} \phi \; ,
\eea
where $\mathcal{L}_k$ represents the Lie derivative
with respect to the Killing vector.
In the neutral case, it is clear that the boundary term representing 
the coupling of the background 
gauge field to the open string can be incorporated in the bulk
action  through the simple 
substitution\footnote{When the charges at both ends of the string
do not add to zero, one cannot get completely rid of the boundary 
term. We shall comment on a similar situation later in the main
text.}
\be
b_{\mu\nu} \rightarrow {B}_{\mu\nu}= b_{\mu\nu} +
 F_{\mu\nu}\; .
\ee
Using the conditions on the background fields, it is easy to show that
\be
\mathcal{L}_k \, B_{\mu\nu} = 0 \; .
\ee
In order to perform the duality transformation, it is convenient to
rewrite the whole action as a redundant gauge system where the isometry  is
gauged. We must introduce a Lagrange multiplier to ensure that the
auxiliary gauge field is flat. Minimal coupling 
is enough to construct the gauged action
\bea
S_{\rm gauged} &=& 
{1\over 4\pi} \int_{\Sigma} \left( g_{\mu\nu}\eta^{ab} +
 i\,{B}_{\mu\nu}\epsilon^{ab}\right) D_a
x^{\mu}D_b x^{\nu}\nn\\
& & +{i\over 4\pi} \int_{\Sigma} 
{\tilde x}^{0}(\partial_a V_b - \partial_b V_a)
\epsilon^{ab}-{i\over 2\pi}\int_{\partial\Sigma}{\tilde x}^0 V \; ,
\label{eq:opstrgauge}
\eea
where $D_a x^{\mu} = \partial_a x^{\mu} + k^{\mu}V_a$ and,
in adapted coordinates, the Killing vector reads
$ k = {\partial\over \partial x^0}$.  
The one-form associated to the gauge field is represented as 
$V = V_a dx^a$.  
The r\^ole of the boundary term is to convey invariance under 
translations of the Lagrange multiplier: ${\tilde x}^0 \rightarrow
{\tilde x}^0 + C$. This was first derived by Dai, Leigh and Polchinski
\cite{dalepo} and using this form in \cite{borlaf}.
\par
Boundary conditions are restricted by several physical requirements.
The gauge parameter must have the same boundary conditions
as the world-sheet fields ({\it i.e.} Neumann), in order for the isometry
to be realized on the boundary of the Riemann surface.
This has the obvious consequence that, if the gauge 
$V=0$ were needed, it would be neccessary to impose on 
the gauge fields the boundary condition $ n^{a}V_{a} \equiv V_n =0$
(because this
component can never be eliminated with gauge transformations
obeying Neumann boundary conditions). It turns out, however, that
in order to show the equivalence of (\ref{eq:opstrgauge})
with the original model
(\ref{eq:actopstr}), the behaviour of $V|\,_{\partial\Sigma}$ is immaterial.
The only way the previous action  can now lead to the {\it unique} 
restriction $dV=0$ on the gauge field is to restrict the variations
of the Lagrange multiplier in such a way that $\delta {\tilde x}^0|\,_{\partial \Sigma}=0$.
In this way we are forced to impose Dirichlet boundary conditions
${\tilde x}^0 = C$ on the multiplier. Since the rest of 
the coordinates remain Neumann, a Dirichlet $(d-2)$-brane is obtained.
Besides, this ensures gauge invariance.
\par
The last two terms can be combined into
\be
{i\over 2\pi}\int_{\Sigma}-d{\tilde x}^0 \wedge V.
\ee
This means that the gauge field enters only algebraically 
in the action,
and it can be replaced by its classical value
(performing the Gaussian integration only modifies 
the dilaton terms) 
\be
V_a^{cl}=-{1\over k^2}\left(
k^{\mu} g_{\mu\nu}\partial_a x^{\nu}
+i \epsilon_a\,^b \partial_b {\tilde x}^0 + i\, \epsilon_a\,^b k^{\mu}
B_{\mu\nu} \partial_b x^{\nu}\right).
\ee
Particularizing now to adapted coordinates, $k = {\partial\over\partial x^0}$
and choosing the dual gauge $x^0 = 0$, we get
the dual model, whose functional form is exactly like the former,
but with the backgrounds ${\tilde G}_{\mu\nu}$, ${\tilde B}_{\mu\nu}$
given in terms of the original ones 
through Buscher's formulas
\bea
{\tilde G}_{00} &=& {\tilde g}_{00} = {1\over g_{00}}\; ,\nn\\
{\tilde G }_{0i} &=& {B_{0i}\over g_{00}}  \; ,\nn \\
{\tilde G}_{ij} &=& g_{ij} -{g_{0i}g_{0j}-B_{0i}B_{0j}\over g_{00}} \; ,\nn\\
{\tilde B}_{0i}&=& {\tilde b}_{0i} = {g_{0i}\over g_{00}} \; ,\nn\\  
{\tilde B}_{ij}&=& B_{ij}-{g_{0i}B_{0j}-B_{0i}g_{0j}\over g_{00}}\; .
\eea
\par
In deriving these expressions, some care must be
exercised in choosing the appropriate variables in adapted
coordinates. In this frame, the isometry is represented by simple
translations $x^0 \rightarrow x^0 + \epsilon$. This means that the
various backgrounds must be independent of $x^0$, up to target-space
gauge transformations, which in this model are defined by
\be
{ b\rightarrow b+ d\lambda \hspace{.5cm},\hspace{.5cm} A\rightarrow
A-\lambda \; ,}                                            
\ee
where $\lambda$ is an arbitrary one-form, in such a way that $B=b+dA $
is invariant. This gauge ambiguity is responsible for the
occurence of the non-trivial Lie derivatives. In order to 
consistently reach the gauge $x^0 =0$ in the closed string sector   
(world-sheets without                                              
boundaries), the torsion Lie derivative   must be cancelled          
within the local patch of adapted coordinates. In fact, it
is easily seen that                 
the gauge transformation $\lambda$, defined as                     
\be
\mathcal{L}_k \,\lambda = -\omega + d\varphi  \; ,
\ee
cancels                                                  
both the $\omega$ and $\varphi$ terms.
In this gauge, all fields are locally independent of $x^0$.
\par
The behaviour of the dilaton under T-duality is always a subtle
issue. In the present situation this is even more so, due to the fact
that the metric is not a massless background of the open string, and
one has to consider closed string corrections, thus driving the sigma
model away from the conformally invariant point, in order to get a
consistent Fischler-Susskind mechanism. There is, 
however, a neccesary condition for the equivalence
of the two theories, and this is that the effective action must remain
invariant. It turns out that this condition is sufficient to determine
the dual dilaton to the value
\be
{\tilde \phi} = \phi - {1\over 2} \,{\rm log}\, k^2 \; .
\ee
The invariance under translations of the dual model can sometimes be
put to work in our benefit. Let us consider, for simplicity, the
Wilson line $A_0 = {\rm diag} (\theta_1, ..., \theta_N)$, when only
the coordinate $x^0$ is compactified in a circle of length $2\pi R$ in
an otherwise flat background. The Wilson line itself, in the sector of
winding number $n$ is given by
\be
\sum_{a=1}^N e^{2\pi inR\theta_a } \; .
\ee
This term is reproduced in the dual model
by simply  taking into account
the ``total derivative term" coming from the Gaussian
integration of the auxiliary gauge field, namely
\be
\int_{\Sigma} dx^0 \wedge d{\tilde x}^0 \; ,
\ee
which by Stokes' theorem can be written as $-\oint {\tilde
x}^0 \, dx^0 = - C\,2\pi nR$, where $C$ is the constant value of the
multiplier on the boundary. This value can depend on the (implicit)
Chan-Paton indices of the world-sheet fields, and we recover with our
techniques the result of Polchinski
that the Wilson line considered
induces in the dual model a series of D-branes with fixed positions
determined by the $\theta_a$ parameters. 
\par
Given any two different branes, there are open strings which can have one endpoint
in each brane. The mass of those states is proportional to the distance 
between branes; this means that there is an enhancement of symmetry for
coincident branes, which then should be described by some kind
of non-Abelian generalization of DBI.
\par
An interesting observation is that the collective motion of the
D-brane is already encoded in Buscher's formulas. To see this, note
that the dual backgrounds differ from the standard duals
without gauge fields by the terms
\bea
{\tilde G}_{0i} &= & \tilde{g}_{0i} -\tilde{g}_{00} \, \partial_i A_0 \; ,\nn\\
\tilde{G}_{ij} &=& \tilde{g}_{ij} + \tilde{g}_{00} \, 
  \partial_i A_0 \partial_j A_0 -
\tilde{g}_{0i} \,\partial_j A_0 -\tilde{g}_{0j}\,\partial_i A_0 \, ,\nn\\
\tilde{B}_{ij} &=& \tilde{b}_{ij} + F_{ij} 
+\tilde{b}_{0i} \, \partial_j A_0 - \tilde{b}_{0j}
\, \partial_i A_0 \, . 
\eea  
Using these formulas it can be checked that the dual
sigma model reduces to the standard one  in terms of the backgrounds
$\tilde{g}_{\mu\nu}$ and $\tilde{b}_{\mu\nu}$,
provided we make the replacement
\be
\tilde{x}^0 \rightarrow \tilde{x}^0 + A_0 (x^i).
\ee
Thus, the gauge field component $A_0$ acquires the dual interpretation
of the transverse position of the D-brane, as a function of $x^i$,  
which become longitudinal world-volume coordinates.
The same result
follows from a careful consideration of the boundary conditions. 
\par
In the case of unoriented strings, the change from Neumann to
Dirichlet conditions is supplemented by an orbifold projection in the
spacetime which reverses the orientation of the world-sheet (the 
orientifold). This result follows easily  from the
T-duality mapping at the level of the conformal field theory.
$\partial_z x(z) \rightarrow \partial_z x(z)$, $\partial_{\bar z}
x({\bar z}) \rightarrow - \partial_{\bar z} x({\bar z})$. In order to
study this question in a curved background, let us consider the 
lowest order unoriented  topology,
namely the crosscap  $P_2(\mathbb{R})$. It is enough for our purposes to make
in the boundary of the unit disc the identification of opposite points:
$x^{\mu} (\sigma) = x^{\mu} (\sigma+\pi)$. (Here $\sigma \in (0,2\pi)$
just parametrizes the boundary.)
Dropping the zero mode,
this yields the conditions 
(where the vectors $n$ and $t$ are the normalized outer
normal and tangent vector
to the world-sheet boundary   and,
correspondingly, $\partial_n \equiv n^a \partial_a$,
and $\partial_t \equiv t^b \partial_b$)
\footnote{Actually, in \cite{clny} the {\em boundary state} representing 
a closed string disappearing into the vacuum is constructed obeying
the boundary conditions
\be
\left. \frac{\partial}{\partial\tau} x(\sigma,\tau)\right|_{\tau=T} = 0
\ee
The {\em crosscap} boundary state is a modification of it, obtained by the 
appropriate identification on the boundary, {\it i.e.}
\bea
x(\sigma + \pi, \tau) &=& x(\sigma,\tau)|_{\tau=T} \; ,\nn\\
\frac{\partial}{\partial\tau} x(\sigma+\pi,\tau) &=& - 
\left. \frac{\partial}{\partial\tau} x(\sigma,\tau)\right|_{\tau=T}
\eea
Incidentally, it is quite easy to show that there are no
solutions with a plus sign
instead of a minus.}
\bea
\partial_n \, x^{\mu}(\sigma)&=& -\partial_n \,
x^{\mu}(\sigma + \pi)  \; ,\nn \\
\partial_t\,  x^{\mu}(\sigma) &=&\,  \partial_t 
\, x^{\mu}(\sigma + \pi) \; .
\eea
\par
When gauging the isometry, there 
is a covariant generalization of these conditions, namely
\bea
D_n x^{\mu}(\sigma) &=& - D_n x^{\mu}(\sigma + \pi)
\; ,\nn\\
D_t x^{\mu}(\sigma) &=&\, D_t x^{\mu}(\sigma + \pi) \; .
\eea
Using the value of $V_a^{cl}$ obtained above, we easily find, 
after fixing the $x^0 =0$ gauge
\be
D_n x^0 = - {1\over k^2}\partial_n x^i k_i +i\, \partial_t {\tilde x}^0
\; .
\ee
The antisymmetric tensor and Abelian gauge field backgrounds are
projected out from the physical spectrum of unoriented strings in the
weak field limit, and so we only consider a nontrivial metric
background. This then
yields the dual boundary conditions in the form
\be
\left(i\,\partial_t \tilde{x}^0 -{1\over k^2} k_j \partial_n x^j\right)
(\sigma+\pi)=
 -\left( i\,
\partial_t{\tilde x}^0 - {1\over k^2} k_j \partial_n x^j\right)
(\sigma) \, .
\ee
The terms containing the Killing vector cancel away
owing to the boundary
conditions of the original model;
the rest reduces to the orientifold condition on $\tilde{x}^0$,
\bea
\partial_t  {\tilde x}^0 (\sigma) &=&
-\partial_t  {\tilde x}^0 (\sigma + \pi) \; ,\nn \\  
\partial_n \tilde{x}^0 (\sigma) &=& \partial_n \tilde{x}^0 (\sigma + \pi) \, . 
\eea
The first equation implies $\tilde{x}^0 (\sigma) + \tilde{x}^0 (\sigma + \pi) =
{\rm constant}$, so that the crosscap is embedded as a twisted state
of the orbifold. 
This is quite important, because it implies that 
the orbifold character of the dual target-space is
a generic phenomenon, and not
a curious peculiarity of toroidal backgrounds.
It is curious to remark that the dual manifold always enjoys
parity ${\tilde x}^0 \rightarrow -{\tilde x}^0$ as an
isometry, because ${\tilde g}_{0i} =0$.
\par
The rest of the coordinates still satisfy standard crosscap
conditions. An important consequence of this
is that
at least two points of the boundary 
are mapped to the orientifold fixed points
in the target, which means that local contributions of 
non-orientable world-sheets 
are concentrated at the orientifold location;  
in the bulk of spacetime the
dual theory is orientable along the direction $\tilde{x}^0$ .  
This is compatible with the appearance of a non vanishing dual
antisymmetric tensor $\tilde{b}_{0i} = g_{0i} /g_{00}$ as long as the
original background has a ``boost" component. The effects of this
background field are supressed only for world-sheets mapped to the
fixed point. Another observation is that, in the absence
of a $U(1)$ gauge field, there is no collective coordinate for the
orientifold, which becomes a rigid object. Indeed, according to the
previous formulas, the induced backgrounds are exactly the
same as the vacuum dual backgrounds.
\par
The theory dual to Type I compactified
on a circle of radius R (often called $\tilde{I}$ or $I'$)
is then characterized by two orientifold planes and 16 8-branes.
Off the D-branes the orientation projection in $\tilde{I}$ does not
constrain the local state of the string, meaning that we have a Type II theory,
so that the vaccum without branes enjoys N=2 supersymmetry.
\par
On the other hand, the state containing the D-brane is only invariant under 
N=1 supersymmetry, so that it must be a BPS state. This means that it 
necessarily carries a conserved charge,
which in its turn is only possible
if this charge is of the RR type, consistent
in turn with the $\frac{1}{g}$ behaviour of the D-brane tension.
It was the realization of this fact by Polchinski that opened up all 
recent developments.
\par
Please note that from the involutive property of T-duality, $T^2 =1$,
and by interpreting that the 10-dimensional SO(32) Type I strings
have 32 D-9-branes, (filling the space) 
on which open strings can end we get the general rule that
\par
{\tt T-duality along a tangent direction maps D-p-branes into D-(p-1)-branes}
\par
whereas
\par
{\tt T-duality along a normal direction maps D-(p+1)-branes into D-p-branes}.

\noindent This then means that all branes are, in principle, related 
through T-duality.
\subsection{Physics on the brane (Born-Infeld) versus branes as sources}
It is not difficult to study the conditions for conformal invariance of
string theory in arbitrary backgrounds with k Dirichlet boundary conditions,
which we represent, following Leigh \cite{leigh}, as
\be
x^{\m}|_{\partial \Sigma} = f^{\m}(\eta^A)
\hspace{2cm}(A=1\ldots 26 - k) \; ,
\ee
where the boundary of the worldsheet, $\partial\Sigma$, is imbedded into a 
$(26-k)$ submanifold ${\cal M}$, with coordinates $\eta^{A}$, of the
target space. This condition is sufficient to ensure that the variation
$\delta x^{\mu}\mid_{\partial\Sigma}$ is tangent to ${\cal M}$, thus
imposing the $k$ Dirichlet conditions.
The action, in the conformal gauge, reads
\be
S = 
{1\over 4\pi
}\int_{\Sigma} (g_{\mu\nu}\eta^{ab} +i
\, b_{\mu\nu}\epsilon^{ab})\, \partial_a
x^{\mu}\partial_b x^{\nu} + \partial^a \rho \partial_a x^{\m} \partial_{\m}\phi(x)+
{i\over 2\pi}
\int_{\partial \Sigma}(\nu_{\m}(x)\partial_n x^{\m} - i A_B(\eta) 
\partial_{\tau}\eta^B),
\ee
where $\rho$ is the conformal factor of the 2-dimensional metric, 
the $\nu$'s are $k$ fields perpendicular to ${\cal M}$, as defined
by $\nu_{\mu}f^{\mu}(\eta)=0$, $A$ is a $U(1)$ field tangent\footnote{Although
it is introduced here as a Lagrange multiplier field, it is nowadays seen as
the background field of the open string sector
living on the D-brane.} to ${\cal M}$
and $\phi$, $g_{\mu\nu}$ and $b_{\mu\nu}$ are the usual background 
dilaton, metric and Kalb-Ramond field.
\par
By using Riemann normal coordinates $\xi^{\m}$ on spacetime, and $\zeta^A$
on M, the $\beta$-functions can be obtained by a slight modification of the
calculation in \cite{clny}
\bea
\beta_B (A) &=& -\frac{1}{2} (B +F)_B^C 
   \partial_C \phi + J^{AC}(B + F)_{AB;C}\nn\\
&&+ J^{AC}(\frac{1}{2} (B + F)_B^D H_{DAE} 
  (B + F)^E_C + K^{\m}_{BC}b_{\m\n} f^{\n}_{,C}) \; .
\eea
Here $ J^{AB} = (h - (B +F)^2)^{-1\mid AB}$,
and the normal coordinates expand as
$\xi^{\m}|_{\partial \Sigma} = f^{\m}_A \zeta^A + \frac{1}{2} K^{\m}_{AB} 
\zeta^A\zeta^B + \ldots $
The other $\beta$-functions are
\be
\beta_{\m}(\nu) = \frac{1}{2}\partial_{\m}\phi + 
J^{AC}(\frac{1}{2} (B + F)_C^B H_{\m AB} - K_{\m AC})\; .
\ee
It can be shown that the Dirac-Born-Infeld action \cite{art:ACNY}
\be
S_{DBI} = T' \int d^{26 - k} \eta e^{-\phi/2} \sqrt{det(h+B+F)} \; ,
\ee
gives equations of motion which are proportional to the beta functions above.
This can be easily generalized to the supersymmetric case.
\par
The non-Abelian generalization (corresponding to $N$ coincident D-branes)
is not known, but it is believed that in the 
low energy limit, the effective non-Abelian theory on the brane should be
the dimensional reduction of $N=1$ Super-Yang-Mills with gauge group $SU(N)$ 
in $d=10$ dimensions, 
to the appropriate world-volume of the brane, $W_p$.
To be specific
\be
S_p = - T_p\int tr \int_{W_p} d^P \xi e^{-\phi/2} \left( 
F_{\m\n}^2 + 2 (D_{\m}X^I)^2
+ [X^I,X^J]^2 \right) \; ,
\ee
All fields are $N\times N$ matrices; the eigenvalues of the matrices $X^I$
represent the positions of the $N$ $D-(p-1)$ branes, and the $U(1)$ describes
the overall center of mass.
\par
When studying the field equations of the effective supergravity theory, 
there are all kinds of extended solutions,
which can be grossly classified into
{\em elementary} (if they do not depend on the string coupling $g_s$ at all),
or {\em  solitonic} (if the energy goes as $\frac{1}{g_s^2}$); or, finally,
{\em Dirichlet} if their energy scales like $\frac{1}{g_s}$.
\par
There is a Hodge duality between these solutions.
{}For example: For N=1 supergravity
in d=10 there is the {\em elementary string} of Dabholkar
{\it et. al.} \cite{dabhol},
with a ten dimensional
(Einstein frame) metric of the type
\begin{eqnarray}
ds^2 &=& e^{\textstyle{\frac{3}{4}}(\phi -\phi_{0})} 
         \eta_{\m\n}dx^{\m}dx^{\n} - 
       e^{-\textstyle{\frac{1}{4}}(\phi -\phi_{0})} 
         \delta_{mn}dy^m dy^n \; , \nonumber \\
 B_{01} &=& -e^{\phi -\phi_{0}} \; , \nonumber \\
 e^{-\phi} &=& e^{-\phi_{0}}
                  \left( 1+\frac{k_{2}}{y^{6}}\right) 
 \hspace{1cm} ,\hspace{1cm}
 k_{2}\;=\; \frac{\kappa_{10}^{2}T_{2}}{3\Omega_{7}}e^{3\phi_{0}/4} \; .
\end{eqnarray}
where $\m\in (0,1)$; and $m,n\in (1,\ldots 9)$, $T_{2}$ is the string
tension and $\Omega_{n}$ is the volume of the $n$-dimensional unit sphere.
Given the fact that the solution has a timelike singularity \cite{stelle},
this BPS solution can be thought 
of having a delta function singularity at $y=0$; so it corresponds to an 
energy-momentum tensor with support on the world-sheet of the string.
\par
Correspondingly, there is also a {\em solitonic fivebrane}; a solution of the
source-free field equations of d=10 supergravity alone.
The Ansatz is as above, but with $\m\in (0\ldots 5)$; and $m,n\in (6,\ldots 9)$
There is now a $dH\neq 0$, which means that
there is a nontrivial magnetic charge
$g_6$. Explicitly the solution reads
\begin{eqnarray}
ds^{2} &=& e^{-(\phi -\phi_{0})/4}\eta_{\mu\nu}dx^{\mu}dx^{\nu} \,-\,
           e^{3(\phi -\phi_{0})/4}\delta_{mn}dy^{m}dy^{n} \; , \nonumber \\
H &=& 2k_{6}e^{\phi_{0}/4}\, \epsilon_{3} \; , \nonumber \\
e^{\phi} &=& e^{\phi_{0}}
                  \left( 1+\frac{k_{6}}{y^{2}}\right) 
 \hspace{1cm} ,\hspace{1cm}
 k_{6}\;=\; \frac{\kappa_{10} g_{6}}{\sqrt{2}\Omega_{3}}e^{-\phi_{0}/4} \; ,
\end{eqnarray}
where $\epsilon_{n}$ is the normalized volume form on $S^{n}$.
\par
There is also a dual version in which the r\^oles of 2 and 5 are reversed
(both in the branes and in the supergravity Lagrangians).
\section{The web of dualities and the strong coupling limit: 
Back to the beginning?}
There is now a certain amount of evidence for different
kinds of dualities (See for example \cite{rev:Mdual}), which
can be classified, following Schwarz, as {\em S-dualities}, {\em T-dualities},
or {\em U-dualities}. We shall say that two (not neccessarily different)
theories, $T_1$ and $T_2$ are T-dual, when $T_1$ compactified at large
Kaluza-Klein volume is physically equivalent to $T_2$ at small
Kaluza-Klein volume. If we call $t$ the modulus associated to global variations
of the Kaluza-Klein volume, by $Vol \sim e^t$, this implies a 
relationship of the general form
\be
t(1) \; =\;  - t(2) \; .
\ee
We have already seen how this comes about
in some simple cases from the sigma model
approach to string perturbation theory.
\par
S-duality, on the other hand, refers to 
the equivalence of $T_1$ at small coupling
with $T_2$ at large coupling. It demands for the dilaton something like
\be
\phi (1) \; = \; - \phi (2) \; ,
\ee
and, by definition, lies beyond the possibilities of verification by 
means of perturbation theory.
\par
U-duality is a kind of mixture of the two, and claims an equivalence of $T_1$
at large coupling with $T_2$ at small Kaluza-Klein volume.
In terms of fields,
\be
\phi (1) \; =\;  \pm t(2) \ .
\ee
\par
In these notes we shall only examine a few representative examples of this web.
It is still too early to assess the real meaning of this enormous symmetry.
\subsection{S-Duality for the heterotic string in $M_4\times T_6$}
Let us summarize here the clear analysis by A. Sen \cite{s}, reporting
mostly on joint work with J. Schwarz. We shall
present the two existing pieces of evidence, namely, the effective 
low-energy field theory, and the spectrum of masses and charges of those states
which are protected by supersymmetry from receiving quantum corrections.
\par
We start from the ten-dimensional action, which is the bosonic part
of the effective field theory of the heterotic string
\bea
S^{het}_{(d=10)} &=& \frac{1}{32\pi}\int d^{10} x 
\sqrt{- G^{(10)}} e^{-\phi^{(10)}}
\left[ R^{(10)} - G^{(10)\mu\nu}\partial_\mu \phi^{(10)}\partial_\nu
\phi^{(10)}\right. \nn\\
&&+\left. \textstyle{\frac{1}{12}} H^{(10)}_{\mu\nu\rho} H^{(10)\mu\nu\rho}
-\textstyle{\frac{1}{4}} F^{(10)I}_{\mu\nu} 
F^{(10)I\mu\nu}\right] \; ,
\eea
where $I = 1\ldots 16$ represent the Abelian fields in the Cartan subalgebra
of either $E_8\times E_8$ or $SO(32)$, which are the only ones which 
{\em generically} will remain massless upon 
compactification to four dimensions.
\par
Upon the simplest toroidal compactification, the effective four-dimensional
theory of the massless modes will be
\bea
S_{(4)} &=& \frac{1}{32\pi} \int d^4 x \sqrt{-G} e^{-\phi} 
\lbrack R(G) - G^{\m\n}\partial_{\m}\phi\partial_{\n}\phi 
+\frac{1}{12} G^{\m\a}G^{\n\b}G^{\rho\gamma} H_{\m\n\rho}H_{\a\b\gamma}\nn\\
&&-\textstyle{\frac{1}{4}} G^{\m\a}G^{\n\b}
     F^{(a)}_{\m\n}(LML)_{ab}F^{(b)}_{\a\b}
-\frac{1}{8}G^{\m\n}tr(\partial_{\m}M L\partial_{\n}M L)\rbrack \; .
\eea
Here the indices $a,b = 1\ldots 28$, where the $28$ is gotten from the $16$
that already existed in $d=10$, plus another $6$ coming from the metric 
$G^{(10)}$ 
compactified
on $T_6$, plus another $6$ coming from the $B^{(10)}$. The scalar fields have 
been conveniently packed into a matrix $M\in O(6,22)$, and the numerical
matrix $L$ is given by
\bea
L \;=\; 
\left(\begin{array}{ccc}
0&1_6&0\\
1_6&0&0\\
0&0&- 1_{16}
\end{array}\right)
\eea
\par
T-duality in this language is particularly transparent: Any $g\in O(6,22)$
({\it i.e.} such that $g^{T}Lg = L$)
acts by
\bea
M&\rightarrow& g M g^{T}\; ,\nn\\
A^a_{\m}&\rightarrow& {g^a}_b A^b_{\m}
\eea
(The rest of the fields being inert under T-duality).
The preceding four-dimensional action was written in the String frame.
It can be
rewritten in the Einstein frame through the rescaling
\be
g_{\m\n}\equiv e^{-\phi}G_{\m\n} \; .
\ee
It is also convenient to introduce the axion field, the Hodge dual
of the Kalb-Ramond field
\be
H^{\m\n\rho}= -\frac{1}{\sqrt{-g}} e^{2\phi} 
  \epsilon^{\m\n\rho\sigma}\partial_{\sigma}\psi \; .
\ee
The dilaton and the axion together constitute a complex scalar field
\be
\lambda \equiv \psi + i e^{-\phi} \; .
\ee
It is then a simple matter to check that the {\em equations of motion}
are invariant under $g\in SL(2,\mathbb{R})$,
characterized by four real numbers such that $ac- bd =1$, and 
constituting what is called an {\em S-duality} transformation \cite{ibanez}
\bea
\lambda&\rightarrow&\frac{a\lambda + b}{c\lambda + d}\; ,\nn\\
F^{(a)}_{\m\n}&\rightarrow&(c\lambda_1 + d)F^{(a)}_{\m\n} 
+ c\lambda_2 (ML)_{ab} \tilde{F}^{(b)}_{\m\n} \; ,
\eea
with all other fields remaining inert under S-duality,
and we have used $\lambda
\equiv\lambda_1 + i \lambda_2$.
\par
The action in the Einstein frame is given by
\bea
S^{(E)}_{(4)} &=& \frac{1}{32\pi} \int d^4 x \sqrt{- g} 
\left[ R(g) +\frac{1}{2\lambda_2^2} g^{\m\n}\partial_{\m}\lambda
\partial_{\n}\bar{\lambda}
- \textstyle{\frac{1}{4}}\lambda_2 F^{(a)}_{\m\n}(LML)_{ab}F^{(b)\m\n}
\right. \nn\\
&& +
\left. \textstyle{\frac{1}{4}}
  \lambda_1 F^{(a)}_{\m\n} L_{ab} \tilde{F}^{(b)\m\n}
-\frac{1}{8}g^{\m\n}tr(\partial_{\m}M L\partial_{\n}M L)\right] \; .
\eea
\par
It is a characteristic feat that this action is {\em not} S-dual invariant;
only
the equations of motion enjoy this property. It is also characteristic that 
there exists a dual form of the ten-dimensional effective
action (known already
from supergravity), using the seven form $(*H)_7$
instead of the three-form $H_3$.
The effective four-dimensional action it implies is
manifestly S-dual, although
only the equations of motion are then T-duality invariant.
\par
Another important feat is that quantum effects associated to the term
$\lambda_1 F^{(a)}_{\m\n} L_{ab} \tilde{F}^{(b)\m\n}$ break the classical
$SL(2,\mathbb{R})$ down to $SL(2,\mathbb{Z})$ (because $\lambda_1$ acts
as a generalized $\theta$ angle).
\par
Let us now consider the spectrum of charged particles in the theory.
In the presence of a current $J^{(a)}_{\m}$, whose conserved charge is
defined by $e^{(a)}\equiv \int d^3 x \sqrt{g} j^{(a)0}$, the asymptotics of
the radial 
electric fields changes to $F^{(a)}_{0r}\sim\frac{q^{(a)}_{el.}}{r^2}$
Using the equations of motion it can be shown that $q^{(a)}_{el.}=
\frac{1}{\lambda_2^{as} M^{as}_{ab} e^{(b)}}$, where $as$ stands for the
asymptotic value.
\par
We know, on the other hand \cite{narain},
that $e^{(a)}=\a^{(a)}\in\Lambda^{Narain}$, where $\Lambda^{Narain}$ is an even,
sel-dual, Lorentzian lattice with metric $L$.
Elementary strings states do not have any magnetic charge,
but other states will.
The Dirac quantization condition \cite{dirac} then forces 
$q^{(a)}_{mag.} = L_{ab} \beta^{(b)}$
where $\beta^{(b)}\in \Lambda^{Narain}$.
\par
Taking into account the modification of the quantization conditions
in the presence of a $\theta$ angle (Witten effect)\cite{wo}, the final allowed
spectrum is
\be
(q^{(a)}_{el},q^{(a)}_{mag})\equiv(\frac{1}{\lambda_2^{as}}
M_{ab}^{as} (\a^{(b)} + \lambda_1^{as}\b^{(b)}),L_{ab}\beta^{(b)}) \; ,
\ee
which is easily seen to be invariant under both 
$SL(2,\mathbb{Z})$ and $O(6,22;\mathbb{Z})$.
\par
A similar analysis shows that the masses of those particles sitting in
short multiplets of the supersymmetry algebra obey the formula
\bea
m^2\equiv \frac{1}{16}(\a^{(a)} \b^{(a)})\mathcal{M}^{as} (M^{as}
 + L)_{ab} \left(\begin{array}{c}\a^{(b)}\\
\b^{(b)}\end{array}\right) \; ,
\eea
where the matrix $\mathcal{M}$ is given by
\bea
\mathcal{M}\equiv\frac{1}{\lambda_2}\left(\begin{array}{cc}
1&\lambda_1\\
\lambda_1&|\lambda|^2\end{array}\right) \; ,
\eea
which is, again,
invariant under both $SL(2,\mathbb{Z})$ and $O(6,22;\mathbb{Z})$.
\subsection{The strong coupling limit of IIA strings, $SL(2,\mathbb{Z})$ 
duality of IIB strings and heterotic/Type I duality}
1.- If we are willing to make the hypothesis
that supersymmetry is not going to
be broken whilst increasing the coupling constant, $g_s$,
some astonishing conlusions
can be drawn. Assuming this, massless quanta can become massive
as $g_s$ grows only if their number, charges and spins are such that
they can combine into massive multiplets \cite{ht} 
(which are all larger than the irreducible massless ones). 
The only remaining issue, then, is whether
any other massless quanta can appear
at strong coupling.
\par
Now, as we have seen, in the IIA theory there are states associated to the RR
one form, $A_1$, namely the D-0-branes, whose tension goes as 
$m\sim\frac{1}{g_s}$. This clearly gives new massless states in the strong 
coupling limit.
\par
There are reasons\footnote{In particular: The fact
that there is the possibility of a central extension in the IIA algebra,
related to the Kaluza-Klein compactification of the d=11 Supergravity
algebra.} to think that this new massless states  are the first
level of a Kaluza-Klein tower associated to compactification on a circle
of an 11-dimensional theory.
Actually, assuming an 11-dimensional spacetime with an isometry 
$k=\frac{\partial}{\partial y}$, an Ansatz which exactly reproduces the 
dilaton factors of the IIA string is
\be
ds^{2}_{(11)} = e^{\frac{4}{3}\phi}(dy - A^{(1)}_{\m} dx^{\m})^2 + 
e^{-\frac{2}{3}\phi}g_{\m\n}dx^{\m}dx^{\n} \; .
\ee 
Equating the two expressions for the D0 mass,
\be
\frac{1}{g_s} = \frac{1}{R_{11}} \; ,
\ee
leads to $R_{11} = e^{\frac{2}{3}\phi} = g_A^{2/3}$.
\par
2.- All supermultiplets of massive one-particle states of the IIB 
supersymmetry algebra contain states of at least spin 4. 
This means that under the previous set of hypothesis, the set of 
massless states at weak coupling must
be exactly the same as the corresponding set at strong coupling.
This means that there must be a symmetry mapping weak coupling into 
strong coupling.
\par
There is a well-known candidate for this symmetry:
Let us call, as usual, $l$ the RR scalar and $\phi$ the dilaton (NSNS).
We can pack them together into complex scalar
\be
S \;=\;  l \,+\, i e^{-\frac{\phi}{2}} \; .
\ee
The IIB supergravity action in d=10 is invariant under the $SL(2,\mathbb{R})$
transformations
\be
S \rightarrow \frac{a S + b}{ c S + d} \; ,
\ee
if at the same time the two two-forms, $B_{\m\n}$ (the usual, ever-present,
NS field), and $A^{(2)}$, the RR field transform as
\bea
\left( \begin{array}{c}
B\\
A^{(2)}
\end{array}\right ) \rightarrow 
\left( \begin{array}{cc}
d & - c \\
- b & a\end{array}\right )
\left ( \begin{array}{c}
B\\
A^{(2)}
\end{array}\right) \; , 
\eea
Both the, Einstein frame, metric $g_{\m\n}$
and the four-form $A^{(4)}$ are inert under this
$SL(2,\mathbb{R})$ transformation.
\par
A discrete subgroup $SL(2,\mathbb{Z})$ of the full classical $SL(2,\mathbb{R})$
is believed to be an exact symmetry of the full string theory. 
The exact imbedding of the discrete subgroup in the full $SL(2,\mathbb{R})$ 
depends on the
vacuum expectation value of the RR scalar.
\par
The particular transformation
\bea
g = \left(
\begin{array}{cc}
 0& 1\\
- 1& 0\end{array}
\right) \; ,
\eea
maps $\phi$ into $ - \phi$ (when $l =0$),
and $B$ into $A^{(2)}$. This is then an 
S-duality type of transformation, mapping the ordinary string with NS charge,
to another string with RR charge (which then must be a D-1-brane,
and is correspondingly called a {\em D-string}), and, from there, 
is connected to all other D-branes by T-duality.
\par
Using the fact that upon compactification on $S^1$, IIA at $R_A$ is
equivalent to IIB at $R_B\equiv 1/R_A$, and the fact that the effective action
carries a factor of $e^{-2\phi}$ we get
\be
R_A g_B^2 = R_B g_A^2 \; ,
\ee
which combined with our previous result, $g_A = R_{11}^{3/2}$ implies 
that $ g_B = \frac{R_{11}^{3/2}}{R_A}$.
Now the Kaluza-Klein Ansatz implies that from the eleven dimensional 
viewpoint the compactification radius is measured as
\be
R_{10}^2 \equiv R_A^2 e^{- 2\phi/3} \; ,
\ee
yielding
\be
g_B = \frac{R_{11}}{R_{10}} \; .
\ee
\par
3.- From the effective actions written above it is easy to check 
that there is a (S-duality type) field transformation mapping
the SO(32) Type I open string into the SO(32)
Heterotic one namely
\bea
g_{\m\n}&&\rightarrow e^{-\phi} g_{\m\n}^{Het} \; ,\nn\\
\phi&&\rightarrow - \phi \; ,\nn\\
B'&&\rightarrow B \; .
\eea
This means that physically there is a strong/weak coupling duality, because 
coupling constants of the compactified theories  would
be related by
\bea
g_{het}&& = 1/g_I \; ,\nn\\
R_{het}&& = R_I/g_I^{1/2} \; .
\eea
\subsection{Statistical Interpretation of the Black Hole Entropy}
The fact that the area of (the horizon of) a black hole can be interpreted as a
kind of entropy
was actually first discovered through an
analogy between the equations of black hole physics
and the equations of ordinary thermodynamics \cite{b}.
\par
Hawking's astonishing discovery that even the `{\em dead}' Schwarzschild black
hole
radiates with a black body spectrum
led to a much firmer identification of the entropy as
\be
S_{BH}\equiv \frac{1}{4}\frac{A c^3}{G_4\hbar}
\ee
where $A$ is the area of the black hole's horizon and $G_4$ is the
four-dimensional Newton constant. Furthermore one finds in case of the
four-dimensional Schwarzschild black hole that
\be
A\equiv 4\pi(2 G_4 M)^2 \, ,
\ee
where $M$ is the mass of the black hole.
\par
The problem as to whether a statistical interpretation of
this entropy (as the logarithm
of a corresponding density of states) exists at all
is undoubtebly one of the most important open problems
in the whole topic of gravitational physics.
\par
Recently ({\it cf.} \cite{h} for an introductory review),
there has been some progress
in understanding the counting of states, albeit not in
the physically most interesting cases,
but rather for {\em extremal} black holes;
that is, holes such that the charge is as big as it can be
in a way consistent with the Cosmic Censorship Hypothesis
(that is, without creating a
naked singularity).
These black holes (which can often be considered BPS states)
are usually uninteresting, because they have
zero Hawking temperature and, in addition,
those which can be embedded into the low
energy limit of string theory, usually have singular dilaton behaviour.
But it was pointed out in \cite{tomas} that
in some cases, with several charges, the horizon stays nonsingular.
It is exactly for this
case that one can give a {\em microscopical} interpretation of the 
black hole entropy.
\par
The main idea which makes the counting feasible is first of all,
the fact that, as stems easily from Eq. (2.100),
the Newton constant in $d$ dimensions is given by
\be
G_d \;\sim \;\frac{g_s^2 l_s^8}{V_{10-d}} \, ,
\ee
where numerical factors have been ignored.
This fact on itself means that the strength of the
gravitational coupling, measured roughly by
\be
G_d M \, \sim \, g_s^2 M
\ee
is small when $g_s\rightarrow 0$ as long as the mass does
not grow faster than $g_s^{-2}$.
This in its turn implies that there must exist some
weak coupling description of these states
which clearly will consist in an appropiate set of D-branes with corresponding
RR charges.\footnote{Technical
complications related to the necessity (briefly alluded to above)
of having a finite-area horizon in the extremal limit
imply that in four dimensions one needs at least four different charges}
Their BPS property
implies that, as we make the string coupling constant $g_s$ grow,
the degeneracy of the states does not change. But in doing so,
we change from a perturbative
description in terms of branes, to a non-perturbative black hole
configuration.\footnote{The turning point being clearly when the
curvature is of the order of $l_s^{-2}$}
\par
This is the first time a statistical interpretation
of the entropy of a black hole in more than three
dimensions is obtained, and as such, is one of the most important
applications of D-brane physics.\footnote{It is ironic
to remark in this respect that
shortly {\em before} the statistical interpretation was proposed, it was
{\em proven} (using euclidean regularity arguments) \cite{curioso}
that extremal black holes should have zero entropy even
in those cases in which they enjoyed non-zero area.}
There are, however, essential complications to treat non extremal black
holes using
this set of ideas (except in the case where they are close to extremality).
A useful, quite detailed, general reference is \cite{malda}.
\section{Concluding remarks}
It is probably fair to say that most fundamental questions on quantum gravity
are still waiting to be answered. Many previously unsuspected relationships
between ordinary gauge theories and gravity are now appearing, however, and, 
everything points in the direction of a much more unified and symmetric 
fundamental theory than was previously thought to be the case.
\par
It can only be hoped that specific and concrete experimental predictions
of the theory can be made in the near future. 
\section*{Acknowledgements}
E.A. is indebted to Luis \'Alvarez-Gaum\'e and 
C\'esar G\'omez for countless discussions all over the years. 
The authors would like to thank Tom\'as Ort\'{\i}n for carefully
test-reading the manuscript and for many helpful comments.
We are also grateful to the referee for a very useful report and for 
pointing out several mistakes.
The list of references is not intended to be comprehensive; it only includes 
those items familiar to us, which we thought could be useful for the
beginner when starting out.
This work has been supported by EU contracts ERBFMRX-CT96-0012 and
ERBFMBI-CT96-0616 and by CICYT grants AEN/96/1664 and AEN/96/1655.
\addcontentsline{toc}{section}{References}

\end{document}